\documentclass[pdflatex,11pt]{article}
\usepackage{graphicx,natbib,rotating,bm,array,amssymb,amsmath}

\newcommand{\blind}{0} 

\usepackage{color} 
\usepackage[pdftex, bookmarksopen=true, bookmarksnumbered=true,
pdfstartview=FitH, breaklinks=true, urlbordercolor={0 1 0},
citebordercolor={0 0 1}]{hyperref}  
    
\begin{document} 
   
\bibliographystyle{natbib} 
\cite{}
    
\def\iid{\buildrel\rm iid\over \sim}       
\def\ind{\stackrel{\rm ind}{\sim}} 
   
\def\ignore{[1]{}}
\def\J{{\cal J}}
\def\I{{\cal I}}
\def\L{{\cal L}}
\def\K{{\cal K}}
\def\cur{^{(t)}}
\def\nex{^{(t+1)}}
\newcommand\bX{{\mathbf X}}

\def\alc{\alpha}
\def\itoN{^N_{i=1}}
\def\iton{^n_{i=1}}
\def\jtoJ{^J_{j=1}}
\def\th{\theta}
\def\inv{^{-1}}
\def\r{\right}
\def\l{\left}
\def\Var{{\rm Var}}
\def\E{{\rm E}}

\newcommand{\pscore}{{\large\sc p-score}}
\newcommand{\pfun}{{\large\sc p-function}}
\newcommand{\gps}{{\large\sc gps}}
\newcommand{\scm}{{\large\sc scm}}
\newcommand{\iw}{\large{\sc iw}}
\newcommand{\drf}{{\large\sc drf}}
\newcommand{\drfs}{{\large\sc drf}{\small s}}
\newcommand{\hi}{\textsc{\small HI}}
\newcommand{\ivd}{\textsc{\small IvD}}
\newcommand{\ffgn}{\textsc{\small FFGN}}
\newcommand{\rr}{{\small\sc RR}}
\newcommand{\mhi}{_{\sc HI}}
\newcommand{\mivd}{_{\sc IvD}}
\newcommand{\mrr}{_{\sc RR}}
\newcommand{\indep}{\mbox{$\perp\!\!\!\perp$}}

\renewcommand\floatpagefraction{.9}
\renewcommand\topfraction{.9}
\renewcommand\bottomfraction{.9}
\renewcommand\textfraction{.1}   
\setcounter{totalnumber}{50}
\setcounter{topnumber}{50}
\setcounter{bottomnumber}{50}

\def\spacingset#1{\renewcommand{\baselinestretch}%
{#1}\small\normalsize}

\title{\bf\Large Causal Inference in Observational Studies with Non-Binary Treatments}
\spacingset{1.1}

\if0\blind
\author{\large
   Shandong Zhao\\
  \normalsize
  Department of Statistics, University of California,  Irvine, CA 92697\\
  \normalsize
  \textit{shandonm@uci.edu}
  \and\large   David A. van Dyk\\ 
\normalsize
Statistics Section, Department of Mathematics, Imperial College London, SW7 2AZ\\
\normalsize
  \textit{dvandyk@imperial.ac.uk}
\and\large Kosuke Imai\\
  \normalsize
  Department of Politics, Princeton University, Princeton, NJ  08544\\
  \normalsize
  \textit{kimai@princeton.edu}
}
  
  \fi
  
\date{
\today}
\maketitle

\abstract Propensity score methods have become a part of the standard
toolkit for applied researchers who wish to ascertain causal effects
from observational data.  While they were originally developed for
binary treatments, several researchers have proposed generalizations
of the propensity score methodology for non-binary treatment regimes.
Such extensions have widened the applicability of propensity score
methods and are indeed becoming increasingly popular themselves.  In
this article, we closely examine the two main generalizations of
propensity score methods, namely, the propensity function (\pfun) of
\citet{imai:vand:04} and the generalized propensity score (\gps) of
\citet{hira:imbe:04}, {along with recent extensions of the \gps~that
  aim to improve its robustness.} We compare the assumptions,
theoretical properties, and empirical performance of these alternative
methodologies.  On a theoretical level, the \gps\ and its extensions
are advantageous in that they can be used to estimate the full dose
response function rather than the simple average treatment effect that
is typically estimated with the \pfun. Unfortunately, our analysis
  shows that in practice response models often used with the original
  \gps\ are less flexible than those typically used with propensity
  score methods and are prone to misspecification. We compare new and
  existing methods that improve the robustness of the \gps\ 
  and propose methods that use the
  \pfun\ to estimate the dose response function. We illustrate our
findings and proposals through simulation studies, including one based
on an empirical application.

\bigskip

  
\spacingset{1.5}

\section{Introduction}
\label{sec:intro}
  
Adjusting for observed confounding variables is one of the most common
strategies used across numerous scientific disciplines when making
causal infererence in observational studies. Researchers find that the
results based on regression adjustments can be sensitive to model
specification when applied to the data where the treatment and control
groups differ substantially in terms of their pre-treatment
covariates. The propensity score methods of \citet{rose:rubi:83},
hereafter \rr, aim to address this fundamental problem by reducing the
covariate imbalance between the two groups. \rr\ showed that under the
assumption of no unmeasured confounding, adjusting for propensity
score, rather than potentially high-dimensional covariates, is
sufficient for unbiased estimation of causal effects and this can be
done by simple nonparametric methods such as matching and
subclassification.

Despite their popularity, one limitation of the original propensity
score methods is that they are only applicable to a binary treatment.
About a decade ago, several researchers proposed generalization of the
propensity score methodology for non-binary treatment regimes
\citep{imbe:00, hira:imbe:04, imai:vand:04}.  Such extensions have
widened the applicability of propensity score methods and are indeed
becoming increasingly popular themselves, Google Scholar citation
counts of the aforementioned papers are 776, 227, and 310,
respectively, as of August 8, 2013). Particularly novel applications
appear in \citet{erte:step:10} and \citet{mood:step:12}.

All of these methods, however, require users to overcome the
challenges of first correctly modeling a treatment variable as a
function of a possibly large number of pre-treatment covariates and
second modeling the response variable.  These represent significant
difficulties in practice.  Standard diagnostics based on the
comparison of the covariate distributions between the treatment and
control groups are not directly applicable to non-binary treatment
regimes and the final inference can be quite sensitive to the choice
of response model. \citet*{flor:flor:gonz:Neum:2012}, hereafter \ffgn,
propose two extensions to the method of Hirano and Imbens that aim to
provide more robust estimation through a move flexible response model.

In this paper, we closely examine the two main generalization of
propensity score methods, namely, the propensity function (\pfun) of
\citet{imai:vand:04}, hereafter \ivd, and the generalized propensity
score (\gps) of \citet{hira:imbe:04}, hereafter \hi, {along with the
  \ffgn\ extensions.}  We compare the assumptions and theoretical
properties of these two alternative methodologies and examine their
empirical performance in practice.  In Section~\ref{sec:meth}, we
review the theoretical properties of the original propensity score
methodology and its generalizations.  The {\hi\ method} has a
theoretical advantage over the {\ivd\ method} in that the former can
be used to estimate the full dose response function (\drf) rather than
the simple average treatment effect, which is often estimated with
{\ivd's method}. 
In Section~\ref{sec:comp}, we compare {the method of \ivd, \hi, and
  \ffgn\ }both theoretically and empirically, using a pair of simple
simulation studies. {We demonstrate that the response model used by
  \hi\ is less flexible than those typically used with propensity
  score methods and that the methods proposed by \ffgn\ to address
  this can exhibit undesirable properties.} In
Section~\ref{sec:improve}, we compare these methods with a new
proposal and show how the method of \ivd\ can be extended for robust
estimation of the full \drf. The efficacy of the proposed methodology
is illustrated through the aforementioned simulation studies in
Section~\ref{sec:improve} and an empirically-based study in
Section~\ref{sec:num}. Section~\ref{sec:disc} offers concluding
remarks {and an Appendix introduces a robust variant of \hi's method.}

\section{Methods}
\label{sec:meth}       
      
Suppose we have a simple random sample of size $n$ with each unit
consisting of a $p$-dimensional column vector of pretreatment
covariates, ${\bm X}_i$, the observed univariate treatment, $T_i$, and
the outcome variable, $Y_i$. Although {\ivd's method} can be
applied to multivariate treatments, here we assume the treatment is
univariate to facilitate comparison with {\hi's method}. We omit the
subscript when referring to generic values of ${\bm X}_i$, $T_i$, and
$Y_i$.

We denote the potential outcomes by $\mathcal Y =\{Y_i(t),t \in
\mathcal T$ for $i=1,\ldots,n\}$, where $\mathcal T$ is a set of
possible treatment values and $Y_i(t)$ is a function that maps a
particular treatment level of unit $i$, to its outcome.  This setup
implies the {\it stable unit treatment value assumption}
\citep{rubi:90} that the potential outcome of each unit is not a
function of treatment level of other units and that the same version
of treatment is applied to all units.  In addition, we assume {\it
  strong ignorability of treamtent assignment}, i.e., $Y(t)\ \indep \
T \mid {\bm X}$ and $p(T = t \mid {\bm X}) > 0$ for all $t \in
\mathcal T$, which implies no unmeasured confounding (\rr).

\subsection{The propensity score with a binary treatment}
\label{sec:pscore}

\rr\ considered the case of treatment variables that take on only two
values, ${\mathcal T} =\{0,1\}$, where $T_i = 1$ ($T_i = 0$) implies
that unit $i$ receives (does not receive) the treatment and defined
the {\it propensity score} to be the conditional probability of
assignment to treatment given the observed covariates, i.e., $e({\bm
  X})=p(T=1\mid {\bm X})$.  In practice, $e({\bm X})$ is typically
estimated using a parametric treatment assignment model
$p_\psi(T=1\mid {\bm X})$ where $\psi$ is a vector of unknown
parameters.  The appropriateness of the fitted model can be assessed
via the celebrated balancing property of $e({\bm X})$, namely, that
covariates should be independent of the treatment conditional on the
propensity score, ${\bm X}\ \indep \ T \ | \ e({\bm X})$. In
particular, the fitted model, ${\hat e}({\bm X}) =
p_{\hat\psi}(T=1|{\bm X})$ should not be accepted unless 
adjusting for $\hat e({\bm X})$ results in adequate balance.
 
In order to estimate causal quantities, we must properly adjust for
${\hat e}({\bm X})$. \rr\ propose three techniques: matching,
subclassification, and covariance adjustment. Here we focus on
subclassification and covariance adjustment because they are more
closely related to the generalization of \ivd\ and \hi. The key advantage
of propensity scores when applying these methods is the dimension
reduction, requiring that we only adjust for a scalar variable ${\hat
  e}({\bm X})$, rather than the entire covariate vector, which is
often of high dimension. 

With subclassification (\rr), we adjust for ${\hat
  e}({\bm X})$ by dividing the observations into several subclasses
based on ${\hat e}({\bm X})$. Individual response models are then
fitted within each subclass, adjusting for ${\hat e}({\bm X})$ and
sometimes ${\bm X}$ along with $T$. The overall causal effect is then
computed as the weighted average of the within-class coefficients of
$T$, with weights proportional to the size of subclass. The standard
error of the causal effect is computed typically by treating the
within-subclass estimates as independent of one another.

With covariance adjustment (\rr), we regress the
response variable on ${\hat e}({\bm X})$ separately for the treatment
and control groups.  Specifically, we divide the data into the
treatment and control groups and fit the regression model, $\E(Y \mid
{\bm X}, T = t) = \alpha_t + \beta_t \cdot {\hat e}({\bm X})$,
seperately for $t=0,1$. The average causal effect is then estimated as
\begin{equation}
  (\hat\alpha_1-\hat\alpha_0)+(\hat\beta_1-\hat\beta_0) \cdot \overline{\hat e({\bm X})}
  \label{eq:cov-adj}
\end{equation}
where $\overline{\hat e({\bm X})}$ is the sample mean of the estimated 
propensity score.    

In addition to the techniques in \rr, inverse propensity score
weighting can be used to estimate causal quantities
\citep[e.g.,][]{rose:87,robi:98,robi:hern:brum:00,imbe:00}.  Because
the following equalities
\[
	{\rm E} \bigg{\{} \frac{TY}{e({\bm X})} \bigg{\}} = {\rm
          E}\{Y(1) \} \quad {\rm
	and} \quad {\rm E} \bigg{\{} \frac{(1-T)Y}{1-e({\bm X})} \bigg{\}}={\rm E}\{Y(0)
	\},
\]
hold and the inverse weighting estimate,
\[
	\sum\limits_{i=1}^{N} \bigg ( \ \frac{T_i Y_i}{\hat e({\bm
	X}_i)}-\frac{(1-T_i)Y_i}{1-\hat e({\bm X_i})} \  \bigg )
\]
is an unbiased estimate of the average causal effect. Covariance
adjustment and inverse weighting must be used cautiously as the scalar
estimated propensity score replaces the full set of the covariates
{\citep{rubi:04}}. In particular, inverse weighting can be quite
unstable in practice \citep[see e.g.,][as well as
Sections~\ref{sec:sim-one} and \ref{sec:sim-two}]{kang:scha:07}.

Matching, subclassification, covariance adjustment, {and inverse
  weighting} all aim to provide robust flexible adjustment for $\hat
e({\bm X})$ in the response model.  As we shall see below, the
flexibility of the response model is important especially for
non-binary treatment regimes.  This is because unlike the treatment
assignment model which has an effective diagnostic tool based on the
balancing property of propensity score, the response model lacks such
diagnostics.

\subsection{Generalizations of propensity score: the {\small GPS} and
  the {\small P-FUNCTION}}

\label{sec:gen_pscore}

Suppose now that $\mathcal{T}$ is a more general set of treatment
values, perhaps categorical or continuous. It is in this setting that
\ivd~introduced the \pfun\ and that \hi~introduced the \gps\
(\ivd~also allow for multi-variate treatments, which we do not discuss
in this paper).  In what follows, we review and compare these
generalizations of propensity score methods.  In particular, we
consider the following aspects of propensity score adjustment with
binary treatments that both \ivd\ and \hi\ generalize:

\begin{enumerate}
\item Treatment assignment model: Model the distribution of the
  treatment assignment given covariates to estimate the propensity
  score, i.e., $\hat{e}({\bm X})$

\item Diagnostics: Validate ${\hat e}({\bm X})$, by checking for
  covariate balance, i.e., $T \indep {\bm X} \mid \hat{e}({\bm X})$

\item Response model: Model the distribution of the response given the
  treatment, adjusting for ${\hat e}({\bm X})$ via matching,
  subclassification, covariance adjustment, or inverse weighting
  
\item Causal quantities of interest: Estimate the causal quantities of
  interest and their standard error based on the fitted response model
\end{enumerate}

\noindent {\bf Treatment assignment model.} As in the case of the
binary treatment, we begin by modeling the distribution of the
observed treatment assignment given the covariates using a parametric
model, $p_{\psi}(T \mid \bm{X})$, {where $\psi$ is the parameter.} Common
choices of $p_\psi(T|{\bm X} )$ include the Gaussian or multinomial regression 
models when the treatment variable is
continuous or categorical, respectively. \hi\ define the \gps\ as $R =
r(T, {\bm X}) = p_{\psi}(T \mid \bm{X})$. That is, the \gps\ is equal
to the treatment assignment model density {\it evaluated} at the
observed treatment variable and covariate for a particular
individual. This is analogous to the propensity score for the binary
treatment, which can be written as $e({\bm X}) = r(1, {\bm X}) = p_\psi(T=1\mid{\bm X})$.

\ivd, on the other hand, define the \pfun\ to be the entire conditional
probability density (or mass) function of the treatment, namely $e_\psi(\cdot
\mid {\bm X})=p_\psi(\cdot \mid {\bm X})$.  This is also analogous to the
propensity score for the binary treatment case because $e_\psi(\cdot \mid {\bm
X})$ is completely determined by $e({\bm X}) = p_\psi(T = 1 \mid {\bm X})$.
In order to summarize the \pfun, \ivd\ introduce the {\it uniquely
  parameterized propensity function} assumption which states that for
every value of ${\bm X}$, there exists a unique finite-dimensional parameter,
${\bm \theta} \in \Theta$, such that $e_\psi(\cdot\mid{\bm
  X})$ depends on ${\bm X}$ only through ${\bm \theta}_\psi({\bm X})$. In other
  words, ${\bm \theta}$ uniquely represents
$e\{\cdot\mid {\bm \theta}_\psi({\bm X)\}}$, which we may therefore write as
$e(\cdot|{\bm \theta})$ or simply ${\bm\theta} = {\bm
  \theta}_{\psi}({\bm X_i})$. For example, if we use a normal linear model for
  the treatment, $T_i\sim {\cal N}({\bm X}_i^\top{\bm\beta},
  \sigma^2)$ with $\psi=({\bm \beta}, \sigma^2)$, then $e_\psi(\cdot\mid{\bm
  X_i})$  is uniquely represented by the scalar ${\bm \theta_i}={\bm
  X}_i^\top{\bm\beta}$. {In practice, $\psi$, $\bm\theta_i$, $R_i$, and $r_i$
  are estimated by $\hat\psi$, $\bm{\hat\theta}_i$, $\hat R_i$, and $\hat r_i$}

\medskip
\noindent
{\bf Diagnostics.}
Diagnostics for the treatment assignment model rely on balancing
propereties of the \pfun\ and \gps. In particular, \ivd\ shows that
the \pfun\ is a balancing score, i.e., $T \ \indep \ \bm X \mid
e(\cdot \mid {\bm\theta})$. \ivd\ suggest checking balance by
regressing each covariate on $T$ and $\hat{\bm\theta}$, e.g., using
Gaussian and/or logistic regression and comparing the distribution of
the $t$-statistics for each of the resulting regression coefficients of
$T$ with the standard normal distribution via a normal quantile
plot. Improvement in balance can be assessed by constructing the plot
again in the same manner except that $\hat{\bm\theta}$ is left out of
each regression. Although not typical performed, this diagnostic is
equally applicable in the binary treatment case, but with
$\hat{\bm\theta}$ replaced by $\hat e(\bm X)$.

\hi, on the other hand, show that $\bm{1}\{T = t\}$ is independent of
${\bm X}$ given $r(t,{\bm X})$, where $\bm{1}\{\cdot\}$ is an
indicator function and the \gps\ is evaluated at $t \in \mathcal{T}$.
Following the covariate balancing property for the binary propensity
score, \hi\ construct a series of binary treatments by coarsening the
original treatment $T$ in the form $\{t_j < T \leq t_{j+1}\}$ for some
$t_1,t_2,\dots,t_J$. Covariate balance is then checked for these
binary treatment variables by first subclassifying units on $\hat
r(\widetilde{T}_j, {\bm X})$, where $\widetilde{T}_j$ is the median of
the treatment variable among units with $\bm{1}\{t_j < T \leq
t_{j+1}\}=1$. Then, two-sample $t$-tests are performed within each
subclass to compare the mean of each covariate among units with
$\bm{1}\{t_j < T \leq t_{j+1}\}=0$ against that among units with
$\bm{1}\{t_j < T \leq t_{j+1}\}=1$. Finally the within-subclass
differences in means and the variances of these differences are
combined to compute a single t-statistic for each covariate. \hi\
suggest this diagnostics be repeated for several choices of $\{t_1,
\dots, t_J\}$ that cover the range of observed treatment assignment
variable $T$.  

In both cases, we note that the failure to reject the null hypothesis
of perfect balance does not necessarily imply the lack of balance and
hence these diagnostics need to be interpreted with great care.  In
fact, it may be the case that covariate balance is not desirable but a
smaller sample size limits the ability to detect imbalance
\citep{imai:king:stua:08}.

\medskip
\noindent {\bf Response model.} The response models proposed by \ivd\
and \hi\ are quite different, with \hi\ relying more heavily on parametric
assumptions. \ivd\ propose two response models. The first is completely
analogous to the subclassification technique proposed by \rr. Individual
response models are fitted within each subclass, adjusting for $\hat{\bm\theta}$
and {typically} ${\bm X}$ along with $T$.  The second is a smooth
coefficient model (\scm), which allows the intercept and slope to vary smoothly as a
function of the \pfun\ 
\begin{equation}  
 E(Y \mid T, \hat{\bm \theta}) \ = \ f({\hat {\bm \theta}})+g({\hat {\bm \theta}})\cdot T,
\label{eq:IvD-smooth-coef}  
\end{equation}
where $f(\cdot)$ and $g(\cdot)$ are unknown but smooth continuous
functions.  {In our numerical illustrations, we} fit this model using
the R package {\tt mgcv} developed by Simon Wood, in which smooth
functions are represented as a weighted sum of known basis functions;
and the likelihood is maximized with an added smoothness penalization
term.  We use penalized cubic regression splines as the basis
functions, with dimension equal to five.
    

In contrast, \hi\ propose to estimate the conditional expectation of
the response as a function of the observed treatment, $T$, and the
\gps, $\hat R$. They recommend using a flexible parametric function of
the two arguments and give the following Gaussian quadratic regression
model,
\begin{equation}
  E(Y\mid T,\hat R) \ = \ \alpha_0+\alpha_1 \cdot T+\alpha_2 \cdot T^2 + \alpha_3
  \cdot \hat R + \alpha_4 \cdot {\hat R}^2 + \alpha_5 \cdot T \cdot \hat R.
\label{eq:gps-r-model}
\end{equation}
This can be viewed as a generalization of \rr's covariance adjustment technique,
which in the binary treatment case involves regressing $Y$ on ${\hat e}({\bm
X})$ separately for the treatment and control groups. \hi, on the other hand,
parametrically estimate the average outcome for all possible treatment levels
simultaneously via the quadratic regression on $T$ given in
\eqref{eq:gps-r-model}.

\medskip
\noindent {\bf Estimating causal quantities of interest.}  \ivd~and
\hi~aim to estimate different causal quantities, \ivd~the average
causal effect and \hi\ the \drf.  Computing the average causal effect
under the \scm, involves averaging $g({\hat {\bm \theta_i}})$ across
all units. Bootstrap standard errors are computed by resampling the
data and refitting both the treatment assignment and response models.
With subclassification, computing the estimated average causal effect
proceeds exactly as in the binary case.  Because a response model is
fit conditional on $T$ within each subclass, we can in principle
average these fitted models and estimate the \drf. While we illustrate
this possibility in our simulations, we advocate a flexible
non-parametric approach in Section~\ref{sec:pfun-drf}.

In contrast, to estimate the \drf, \hi\ computes the
average potential outcome on a grid of treatment values. In
particular, at treatment level $t$, they compute 
\begin{equation}
{\hat E} \{Y(t)\} \ = \ \frac{1}{n} \sum\limits_{i=1}^{n} \Big(
\hat\alpha_0+\hat\alpha_1 \cdot t+\hat\alpha_2 \cdot t^2 + \hat\alpha_3
\cdot \hat r(t,{\bm X_i}) + \hat\alpha_4 \cdot \hat r(t,{\bm X_i})^2 + \hat
\alpha_5 \cdot t \cdot \hat r(t,{\bm X_i}) \Big). 
\label{eq:dose-resp}
\end{equation}
Standard errors can be calculated using the bootstrap, taking into account the
estimation of both the \gps\ and model parameters. In
practice, we are often interested in the {\it relative \drf}, $E \{Y(t) -
Y(0)\}$, which compares the average outcome under each treatment level with that
under the control, i.e., $t=0$. Of course, in some studies there is no control
{\it per se} and we revert to $E \{Y(t)\}$. In our simulation studies we report
the relative \drf\ while in our applied example we report the \drf\ which is
more appropriate in its particular context.

\medskip
\noindent {\bf The {\footnotesize FFGN} extensions to the method of
  {\footnotesize HI}.}  Unfortunately, the quadratic regression in
(\ref{eq:gps-r-model}) is less flexible than either subclassification
or a \scm\ (see Section~\ref{sec:comp-theory}).
{\citet{bia:flor:matt:01} and \ffgn\ } point out that misspecification
of (\ref{eq:gps-r-model}) can result in biased causal quantites and
\ffgn\ proposes two alternatives. The first generalizes
(\ref{eq:gps-r-model}) with,
\begin{equation}
  E(Y\mid T,\hat R) \ = \ \beta(T,\hat R) 
\label{eq:scm-gps-model}
\end{equation}
where $\beta(T,\hat R)$ is a flexible nonparametric model; in our
numerical studies we use a SCM.\footnote{{\ffgn\ propose a
    nonparametric kernel estimator with polynomial regression of order
    1 \citep{fan:gijb:1996}, but we use the \scm\ to facilitate
    comparisons of the methods. As with (\ref{eq:IvD-smooth-coef}), we
    use the {\tt mgcv} package with penalized cubic regression splines
    as the basis functions with dimension equal to five for both $T$
    and $\hat R$ along with a tensor product.} }  {The \drf,
  $\hat{E}\{Y(t)\}$, and its standard errors are computed as in
  (\ref{eq:dose-resp}), but with }
\begin{equation}
\hat E \{Y(t)\} \ = \ \frac{1}{n} \sum\limits_{i=1}^{n} \hat \beta [t,\hat r(t,{\bm X}_i)]
\label{eq:scm-gps-est}
\end{equation}
{Because the \scm~is a
function of the \gps, we refer to this as the \scm(\gps) method.}

The second method involves inverse weighting (\iw) and estimates the \drf\ with
\begin{equation}
	\hat{E}\{Y(t)\} \ = \
	\frac{\sum\limits_{i=1}^{N}{\tilde{K}_{h,X}}(T_i-t)
	\cdot Y_i}{\sum\limits_{i=1}^{N}{\tilde{K}_{h,X}}(T_i-t)} ,
\label{eq:iw-1}
\end{equation}
where ${\tilde{K}_{h,X}}(T_i-t) =K_h(T_i-t)/ \hat{r}(t,\bm{X}_i)$, $K(\cdot)$ is
a kernel function with the usual properties, $h$ is a bandwidth satisfying $h
\rightarrow 0$ and $Nh \rightarrow \infty$ as $N \rightarrow \infty$, and
$K_h(\cdot)=h^{-1}K(\cdot/h)$. This is the local constant regression
(Nadaraya-Watson) estimator but now with each individual's kernel weight being
divided by its \gps~at $t$. To avoid boundary bias and to simplify derivative
estimation, the \iw~approach estimates $E\{Y(t)\}$ using a
local linear regression of $Y$ on $T$ with a weighted kernel function
$\tilde{K}_{h,X}(T_i-t)$, i.e.,
\begin{equation}
	\hat{E}\{Y(t)\} \ = \
	\frac{D_0(t)S_2(t)-D_1(t)S_1(t)}{S_0(t)S_2(t)-S_1^2(t)} ,
\label{eq:iw-2}
\end{equation}
where $S_j(t)=\sum\limits_{i=1}^{N} {\tilde{K}_{h,X}}(T_i-t)(T_i-t)^j$ and
$D_j(t)=\sum\limits_{i=1}^{N} {\tilde{K}_{h,X}}(T_i-t)(T_i-t)^jY_i$. The
global bandwidth can be chosen following the procedure of
\citet{fan:gijb:1996}. We use (\ref{eq:iw-2}) as the 
\iw~estimator in our numerical studies.

\section{Comparing  the {\large GPS} and the {\large P-FUNCTION}}
\label{sec:comp}

In this section, we examine the differences between {the method of \ivd,
\hi, and \ffgn\ using both simulation studies and
theoretical comparisons.} The key differences lie in how each method summarizes
$p(T \mid {\bm X})$: the \gps\ evaluates this density at the observed covariate, whereas the \pfun\
uniquely parameterizes it.  As we show below, this difference leads to
alternative choices for the response model, which can yield markedly divergent
results.

\subsection{Simulation study~I}
\label{sec:sim-one}

In our first simulation study, we have $2,000$ observations, each of
which comes with a single continuous covariate, $X$, a continuous
univariate treatment, $T$, and a response variable, $Y$.  We simulate
$X_i \ind {\cal N}(0.5, 0.25)$ and $T_i \mid X_i \ind {\cal N}(X_i,
0.25)$ and assume that the potential outcome has the following
distribution, $Y_i(t) \mid T_i, X_i \ind {\cal N}(10X_i,1)$ for all $t
\in \mathcal{T}$. In this simulation study the true treatment effect
is zero and the true \drf~is five for all $t$.  We deliberately choose
this simple setting where any reasonable method should perform well.
Fitting a simple linear regression of $Y$ on $T$ yields a
statistically significant treatment effect estimate of roughly
five. However, adjusting for $X$ in the regression model is sufficient
to yield an estimate that is much closer to and is not statistically
different from the true effect of zero.

{Using the correctly specified treatment assignment model, $T_i \mid X_i
\ind {\cal N}(X_i, 0.25)$, we implement the \hi, \ivd, \scm(\gps), and
\iw~methods.} For the response models, we use the quadratic regression given in
\eqref{eq:gps-r-model} with {the method of \hi, regress $Y$ on $T$ within
each of $S$ subclasses with the method of \ivd, and use the default models for
\scm(\gps) and \iw.} For the purposes of illustration, we do not adjust for
$\hat{\bm\theta}$ within each subclass when using \ivd's method. Owing to the
linear structure of the generative model, doing so would dramatically reduce
bias even with a small number of subclasses.
Here we illustrate, instead, how bias can be reduced by increasing the number of
subclass;  we implement \ivd\ with $S= 5, 10$, and $50$ subclasses. For {the
\hi, \scm(\gps), and \iw~methods}, we use a grid of ten equally spaced points
between $-0.5$ and $1.5$, $t_1,\ldots, t_D$ with $D=10$, to compute the 
relative \drf\ and its derivative. Standard errors are computed using 1,000
bootstrap replications. 

\begin{figure}[t]  
  \spacingset{1}
  \centering      
  \includegraphics[width=0.24\textwidth]{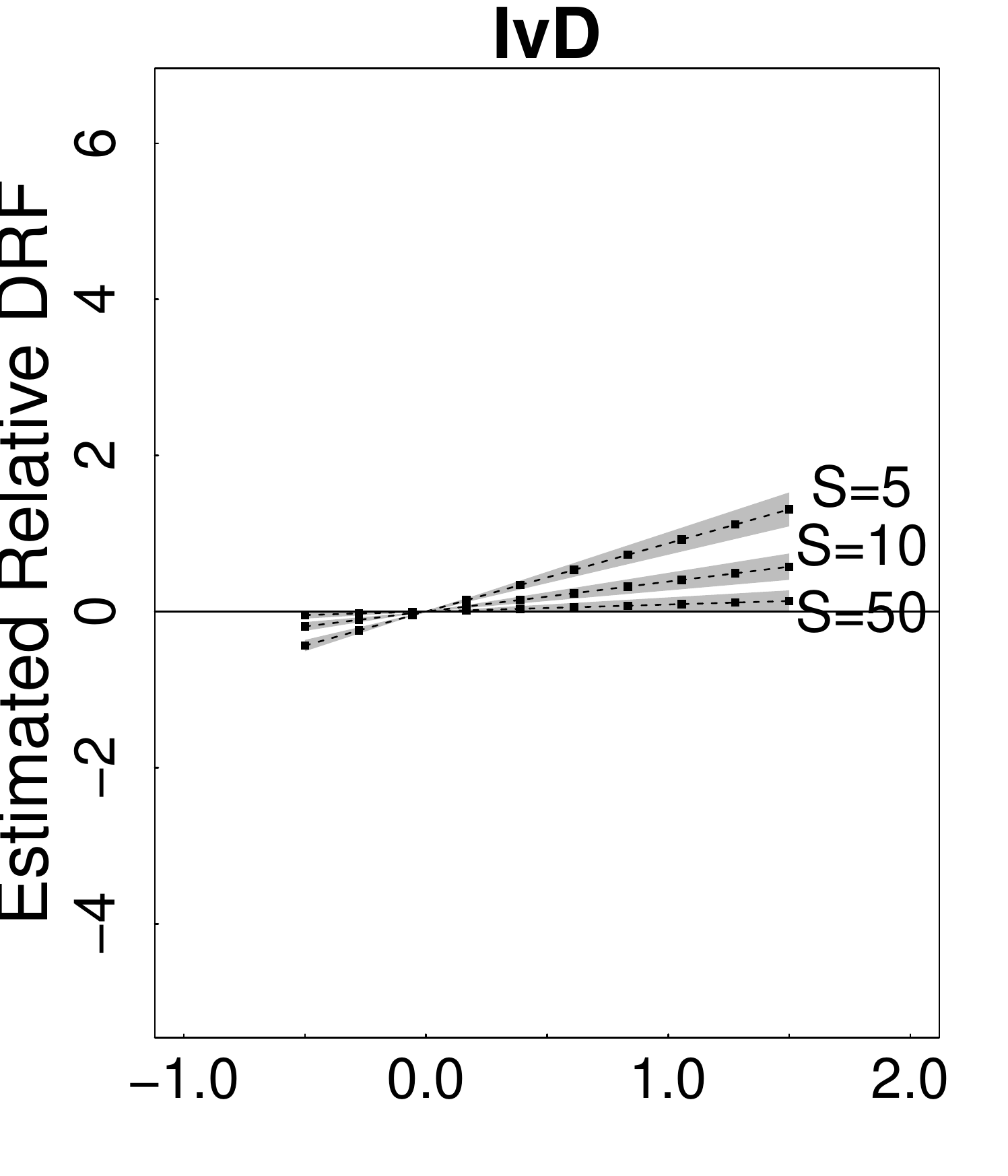}
  \includegraphics[width=0.24\textwidth]{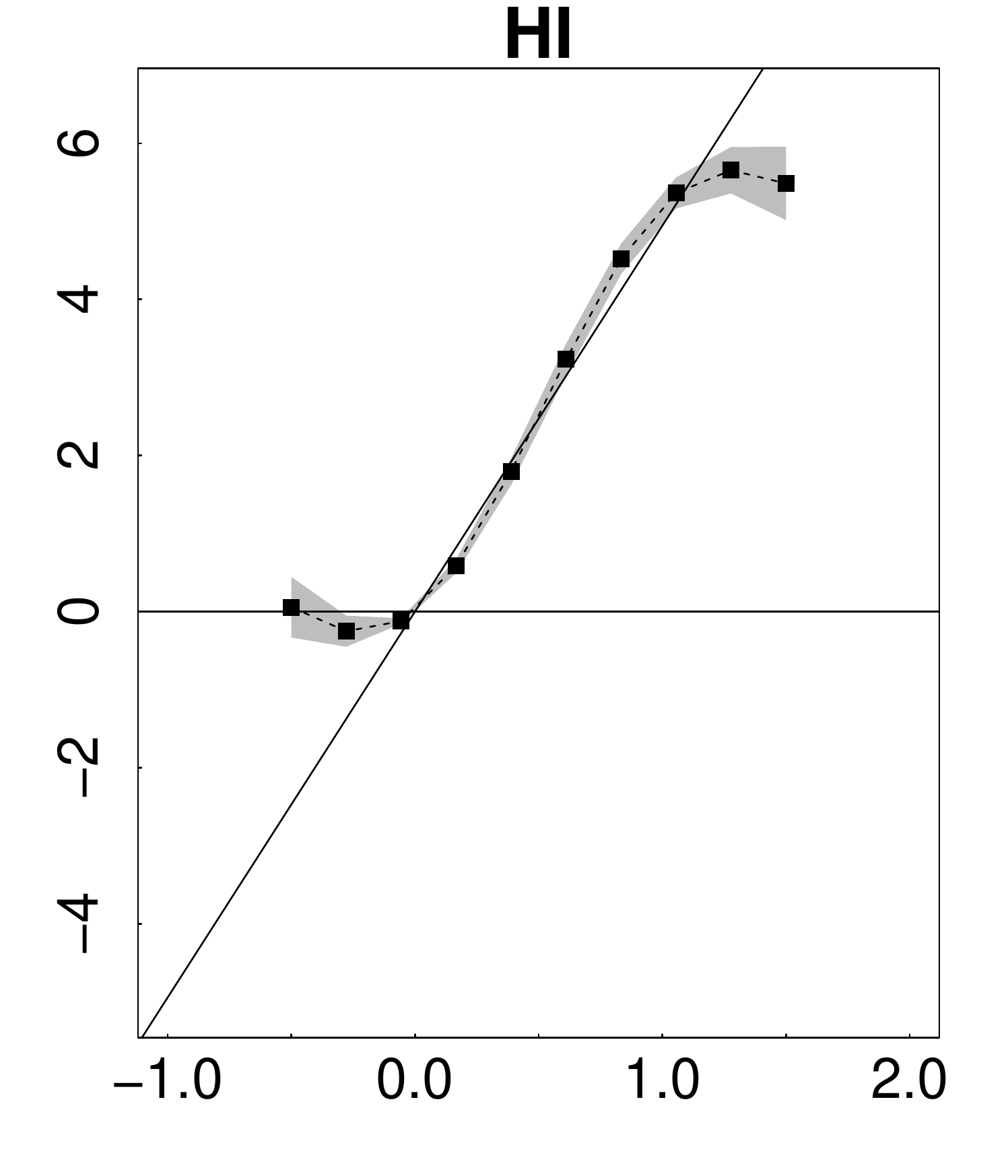}   
  \includegraphics[width=0.24\textwidth]{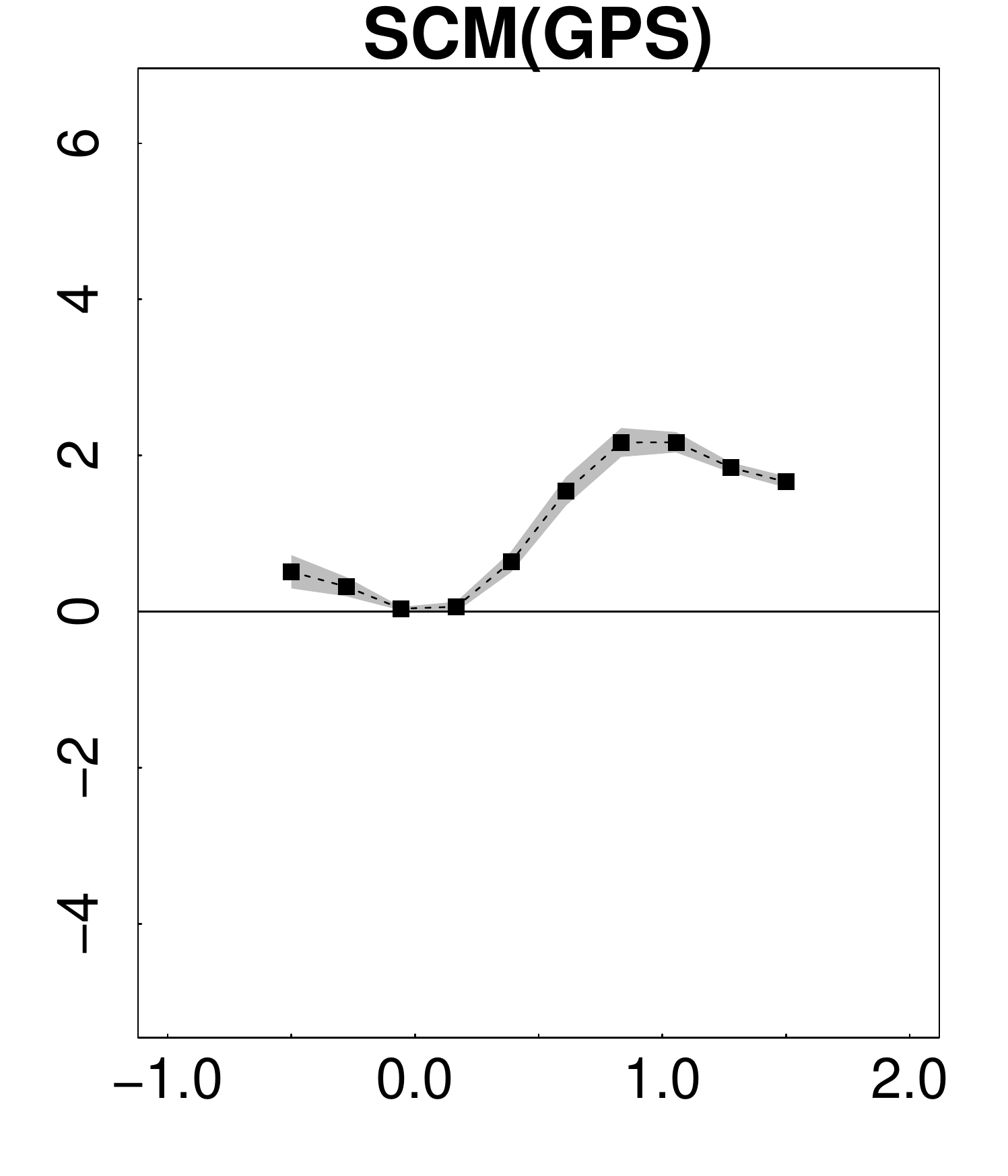}
  \includegraphics[width=0.24\textwidth]{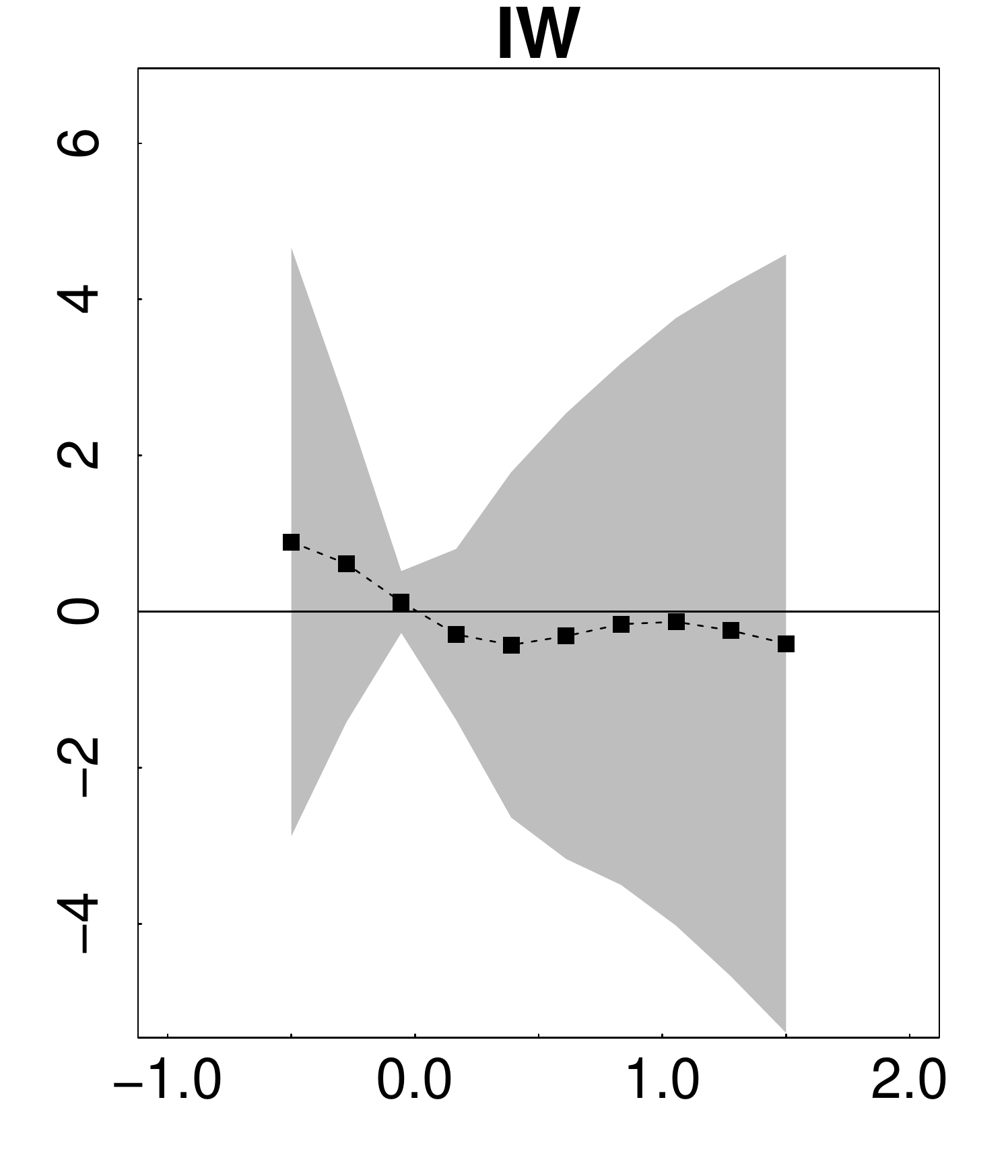}\\
  \includegraphics[width=0.24\textwidth]{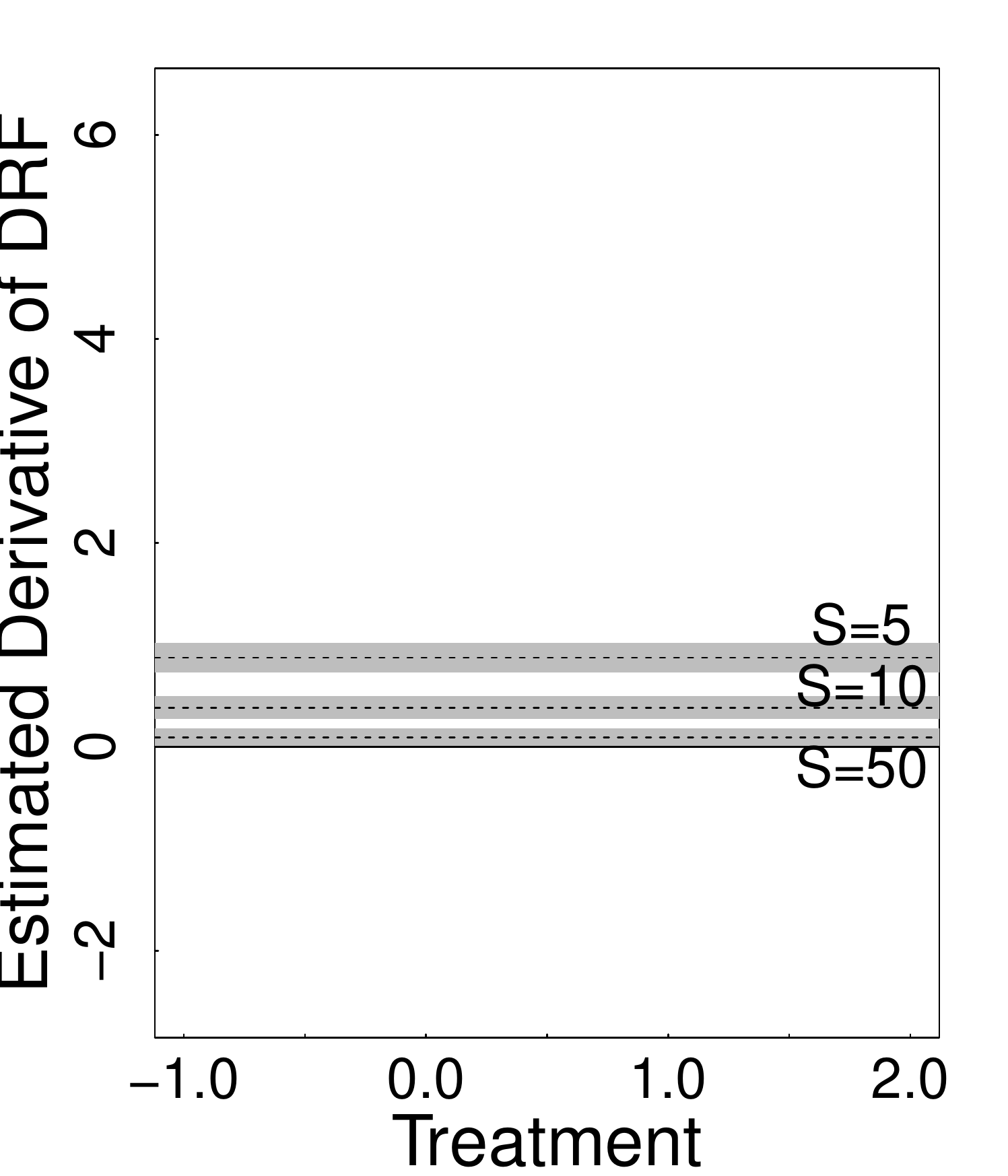} 
  \includegraphics[width=0.24\textwidth]{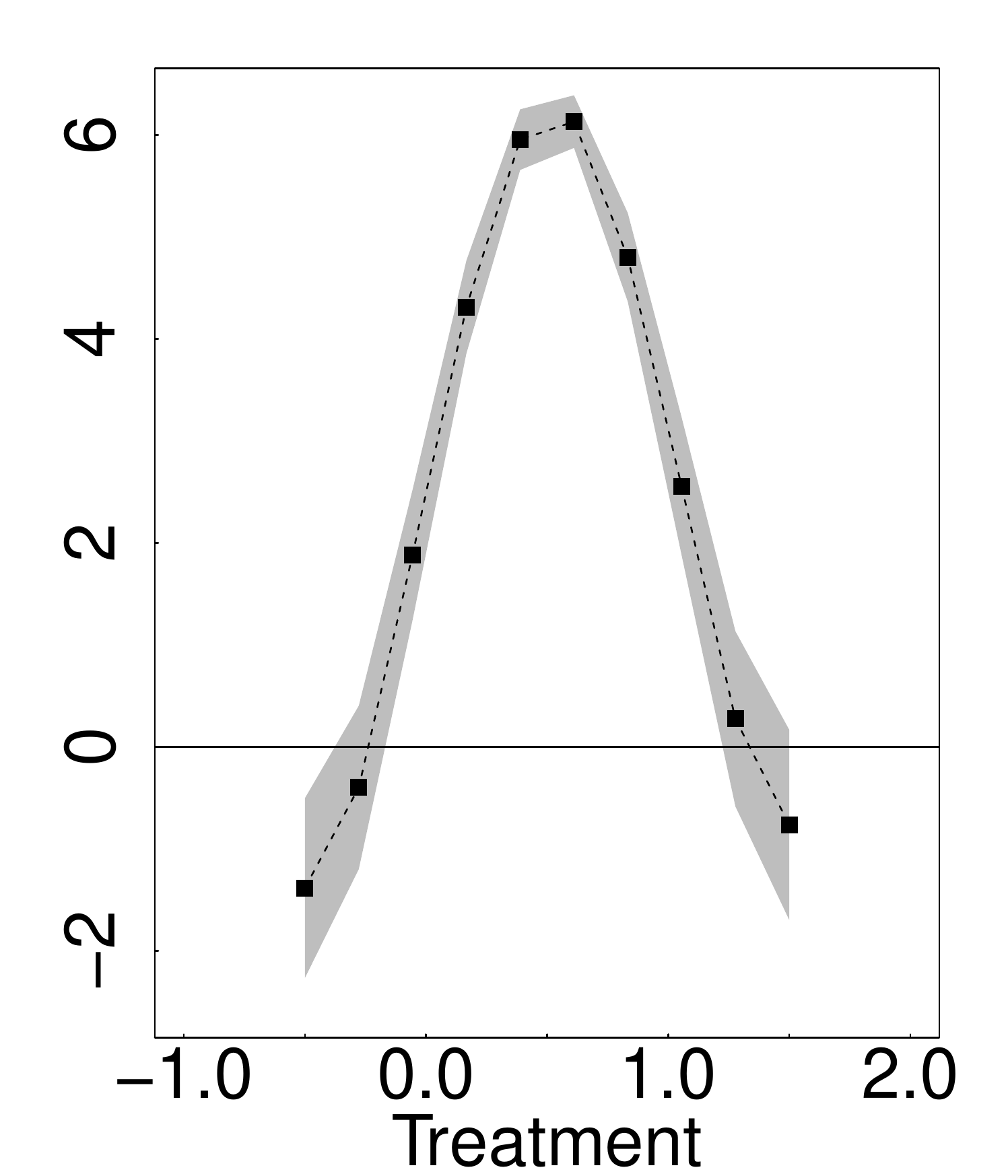}    
  \includegraphics[width=0.24\textwidth]{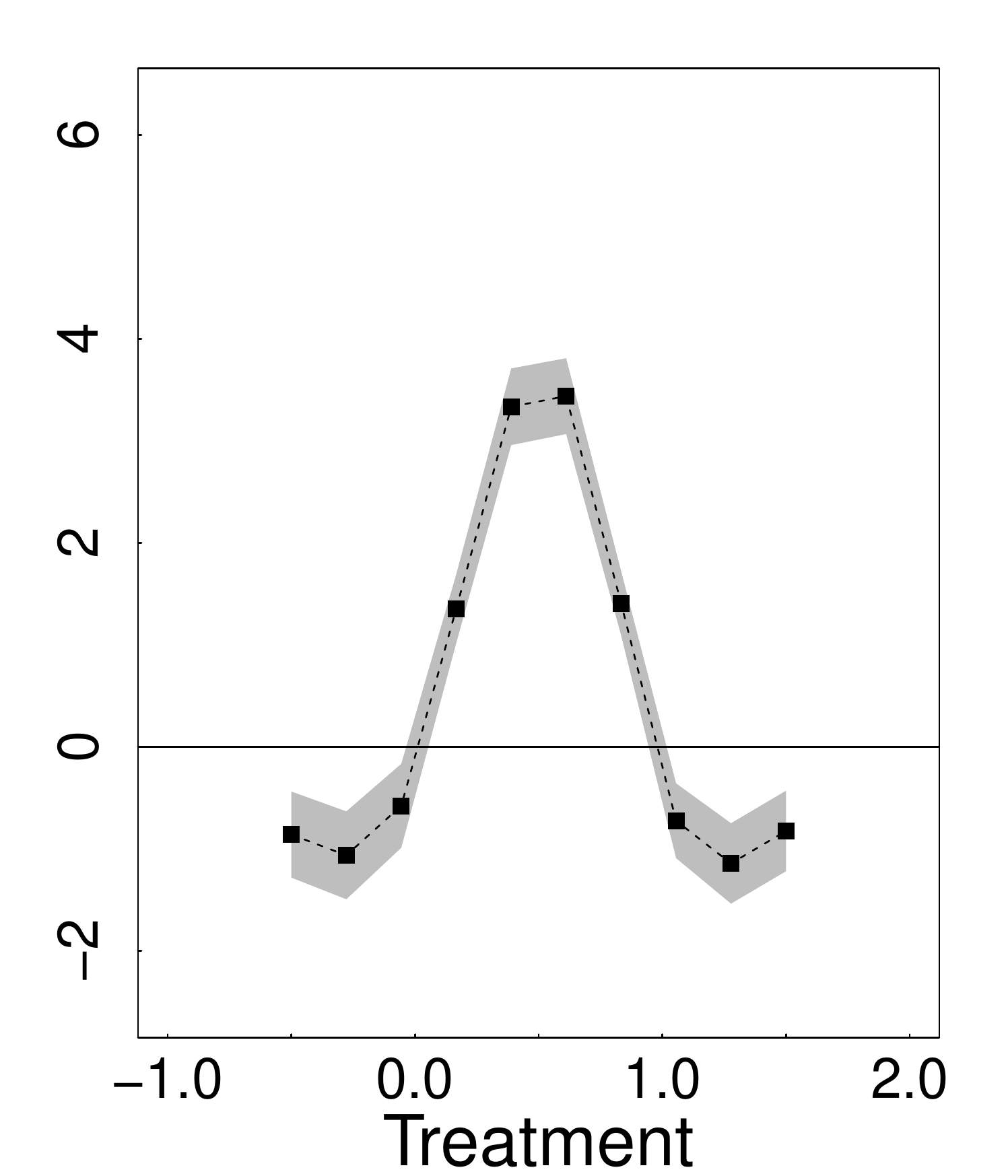}  
  \includegraphics[width=0.24\textwidth]{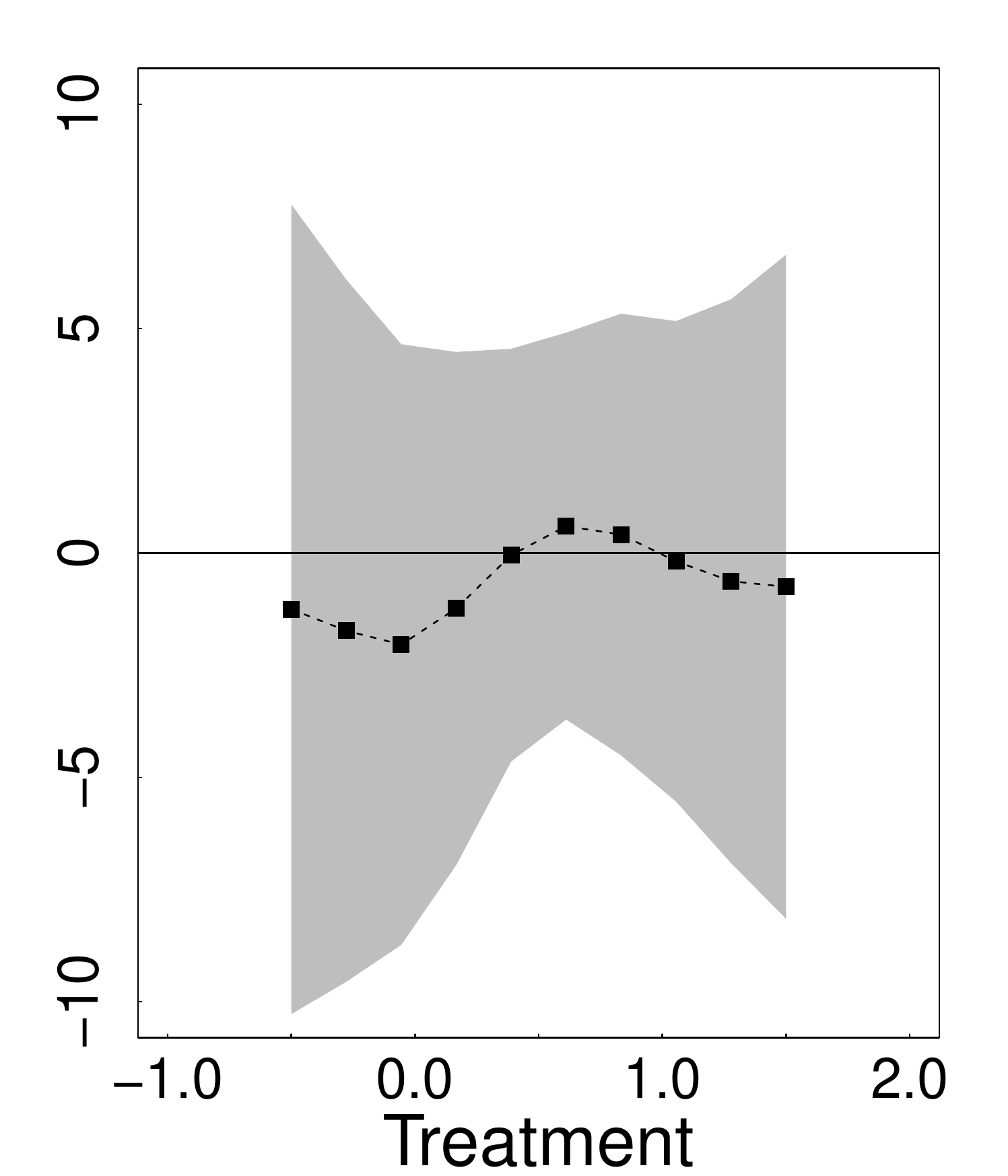} 
  \caption{The Results of Simulation Study~I. {The first row plots the
      estimated relative \drf\ where the horizontal solid black line
      represents the true relative \drf. For \ivd, we use $S=5,10,50$
      subclasses.  The solid black diagonal line for the method of
      \hi\ is the unadjusted regression of $Y$ on $T$. The second row
      plots the estimated derivatives of the dose response function
      (\drf) where the solid black line represents the truth. In both
      rows, the grey shaded areas represent 95\% confidence
      intervals. The estimated derivative for \iw\ is plotted on a
      different scale as its standard error is significantly larger
      than that of the other methods.}
  \label{fig:sim1-1} }
\end{figure}    

Figure~\ref{fig:sim1-1} presents the results. {In the first row, we plot
the estimated relative \drf~while the second row plots the estimated derivative
of the \drf. For \hi, \scm(\gps), and \iw, the derivative is computed as} 
\begin{eqnarray}
{1\over 2}\left[ { \hat E\{
    Y(t_{d+1})\} - \hat E\{ Y(t_d)\} \over t_{d+1} - t_d } + {\hat E
    \{Y(t_{d})\} - \hat E\{ Y(t_{d-1})\} \over t_{d} - t_{d-1} }\right]  
\label{eq:der}
\end{eqnarray}
for $d = 2, \ldots D-1$.  For $d=1$, we simply use the first term in
(\ref{eq:der}) and for $d=10$ we use the second term in
(\ref{eq:der}). {For \ivd, the derivative is the weighted average of
  the within subclass linear regression coefficient; 95\% point-wise
  confidence intervals are shaded gray. }

Figure~\ref{fig:sim1-1} shows that even in this simple simulation,
{all methods except IW} miss the true {relative} \drf\ and its
derivative, albeit to differing degrees. The behavior of \ivd's
estimate improves with more subclasses, a luxury we can afford here
because of the large sample size. \ivd\ makes the general
recommendation that the within subclass models be adjusted for ${\bm
  X}$ or at least for $\hat{\bm\theta}$. Because of the simple
structure of this simulation, doing so would result in a correctly
specified model even with a single subclass, eliminating bias in the
estimated average treatment effect.\footnote{It would also complicate
  estimation of the \drf. Because the treatment assignment mechanism
  is strongly ignorable given the propensity function (\ivd), we aim
  to adjust for the propensity function in a robust manner in the
  response model. Thus, adjusting for $\hat\theta$ within the
  subclasses poses no conceptional problem. Practically, however,
  $\hat\theta$ tends to be fairly constant within subclasses and its
  coefficient tends to be correlated with the intercept. A solution is
  to recenter $\hat\theta$ within each subclass. Because, we propose a
  more robust strategy for estimating the \drf\ using the \pfun\ in
  Section~\ref{sec:pfun-drf}, however, we do not purse such adjustment
  strategies here.} We do not recommend estimating the \drf\ by
averaging the unadjusted within subclass models, but do so here to
facilitate comparisons between the methods. We propose a new estimate
of the \drf\ using the \pfun\ in Section~\ref{sec:pfun-drf}.

{The performance of the \hi\ method is particularly poor; It differs only
slightly from the unadjusted regression.
Although \scm(\gps) offers limited improvement, it also introduces a cyclic artifact into
the fit. We will see this pattern again. The \iw\ method, on the other hand,
results in an unstable fit that is characterized by very large standard erros. The
performance of these methods are especially troubling both because the \gps\ was
expressly designed to estimate the \drf\ and because the current simulation
setup is so simple. Given their performance here, it is difficult to think that
these methods can succeed in more realistic settings. The primary goal of this
paper is to explain why the \gps-based methods can fail and to provide a more
robust estimate of the \drf. }

\begin{figure}
\spacingset{1}
\centering
\includegraphics[width=0.32\textwidth]{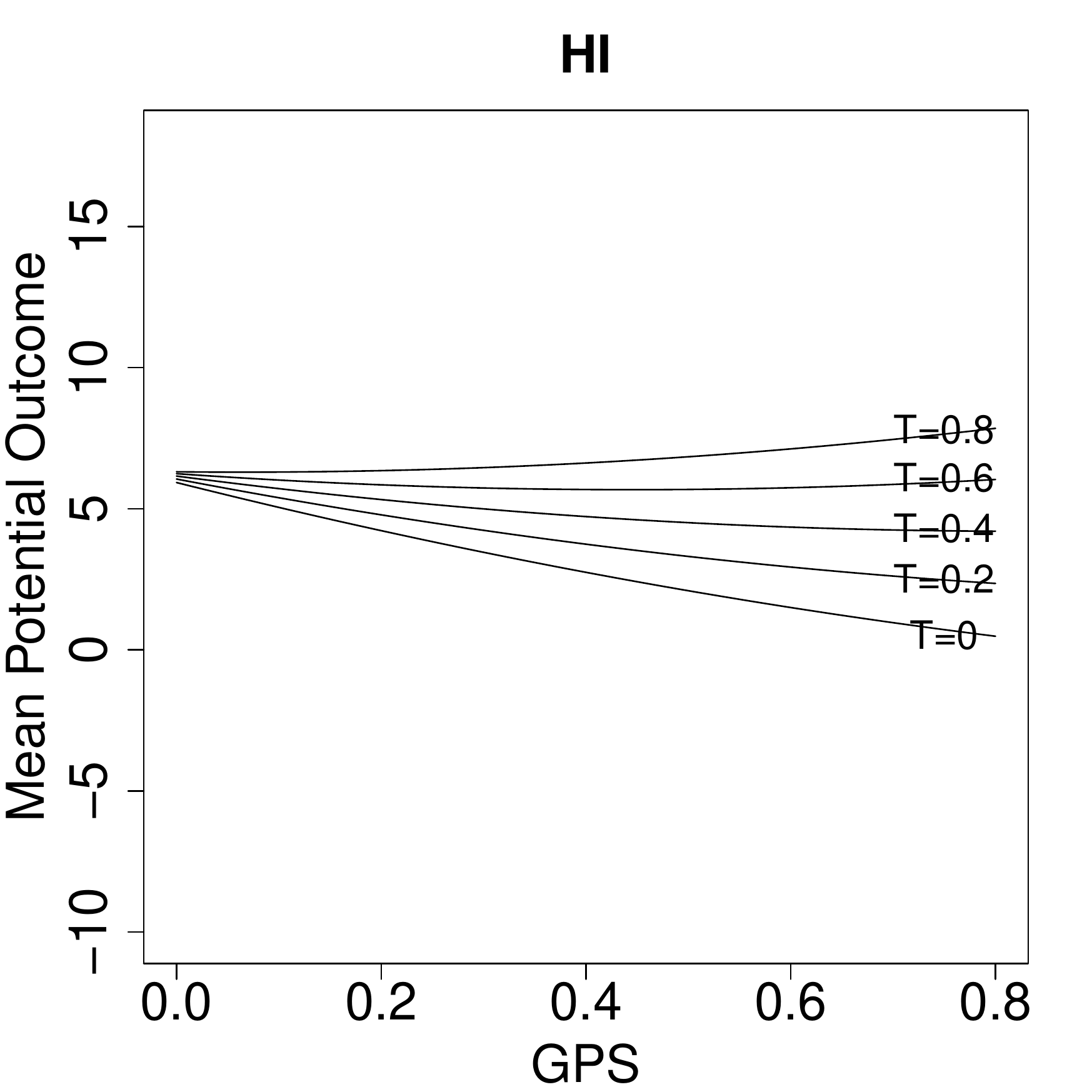}
\includegraphics[width=0.32\textwidth]{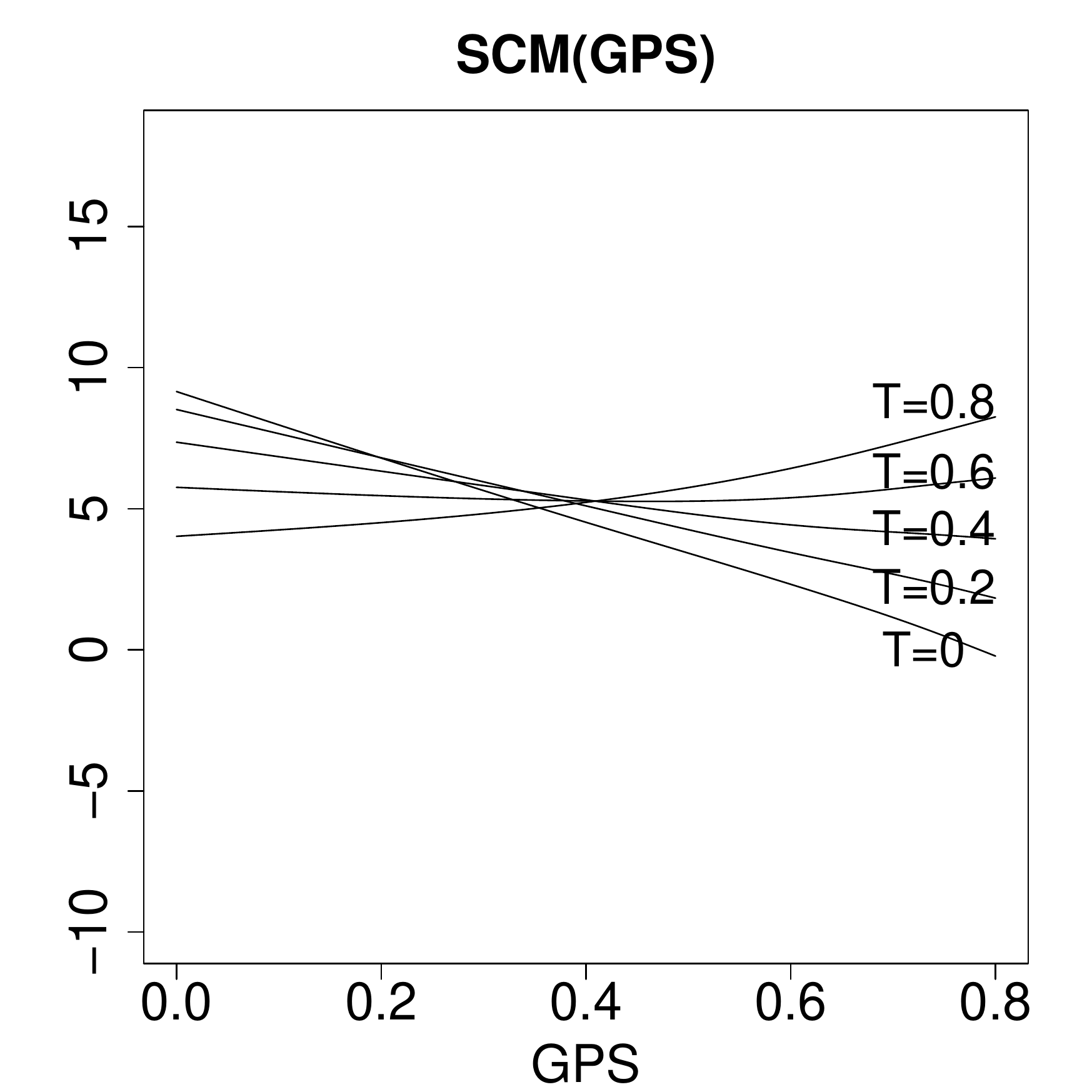}
\includegraphics[width=0.32\textwidth]{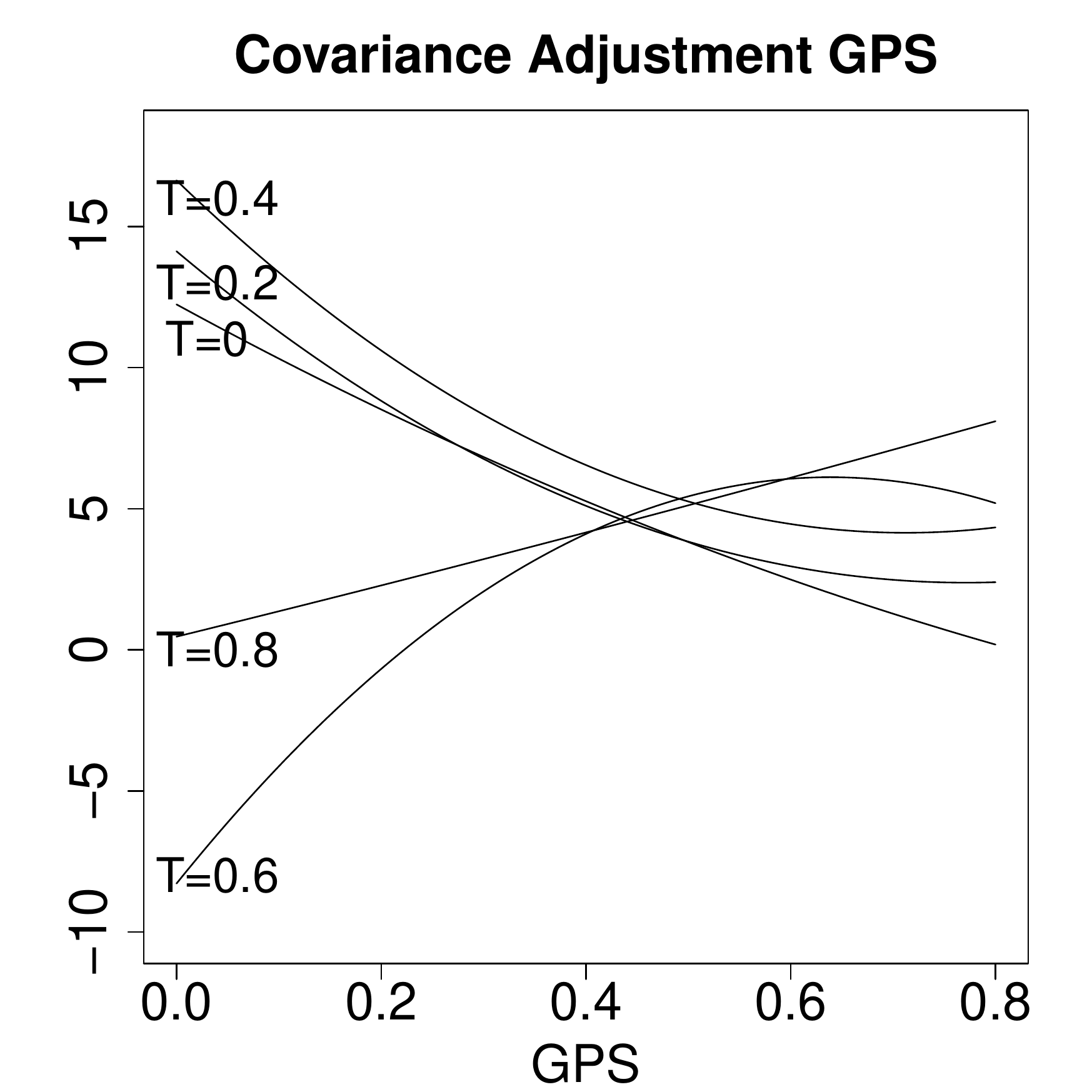}\\
\caption{The Varying Flexibility of the Response Models.  
The plots show the mean potential outcome as a function of the \gps\ and $T$
under \hi's quadratic response model (left panel), fitted \scm(\gps) (middle
panel), and covariance adjustment \gps\ (right panel). Covariance adjustment
\gps\ fits a quadratic regression ($Y \sim R + R^2$) in each of several subclasses
based on $T$. See Appendix~\ref{sec:appA} for details. Here we use 10
subclasses; Only five are shown for ease of plotting. Judging from the plots, subclassification is by far the most flexible of the three response models.
\label{fig:sim-one-b-hi-prob} }    
\end{figure}

{One reason that \gps-based methods can perform poorly is that their
response model are based on overly strong parametric assumptions, especially
(\ref{eq:gps-r-model}). This is illustrated in
Figure~\ref{fig:sim-one-b-hi-prob} which compares the fitted mean potential
outcome as a function of the \gps\ and $T$ under the \hi\ model (left panel) and
under \scm(\gps) (middle panel). The fitted potential outcomes differs
substantially and are considerably more constrained under the quadratic model of
\hi. Although \scm(\gps) is more flexible than \hi, it still exhibits
considerable constraints. To see this we subclassified the data into 10
subclasses based on $T$, and fit a quadratic regression for $Y$ as a function of
the \gps\ seperately within each of the subclasses. Five out of the 10 within
subclass fit are plotted in the right most plot in
Figure~\ref{fig:sim-one-b-hi-prob}. The results differs substantially from \hi\
and reveals the considerable constraint of the quadratic response model.
Subclassifying on $T$ in this way leads to a new response model and a
corresponding new \gps-based estimate of the \drf; this method is discussed in
Appendix~\ref{sec:appA}. 
}

\subsection{Simulation study~II}
\label{sec:sim-two}
 
In Simulation I, although its response model is misspecified in terms
of its adjustment for $X$, the \ivd\ method may benefit from its
assumption that the \drf\ is linear in $T$. We address this in the
second simulation that compares the performance of the methods under
two generative models.  We also explore the frequency properties of
the methods.

Suppose we have a simple random sample of $2,000$ observations that
consist of a univariate covariate $X$, the assigned treatment $T$, and
the response $Y$, respectively. We assume that $X \ind {\mathcal
  N}(0,1)$ and that $T$  depends on $X$ through $T
\mid X \ind {\mathcal N}(X+X^2,1)$. We simulate the potential outcome
$Y(t)$ using two different response models:
\begin{description}
\item[Linear \drf:] $Y(t) \mid t,X \ \ind \ {\cal N}(X+t, 9)$      
\item[Quadratic \drf:] $Y(t) \mid t,X \ \ind \ {\cal N}((X+t)^2, 9)$
\end{description}

To isolate the difference between the {methods}, we correctly
specify the treatment model for all methods. For the response model {under
\hi\ and \ivd's methods}, we consider Gaussian regression models that are linear
and quadratic in $T$.
In particular for the method of \ivd\ we fit (i) $Y \sim T$ and (ii) $Y \sim T+T^2$
within each of 10 equally sized subclasses, and for the method of \hi\ we fit
(i) $Y \sim T+R+R^2+R \cdot T$ and (ii) $Y \sim T+T^2+R+R^2+R \cdot T$.
With \ivd, the relative \drf\ is computed by averaging the
coefficients of the within subclass models. {The default response models are
used for \scm(\gps) and \iw.} 
For the \gps-based methods, the relative \drf\ is evaluated at ten equally spaced values of $t$ between $-1.5$ to $5.5$.
The entire procedure was repeated for all methods
on each of 1000 data sets simulated from each generative model.
 Notice that all of the response models are misspecified in their adjustment for
$X$ and/or $T$, as we expect in practice. Thus, this simulation study
investigates the robustness of the methods to typical misspecification of
the response model.
  
\begin{figure}[t]
\spacingset{1}
\centering
\textsf{\textbf{\large Generative Dose Response Function: Linear}}\\ \smallskip
\includegraphics[width=0.48\textwidth]{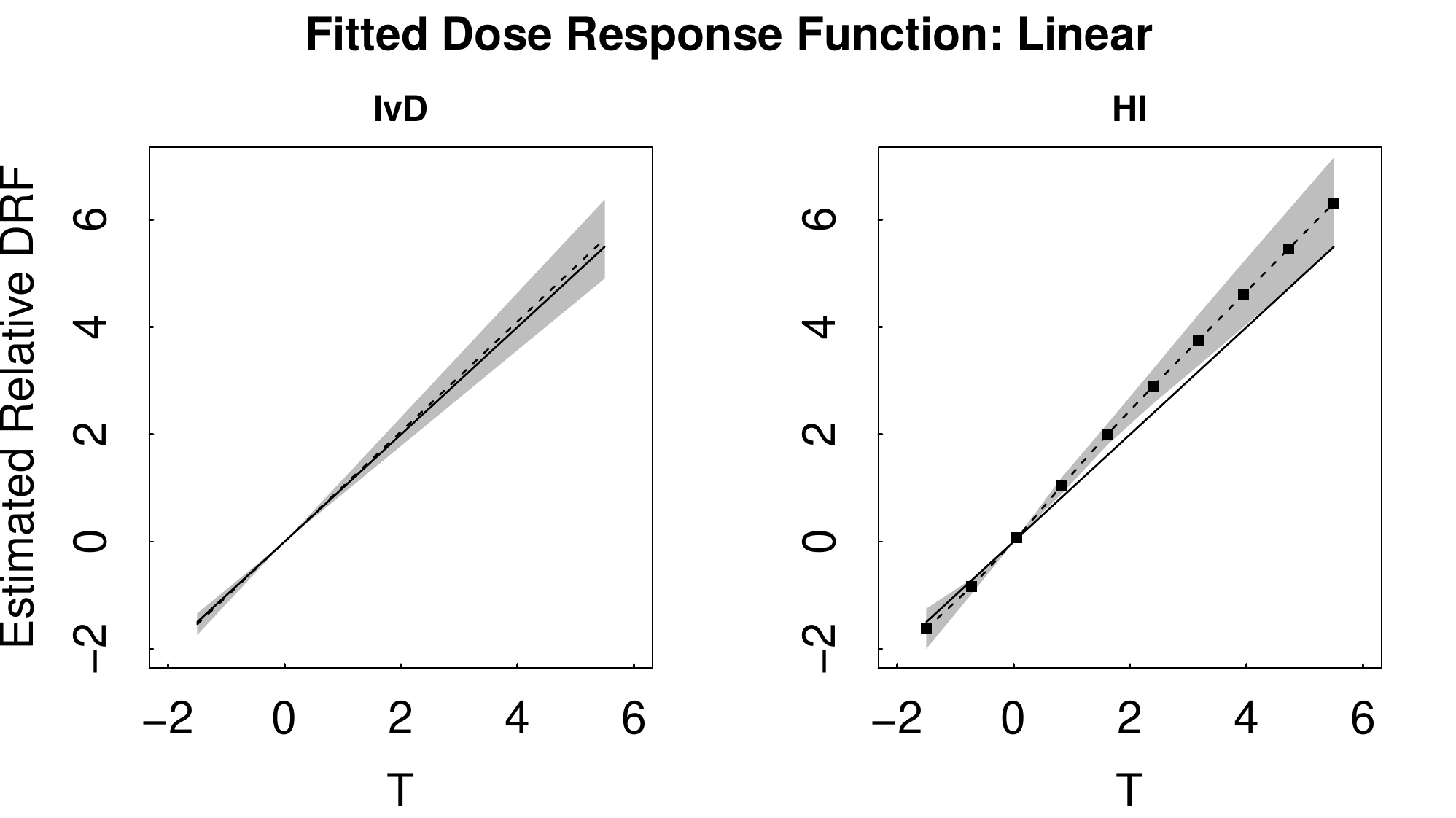}
\includegraphics[width=0.48\textwidth]{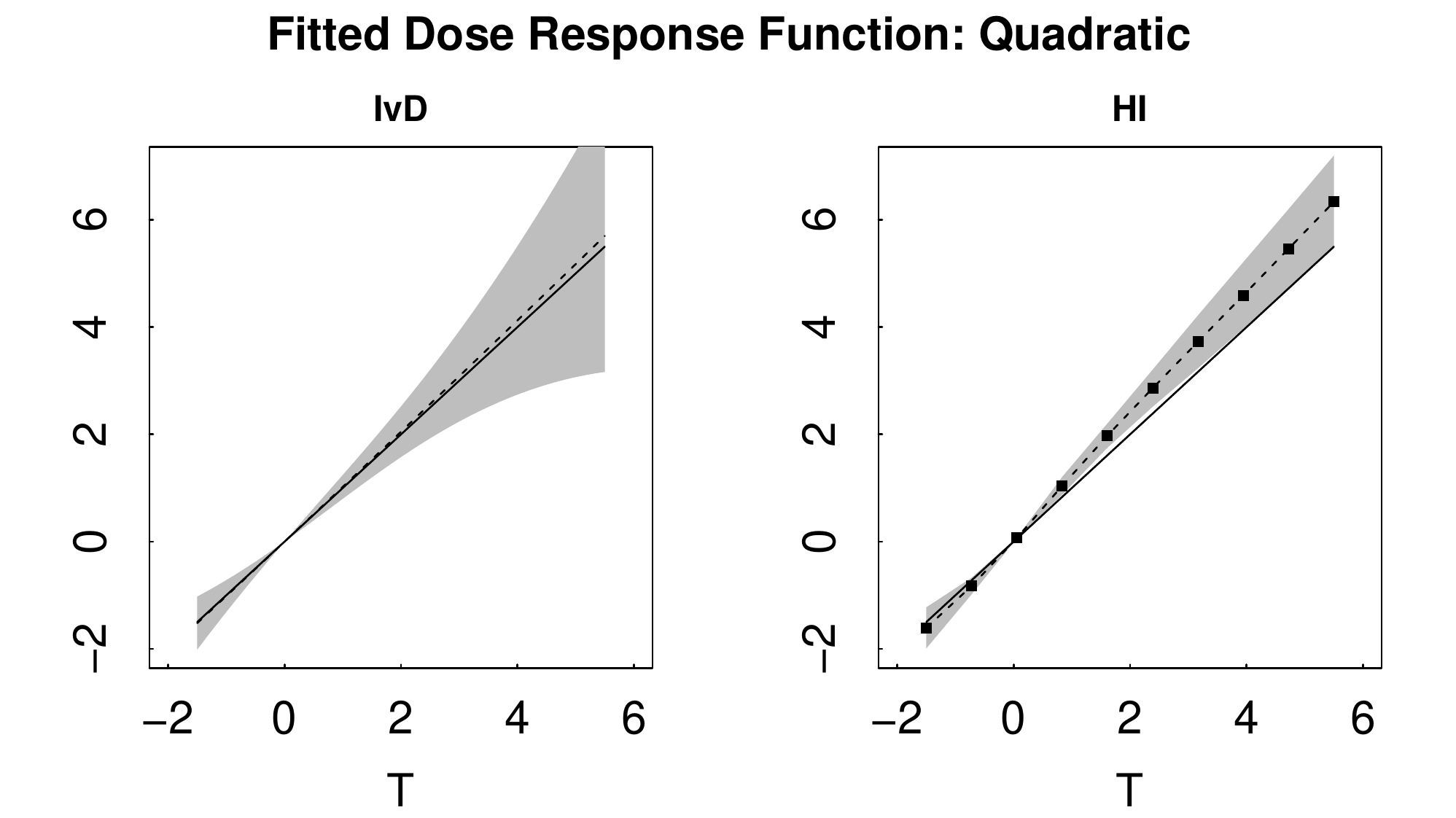}\
\textsf{\textbf{\large Generative  Dose Response Function: Quadratic}}\\ \smallskip
\includegraphics[width=0.48\textwidth]{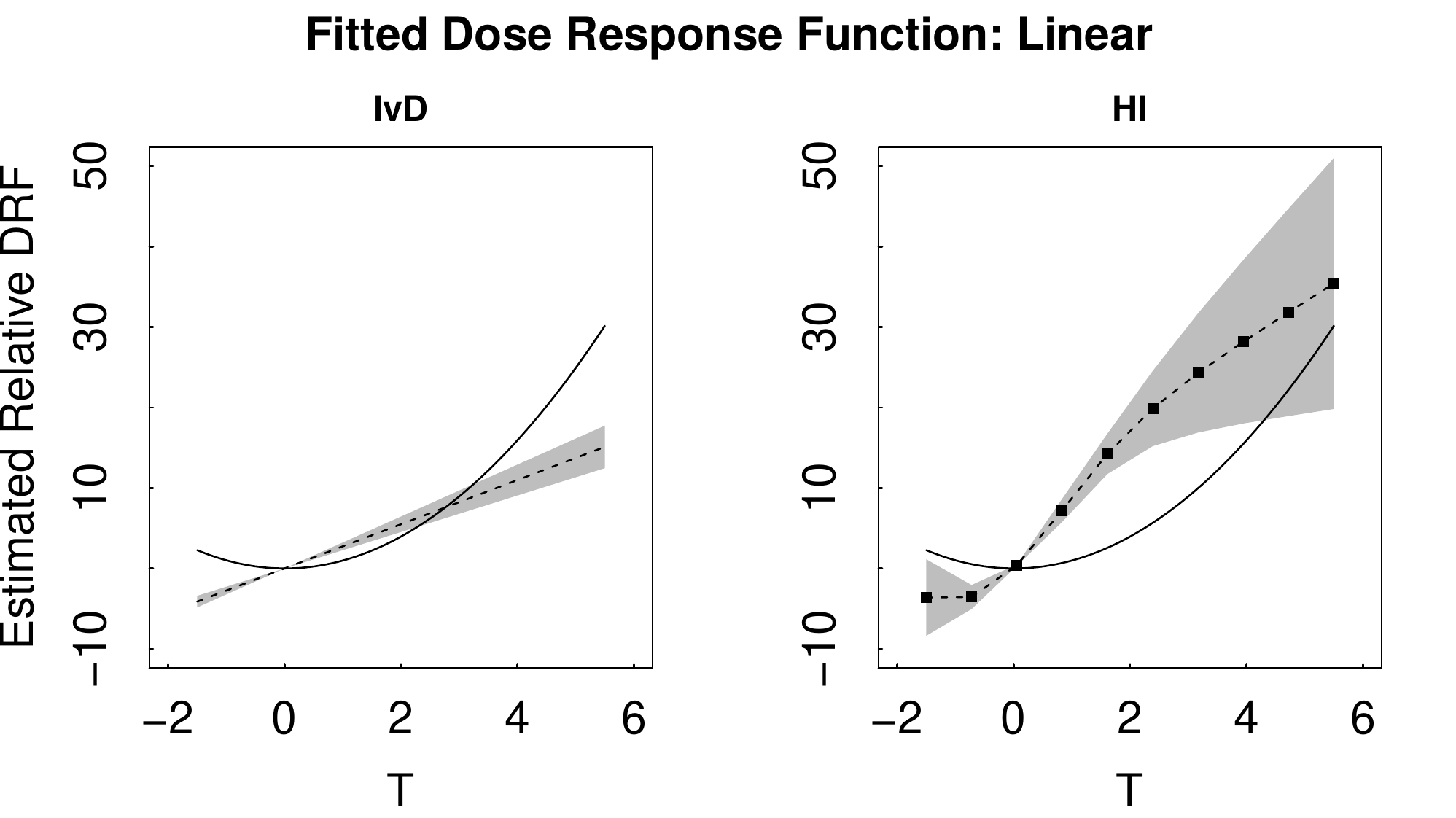}
\includegraphics[width=0.48\textwidth]{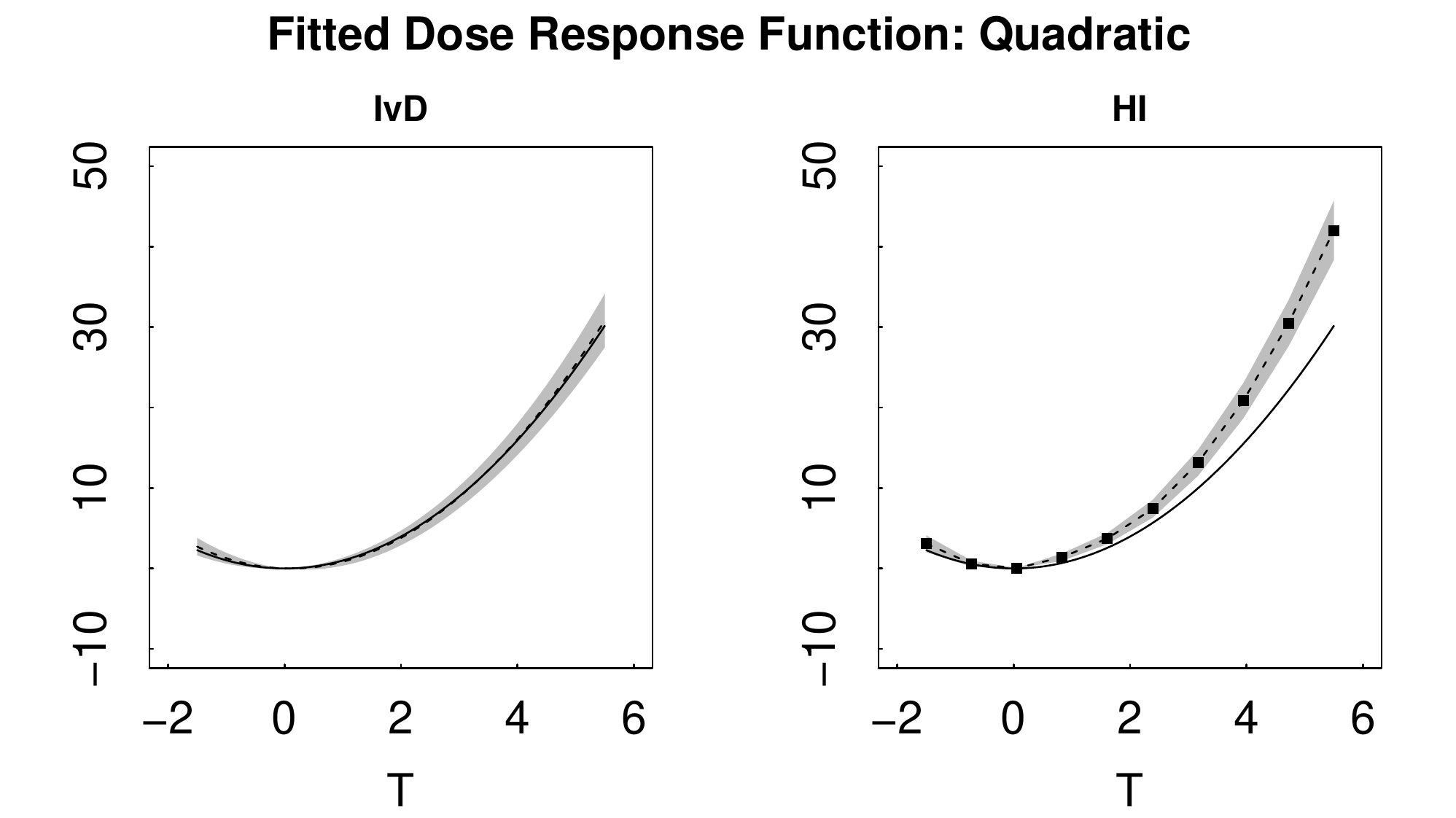}\\
\caption{Estimated Relative \drfs~in Simulation
  Study~II {Using the Methods of \ivd~and \hi}. The solid lines plot
  the true relative \drfs, the dashed lines plot the means of the fitted relative \drfs\
  across 1000 simulations, and the gray shaded regions correspond to two
  standard deviation pointwise intervals across the 1000 fitted relative \drfs.
  The evenly-spaced grid of evaluation points used with \hi\ are also plotted as
  solid circles.
  The method of \hi\ shows appreciable bias with all four combinations of
  generative and fitted response models. The method of \ivd, on the other hand,
  is biased only when the fitted model is of a lower order than the generative
  model.  }
\label{fig:advan_simu}
\end{figure} 

\begin{figure}
\spacingset{1} 
\centering
\textsf{\textbf{\large Generative Dose Response Function: Linear}}\\ \smallskip
\includegraphics[width=0.24\textwidth]{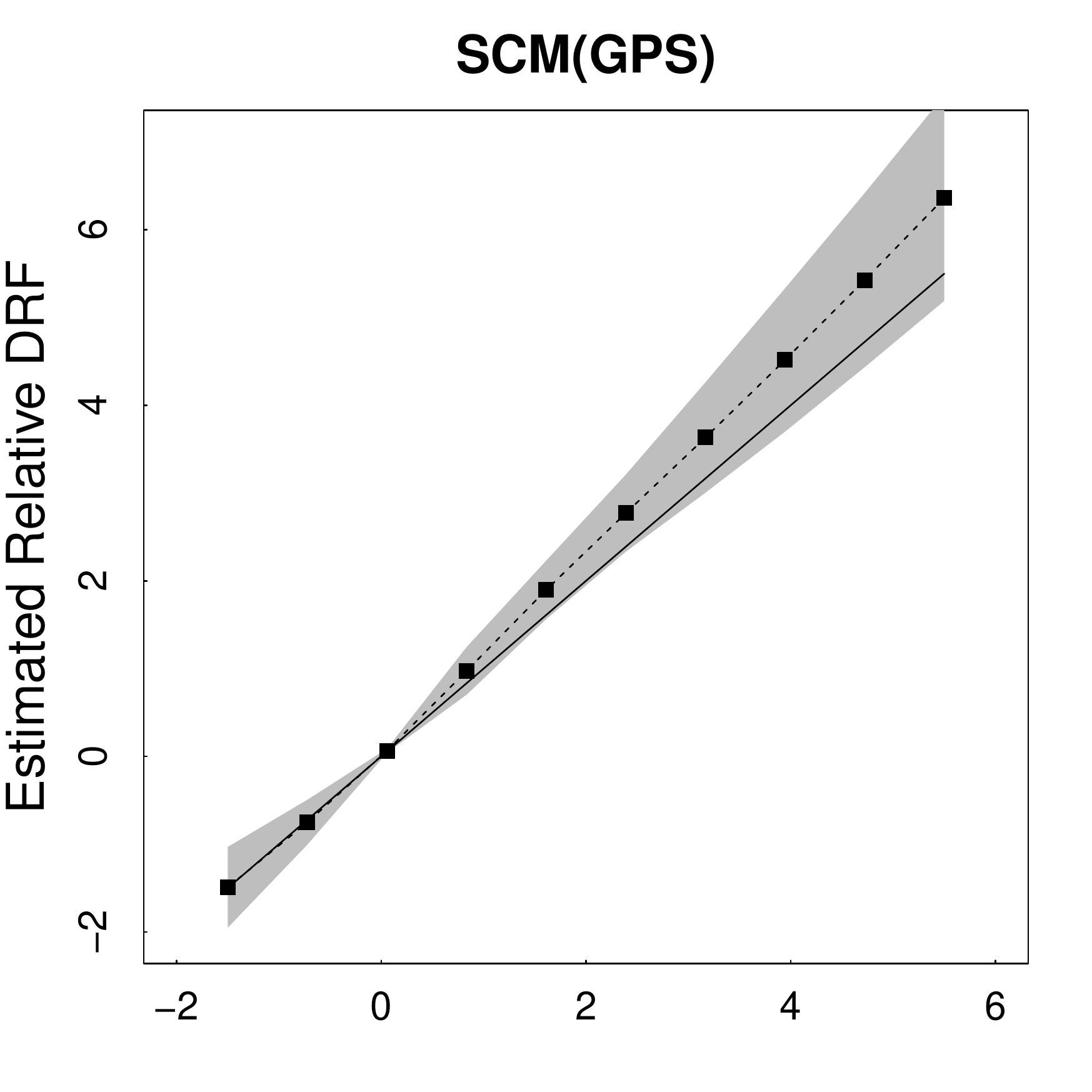}
\includegraphics[width=0.24\textwidth]{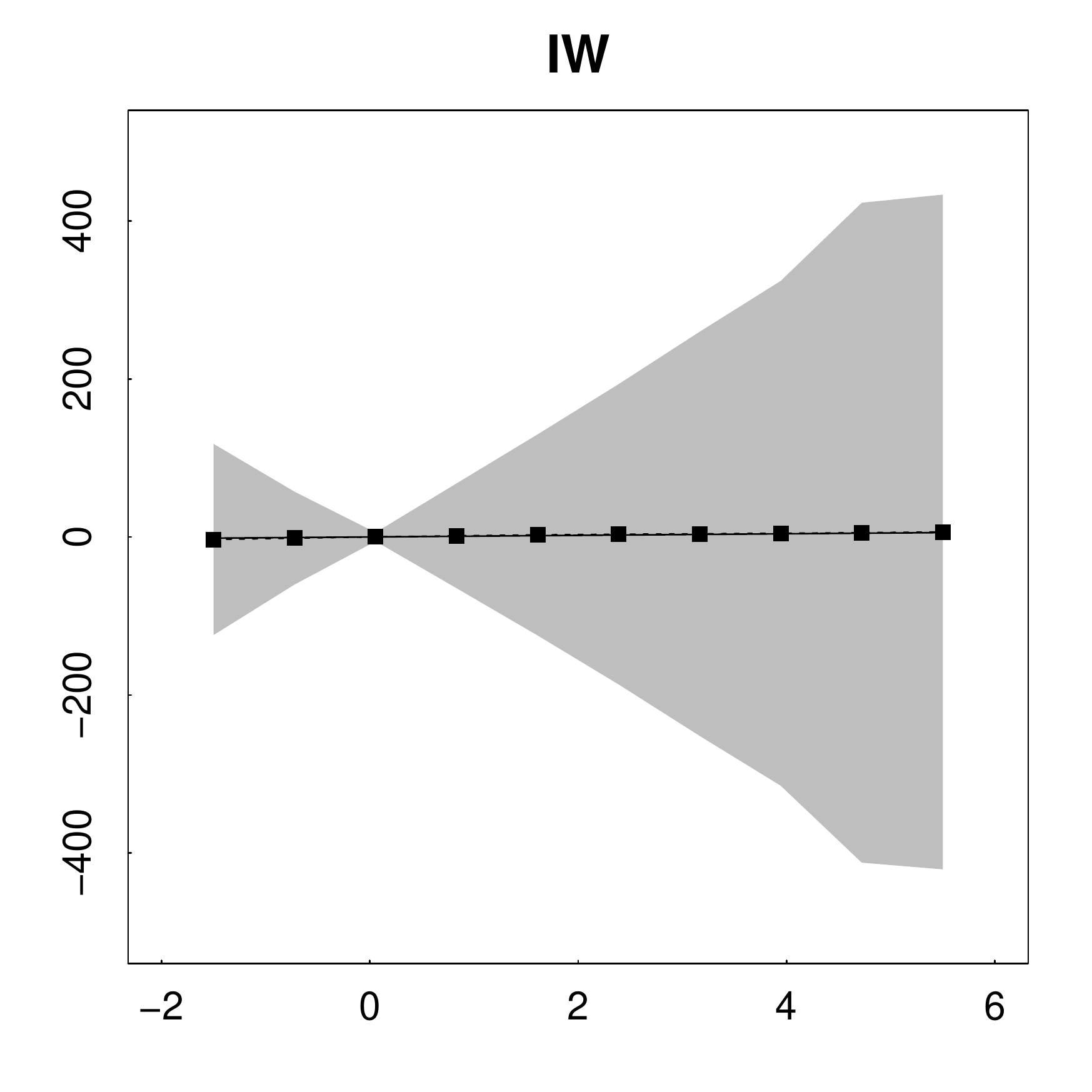}
\includegraphics[width=0.24\textwidth]{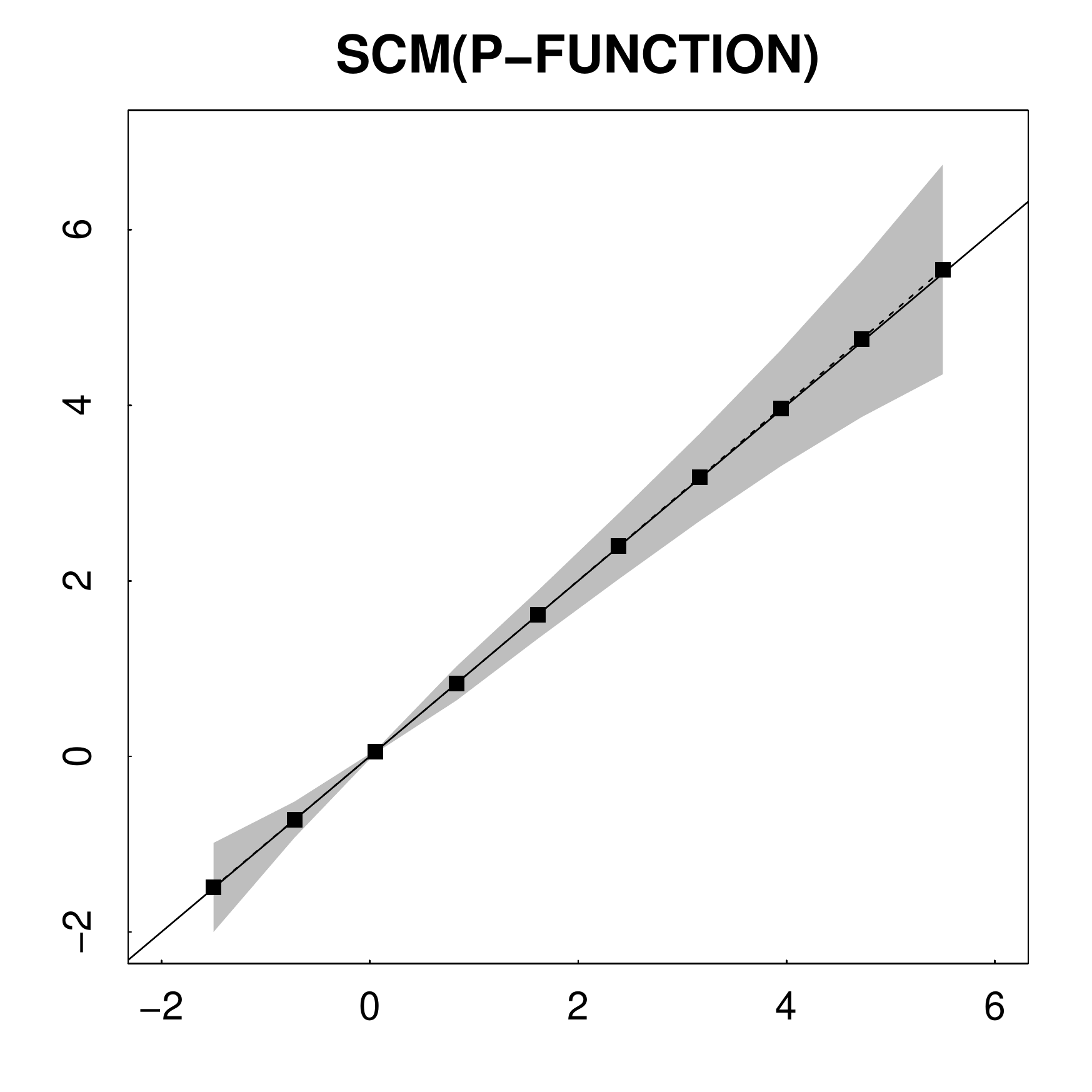}\\
\textsf{\textbf{\large Generative  Dose Response Function: Quadratic}}\\ \smallskip
\includegraphics[width=0.24\textwidth]{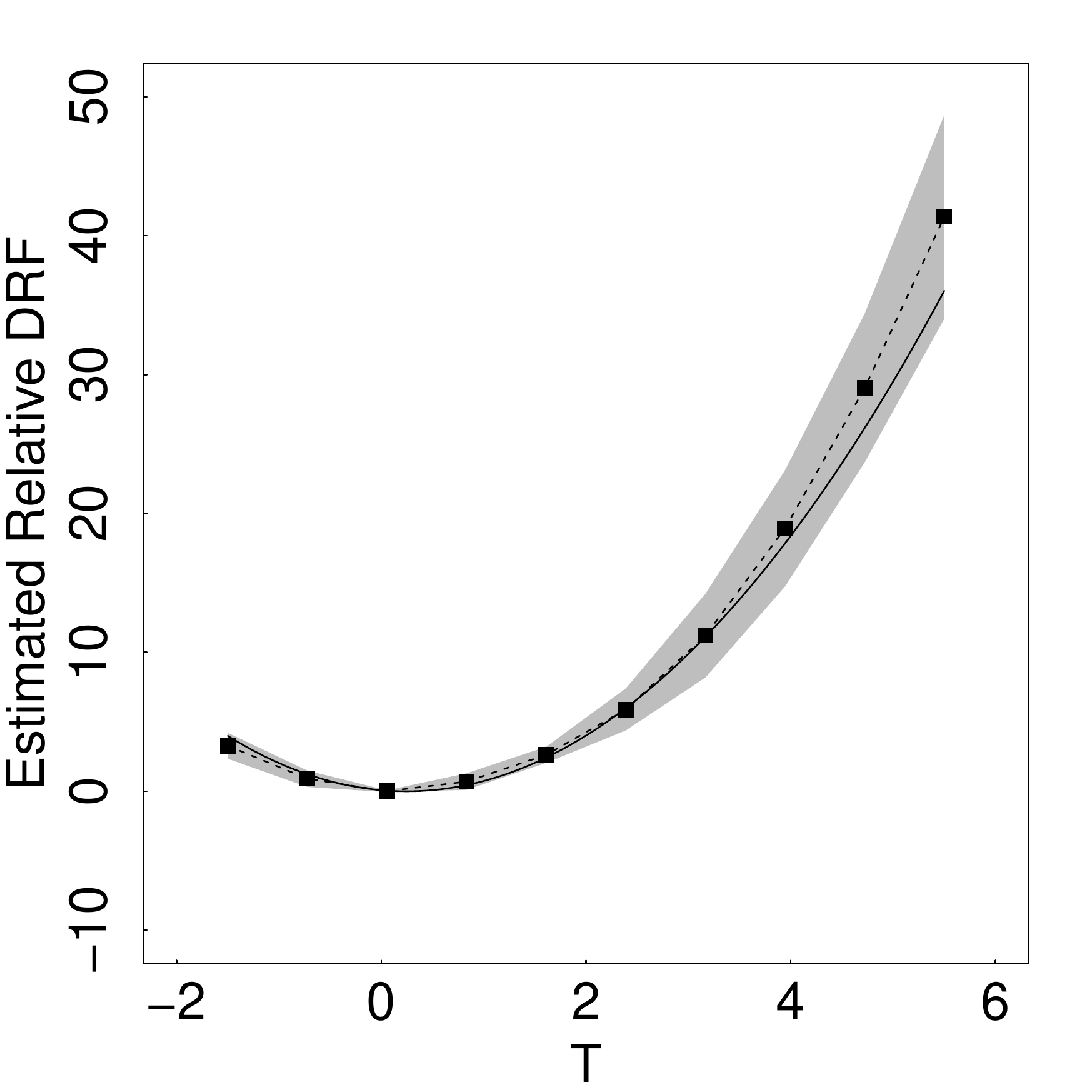} 
\includegraphics[width=0.24\textwidth]{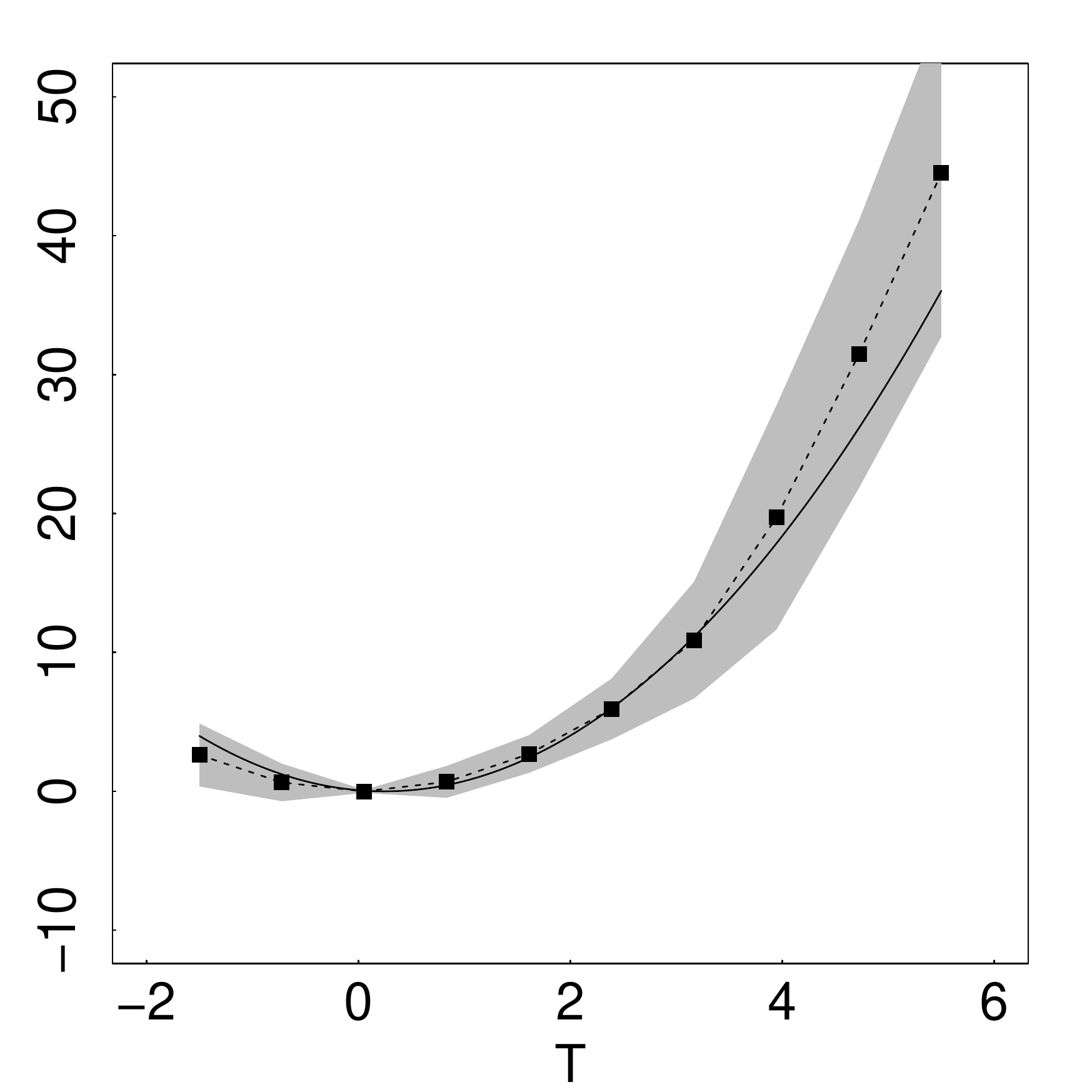}
\includegraphics[width=0.24\textwidth]{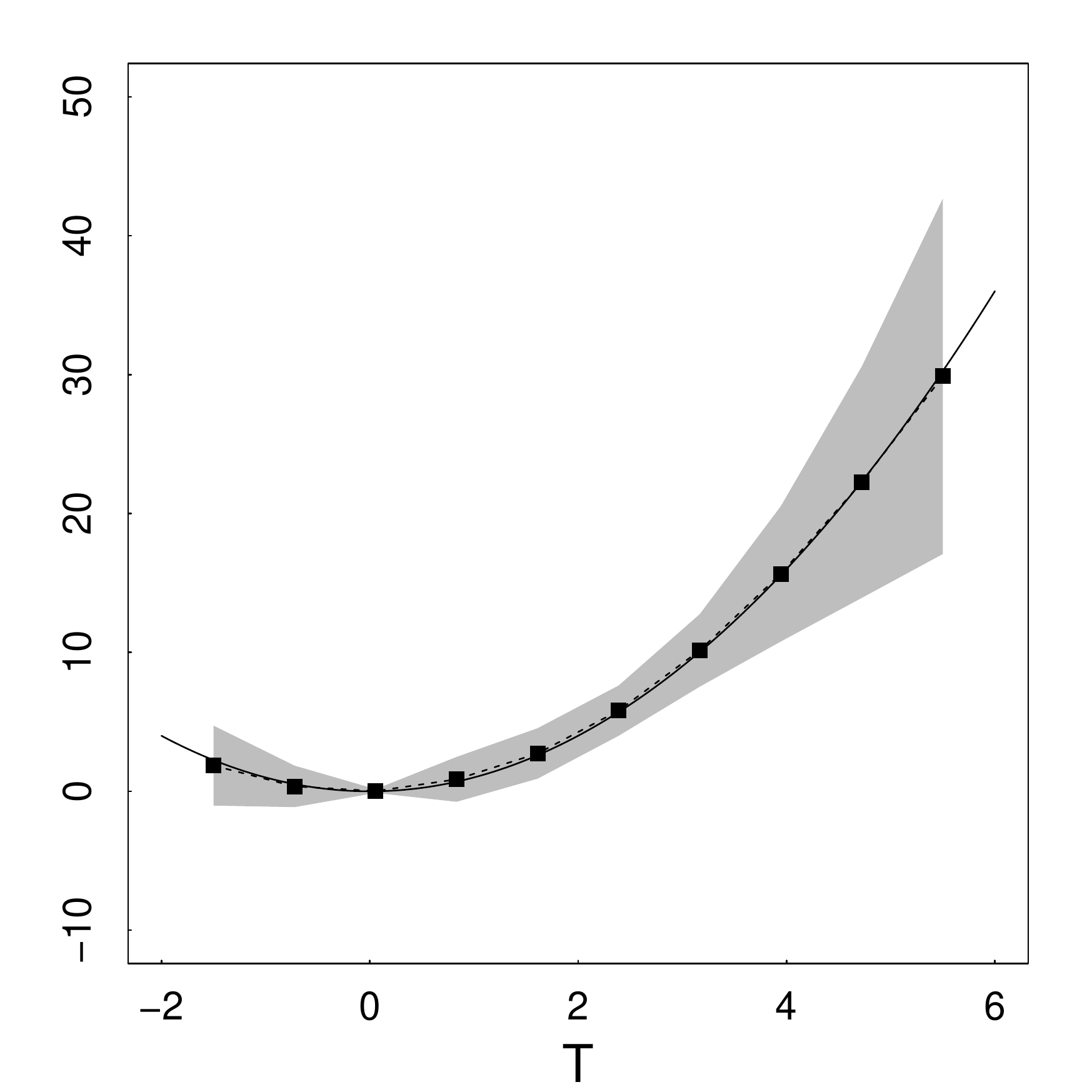}\\
\caption{ {Estimated Relative \drfs~in Simulation
  Study~II for the \scm(\gps), \iw, and 
  \scm(\pfun) Methods. Solid lines represent the true relative \drf\ and dashed
  lines the average of the fitted relative \drfs\ across 1000 simulations.
  Points represent the evenly-spaced grid points. Pointwise
  intervals containing 95\% of the 1000 fitted relative \drfs are shaded grey. Note that the
  scale of the y-axis for the \iw~method under linear generative \drf\ is
  different from others as it has a siginificantly larger standard deviation.
  The \scm(\pfun) method is discussed in Section~{\ref{sec:pfun-drf}}.}
\label{fig:sim-two-b} }
\end{figure}               

Figures~\ref{fig:advan_simu} and \ref{fig:sim-two-b} report the
average of the estimated relative \drfs\ across the simulations
(dashed lines) along with their two standard deviation intervals
(shaded regions). The true relative \drf\ functions are plotted as
solid lines. The first and second rows correspond to the true \drf\
being linear and quadratic, respectively.  The left pair of columns in
Figure~\ref{fig:advan_simu} give results when the fitted model is
linear under the \ivd\ method (column~1) and the \hi\ method
(column~2). The right pair of columns in Figure~\ref{fig:advan_simu}
give results when the fitted model is quadratic under the methods of
\ivd\ (column~3) and \hi\ (column~4).

The \ivd\ method performs reasonably well when the generative model is
linear (row~1).  In this case, the estimated \drf\ is close to the
truth.  When the quadratic model is fitted (column 3), the \drf\ is
estimated with little bias though the estimate has higher variability.
As shown in the lower left quadrant of the figure, the \ivd\ method
exhibits significant bias only when the true \drf\ is of higher order
(i.e., quadratic) than the fitted \drf\ (i.e., linear).  The \hi\
method, on the other hand exhibits appreciable bias even when the
response model matches the true model in its functional dependence on
$t$. Like the \ivd\ method, the bias is most acute when the fitted
model is of lower order than the true model. Unlike with \ivd,
however, the 95\% frequency intervals of \hi\ either skirt or miss the
true value completely across a wide range of treatment values.

{The first two columns of Figure~\ref{fig:sim-two-b} give result for the
\scm(\gps)~and \iw~methods. Both
methods show improvement over \hi~when the generative model is quadratic, although the
variances are larger. For the linear generative model, \scm(\gps) is
comparable to \hi~while \iw~exhibits enormous variance; note the change in
scale of the y-axis. Using the bandwidth suggested by \ffgn\ led to numerical
instability for \iw~in 50 of the 1000 datasets under the linear model. 
Reported results are for the remaining 950 datasets. }


\subsection{Simulation study~III}
\label{sec:sim-three}

\begin{figure}
\spacingset{1}
\centering
\includegraphics[width=0.24\textwidth]{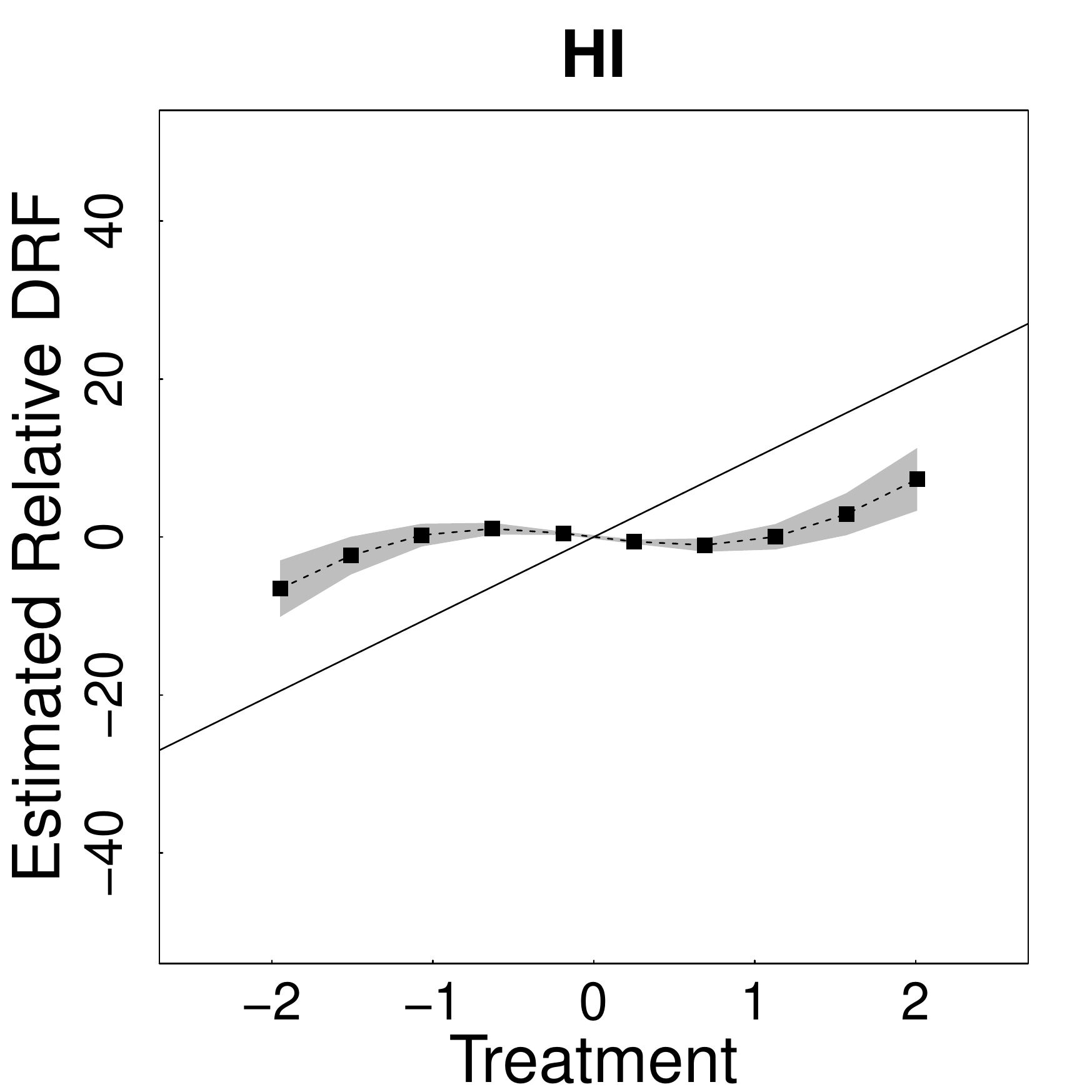}
\includegraphics[width=0.24\textwidth]{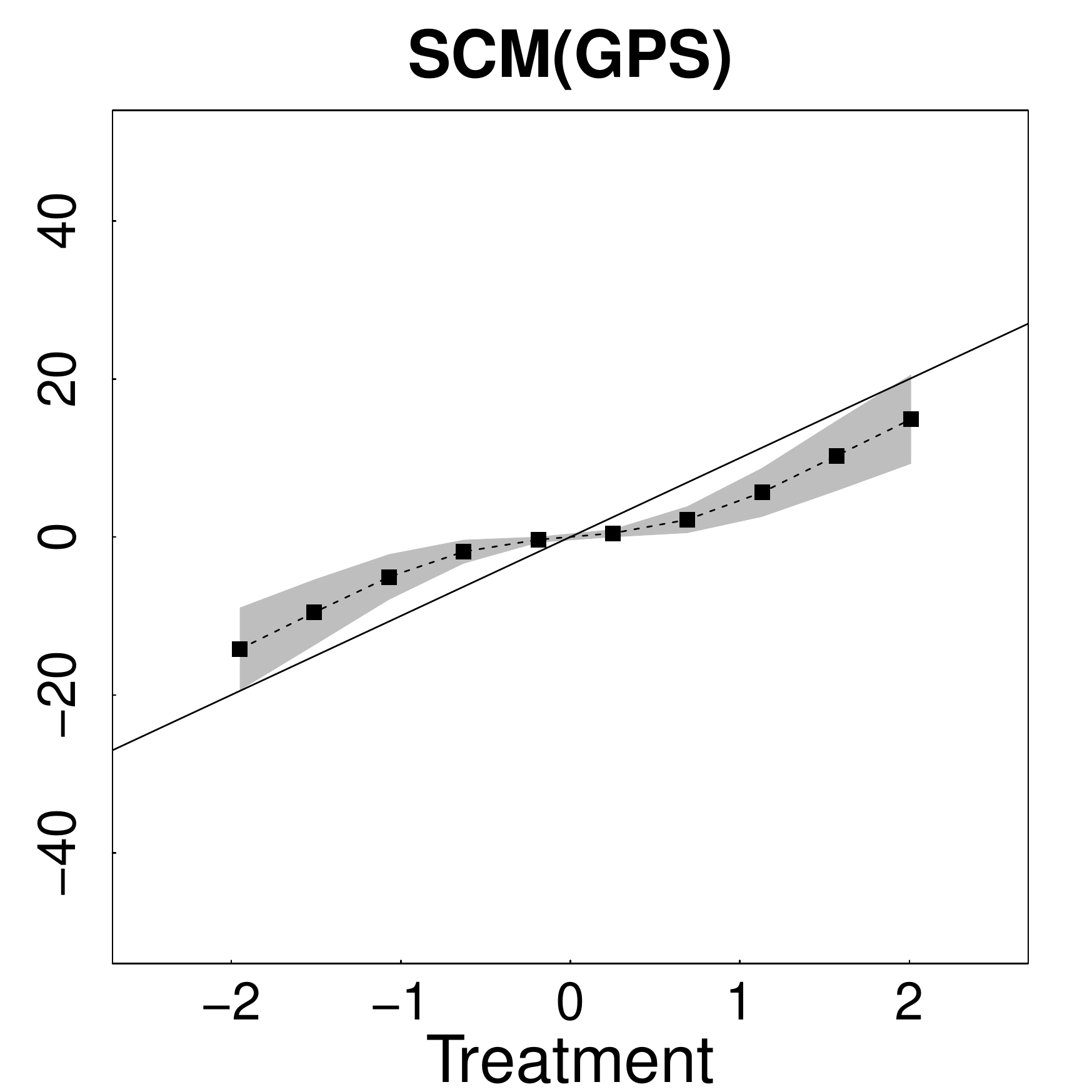}
\includegraphics[width=0.24\textwidth]{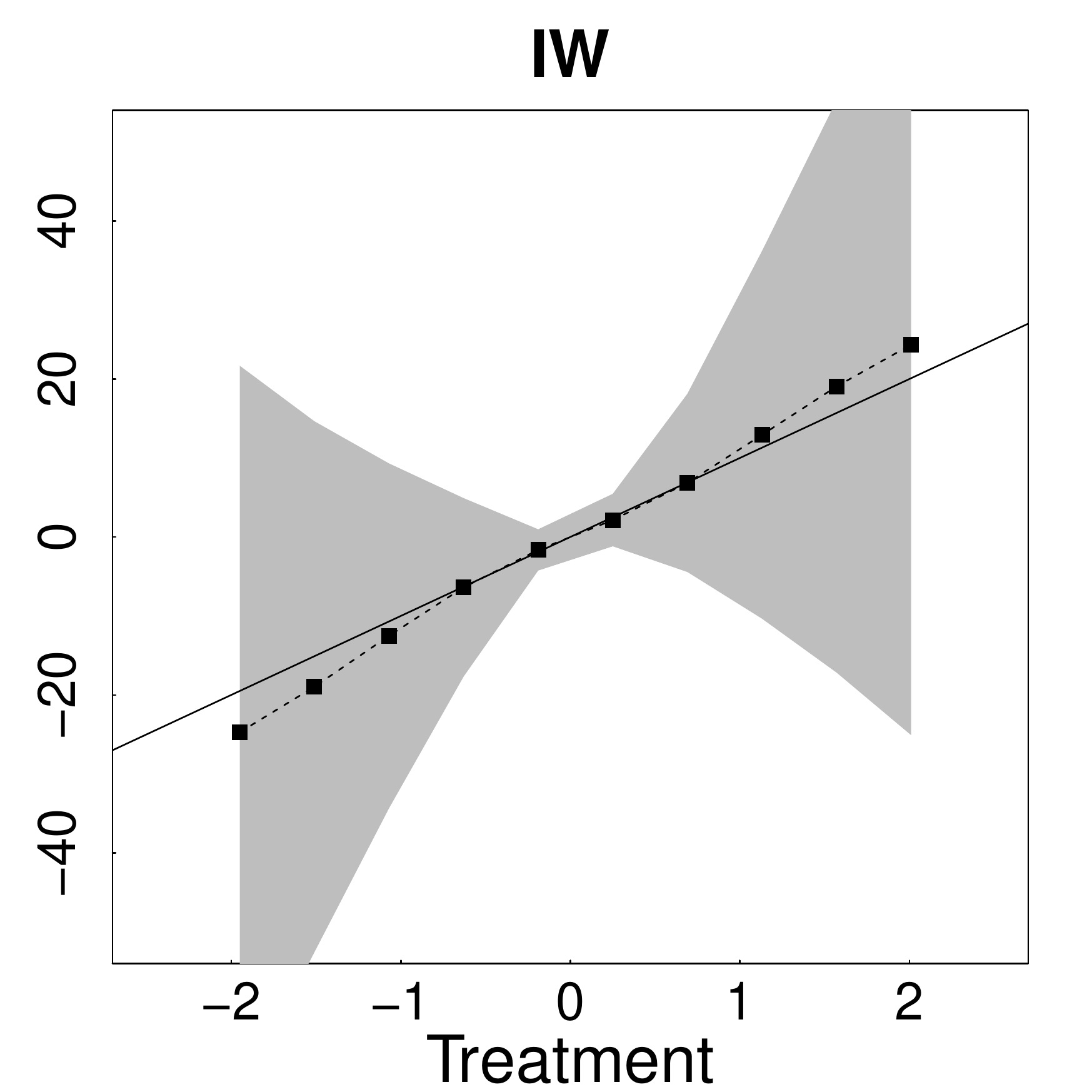}
\includegraphics[width=0.24\textwidth]{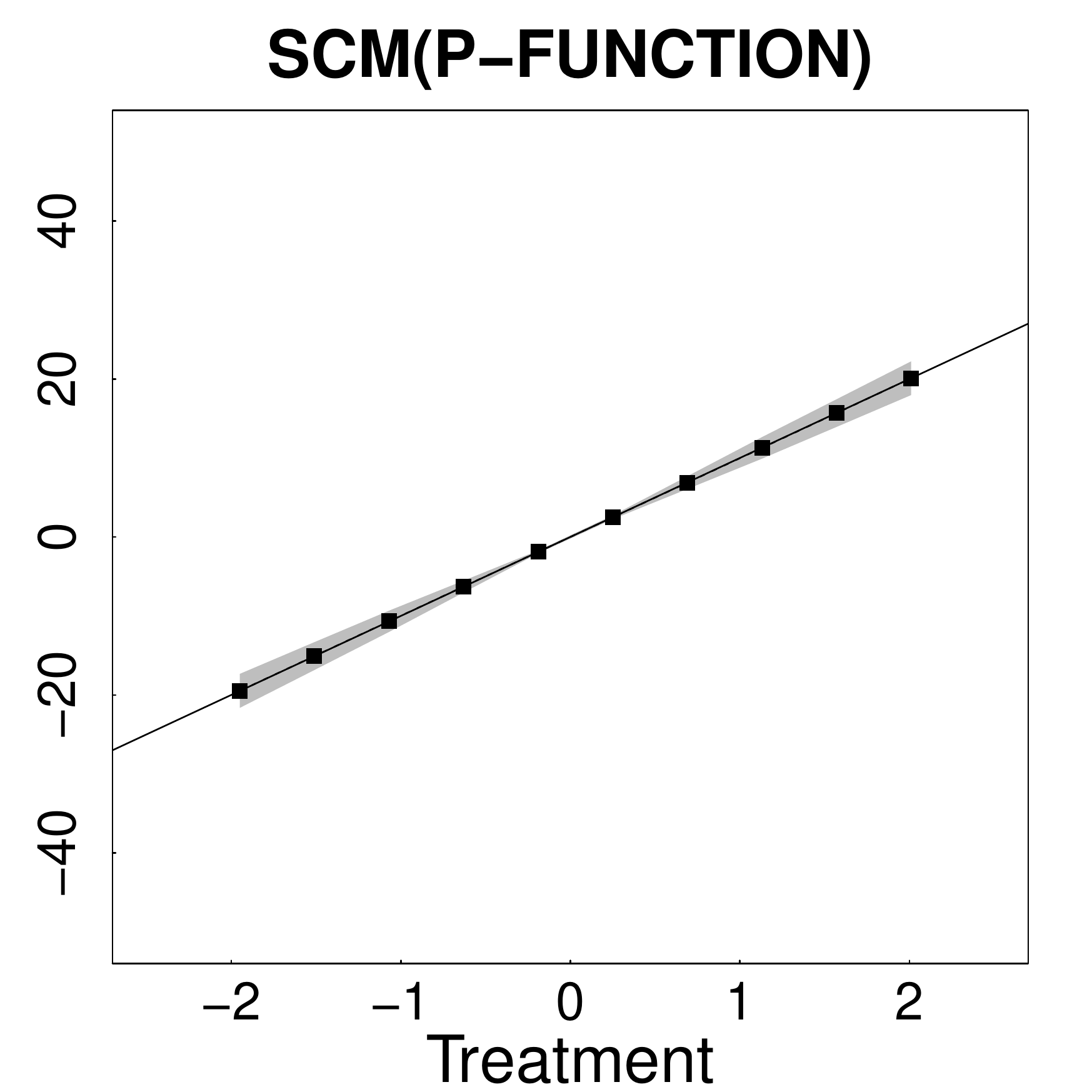}

\caption{{Estimated Relative \drf\ in Simulation III.
Solid lines, dashed lines and  gray regions represent the true relative \drfs\, the means of the 1000 fitted relative
  \drfs\, and  95\%  pointwise intervals. The evaluation points are identical for all plots.  \scm(\gps) exhibits a cyclic artifact and \iw\ is quite unstable. The \scm(\pfun) method proposed in Section~\ref{sec:pfun-drf} again ourperforms the other methods.}}
\label{fig:sim_IV} 
\end{figure}

Although the \scm(\gps) estimate of the \drf\ is biased in Simulation
II, the cyclic bias that it exhibits in Simulation I does not appear
in Figure~\ref{fig:sim-two-b}. To see if the cyclic bias exists in
more complex settings, we extend the well-known simulation study of
\citet{kang:scha:07} to a continuous treatment. This study has a
design similar to Simulation II but with a more realistic set of
covariates. In particular, we independently simulate $Z_{ij} \sim
N(0,1)$ for $i=1,\ldots,2000$ and $j=1,\ldots,4$. We then use the
following generative models for the treatment and outcome variables,
\[
T_i = -Z_{i1}+0.5Z_{i2}-0.25Z_{i3}-0.1Z_{i4}+\sigma_i,
\]
where $\sigma_i \sim N(0,1)$, and
\[
Y_i = 210+27.4Z_{i1}+13.7Z_{i2}+13.7Z_{i3}+13.7Z_{i4}+T_i + \epsilon_i
\]
where $\epsilon_i \sim N(0,1)$. We estimate the relative \drf\ using the methods
of \hi, \scm(\gps), and \iw, on each of 1000 replicated data sets.
In all cases, the treatment model is correctly specified.
Figure~{\ref{fig:sim_IV}} shows that the cyclic bias remains a problem for
\scm(\gps) and that large variances continue to plague \iw.

\subsection{Theoretical considerations and methodological implications}
\label{sec:comp-theory}

To understand the simulation results, we consider the tradeoff in
assumptions required by the \gps\ and \pfun.  In particular, while
\ivd~make a stronger theoretical assumption of a uniquely
parameterized propensity function, \hi\ effectively make the same
theoretical assumption {through their choice of a parametric treatment
  model.}

To flesh this out, we return to the observation that both the
\pfun~and the \gps~can be viewed as generalizations of the \pscore~of
\rr. In the binary case, $p(T \mid {\bm X})$ is uniquely determined by
$e({\bm X}) = p(T=1 \mid X)$.  \ivd\ focus on {\it uniquely}
determining the full conditional distribution of $T$ given ${\bm X}$,
and assume this conditional distribution is parameterized in such a
way that it can be uniquely represented by $\bm\theta$. \hi, on the
other hand, do not constrain the treatment assignment model in this
way and instead focus on the binary \pscore~as the evaluation of $p(T
\mid {\bm X})$ at a particular value of $T$.  It is important to note
that the \gps~does {\it not} uniquely determine $p(T \mid {\bm X})$.
There may be multiple distributions that when evaluated at a
particular $T$ are equal.
The assumption of a uniquely parameterized propensity
function constrains the choice of treatment assignment model that can
be used for a \pfun. In practice, however, the same treatment
assignment models are typically used by both methods.


The assumption of \ivd\ allows  a stronger form of {\it
  strong ignorability of the treatment assignment given the propensity
  function}. In particular, Result~2 of \ivd\ states
\begin{description}
\item[Ignorability of \ivd:] $p\{Y(t) \mid T, e(\cdot \mid {\bm \theta}) \}=  
p\{Y(t)  \mid  e(\cdot \mid {\bm \theta})\}$,
\end{description}  
Whereas, in their Theorem~2.1., \hi\ show
\begin{description}
\item[Ignorability of \hi:] $p_T\{t \mid r(t, {\bm X}), Y (t)\} =
  p_T\{t \mid r(t, {\bm X})\}$ for every $t$.
\end{description}      
In the case where $T$ is categorical, \hi's ignorablity implies that
$\mathbf{1}\{T = t\}$ and $Y(t)$ are independent given $r(t, {\bm
  X})$, where the \gps~is evaluated at the particular value of $t$ in
the indicator function. Although achieving conditional independence of
$Y(t)$ and $T$ would require conditioning on a family of {\gps}, \hi\
provide an insightful moment calculation to show how the response
model described in Section~\ref{sec:gen_pscore} can be used to compute
the \drf. Nonetheless conditioning on either $R$ or $r(t,{\bm X})$,
for any particular value of $t$ does not guarantee that $T$ will be
uncorrelated with the potential outcomes. This restricts the response
models that can be used. Subclassification, for example, is not
feasible unless the classifying variable is low dimensional.  The
advantage of \ivd\ is that independence between $Y(t)$ and $T$ is
achieved by conditioning on a low-dimensional \pfun, as parameterized
by ${\bm \theta}$, enabling the use of a wide-range of response
models.

\subsection{Simulation study IV: The potential cyclic bias of \footnotesize{SCM(GPS)}}
\label{sec:sim-four}

\begin{figure}
\spacingset{1}
\centering
\includegraphics[width=0.28\textwidth]{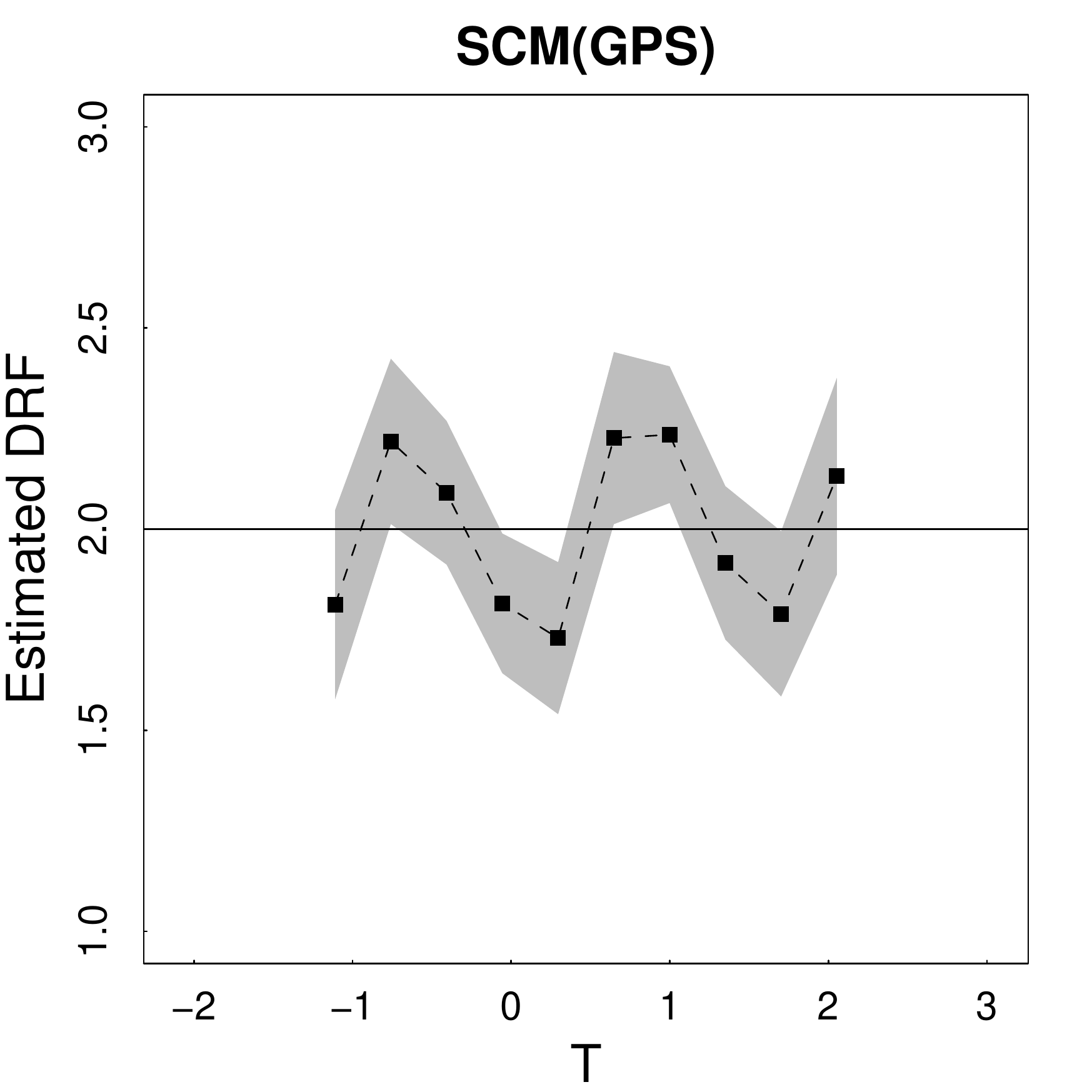}
\includegraphics[width=0.28\textwidth]{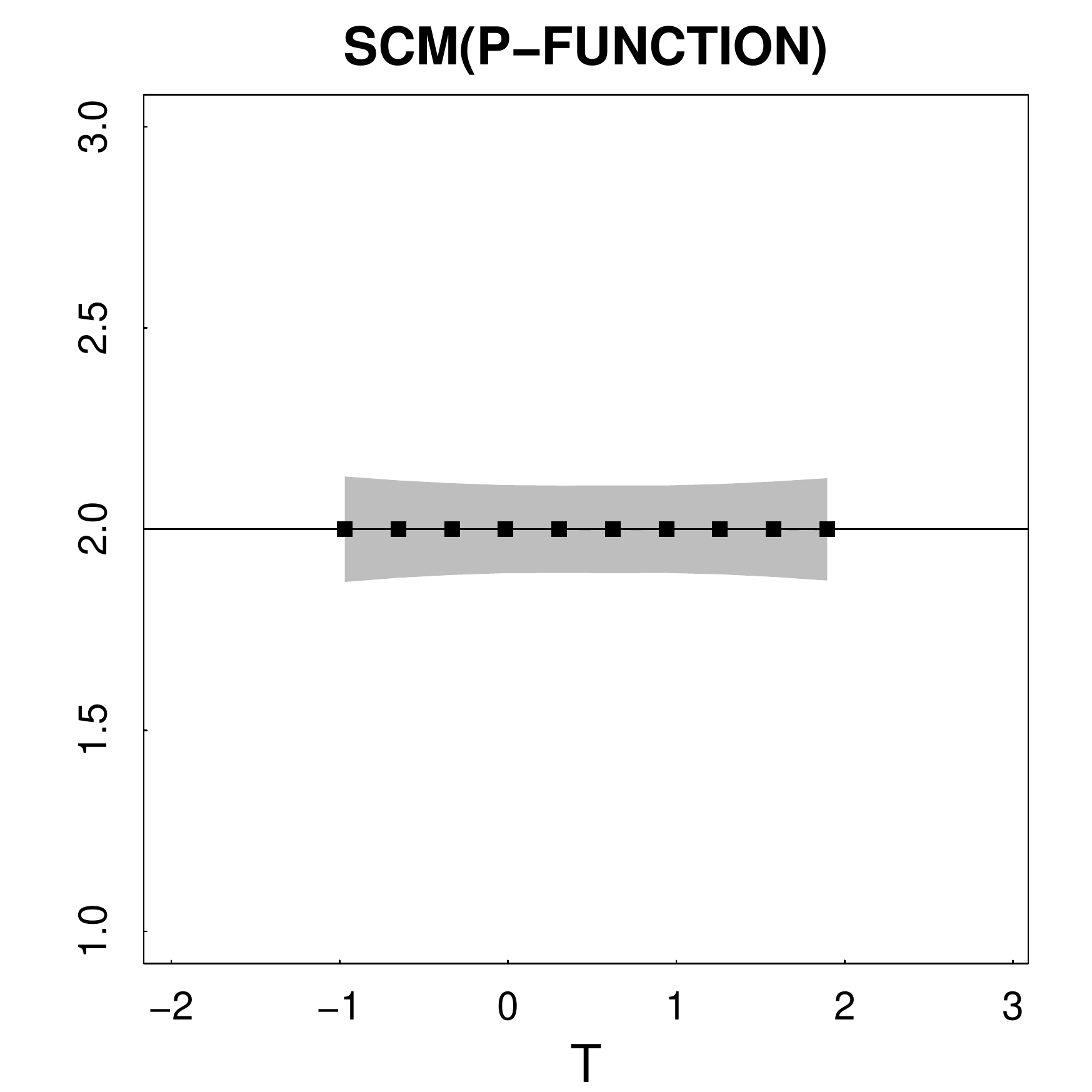}\\
\caption{ Estimated {\small DRF} for Simulation IV. The left
plot correspond to the \scm(\gps) method and {the right plot to a new
method, \scm(\pfun), that is introduced in Section~\ref{sec:improve}. }
Solid lines, dashed lines and  gray regions represent the true relative \drfs\, the means of the 1000 fitted relative
  \drfs\, and  95\%  pointwise intervals. The evaluation points are identical for both plots. 
The \scm(\pfun) method
  does not exhibit the cyclic bias of \scm(\gps).}
\label{fig:sim-four-dr}  
\end{figure}

The \drf\ fitted with \scm(\gps) in Simulations I and III exhibits a
cyclic artifact that does not exist in the underlying \drf. Here we 
present a simulation study {that investigates the origin of this cyclic
bias.} { In particular, we independently generate
$Z_i \sim \rm{Bernoulli}(0.5)$, $X_i \sim {\cal N}(Z_i, 0.01)$, $T_i \sim {\cal
N}(X_i,1)$, and $Y_i \sim {\cal N}(4Z_i,1)$, for $i=1,\ldots,2000$. Using the
correct treatment model, we estimate the \drf\ using
\scm(\gps) at ten evenly spaced theoretical percentiles of $T$.}
We repeat the entire fitting procedure on each of 1000 replicated data sets and
plot the average of the estimated \drf~and their pointwise two
standard deviation intervals in Figure~\ref{fig:sim-four-dr}.
The cyclic bias of the fit is evident.

\begin{figure}
\spacingset{1}
\centering
\includegraphics[width=0.24\textwidth]{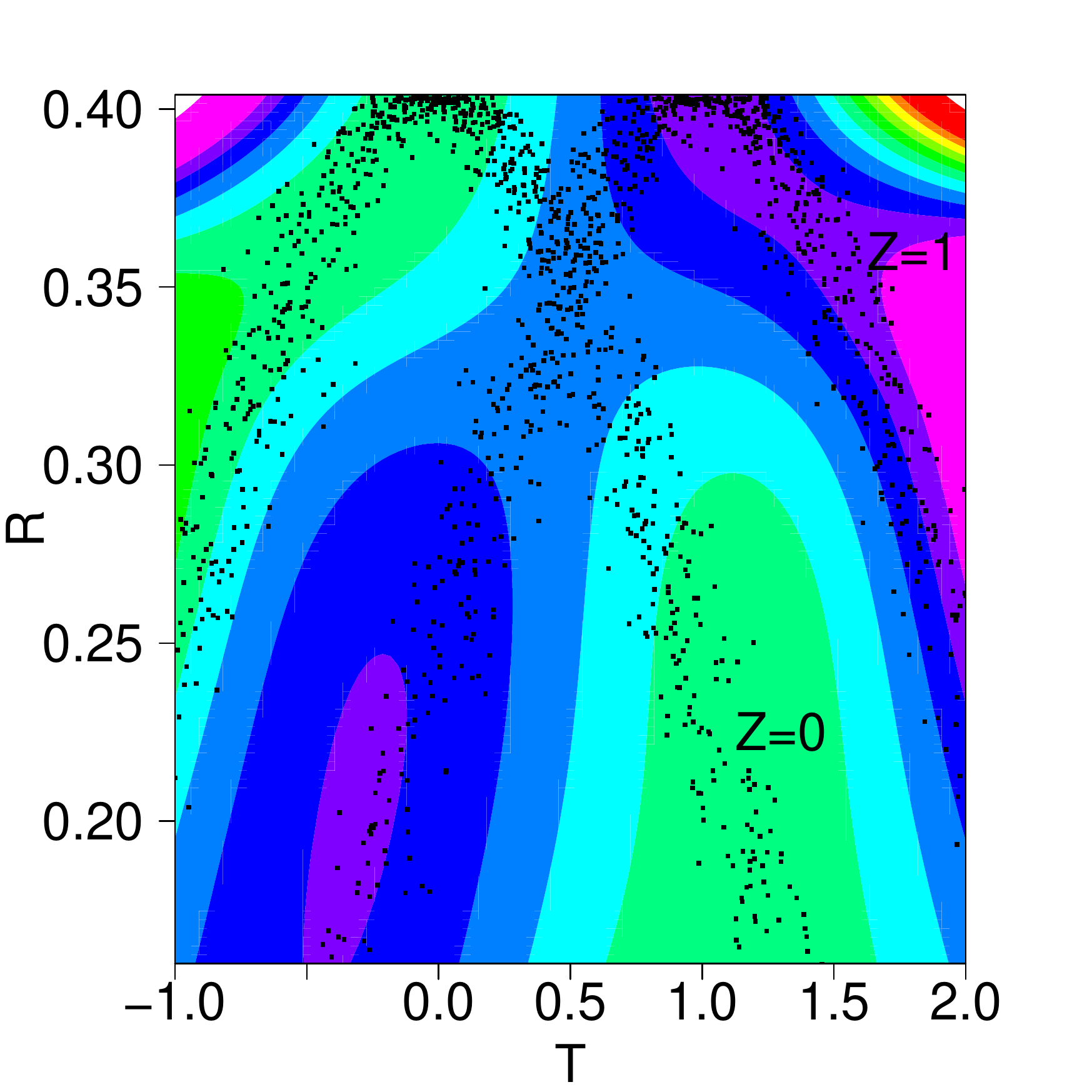}
\includegraphics[width=0.24\textwidth]{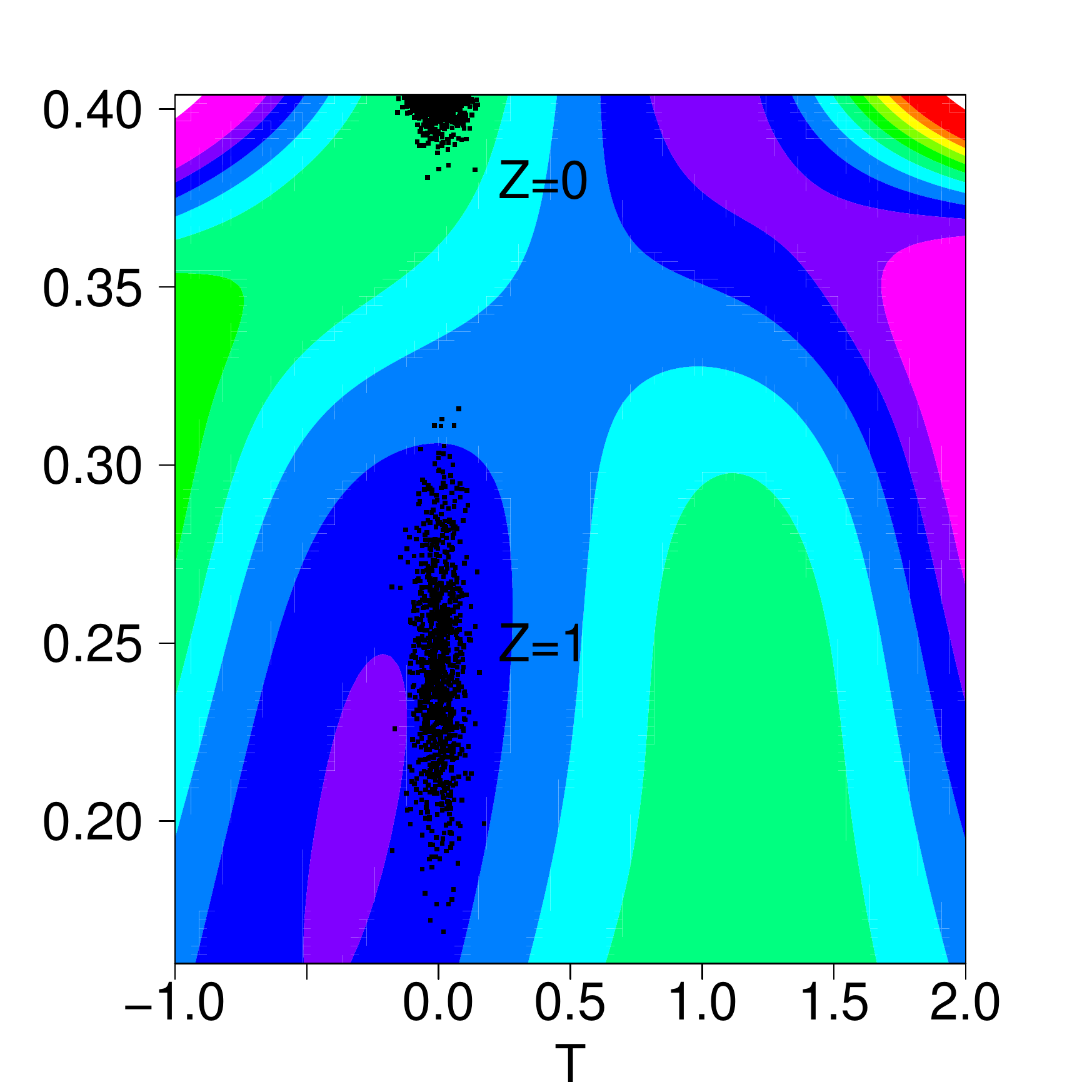}
\includegraphics[width=0.24\textwidth]{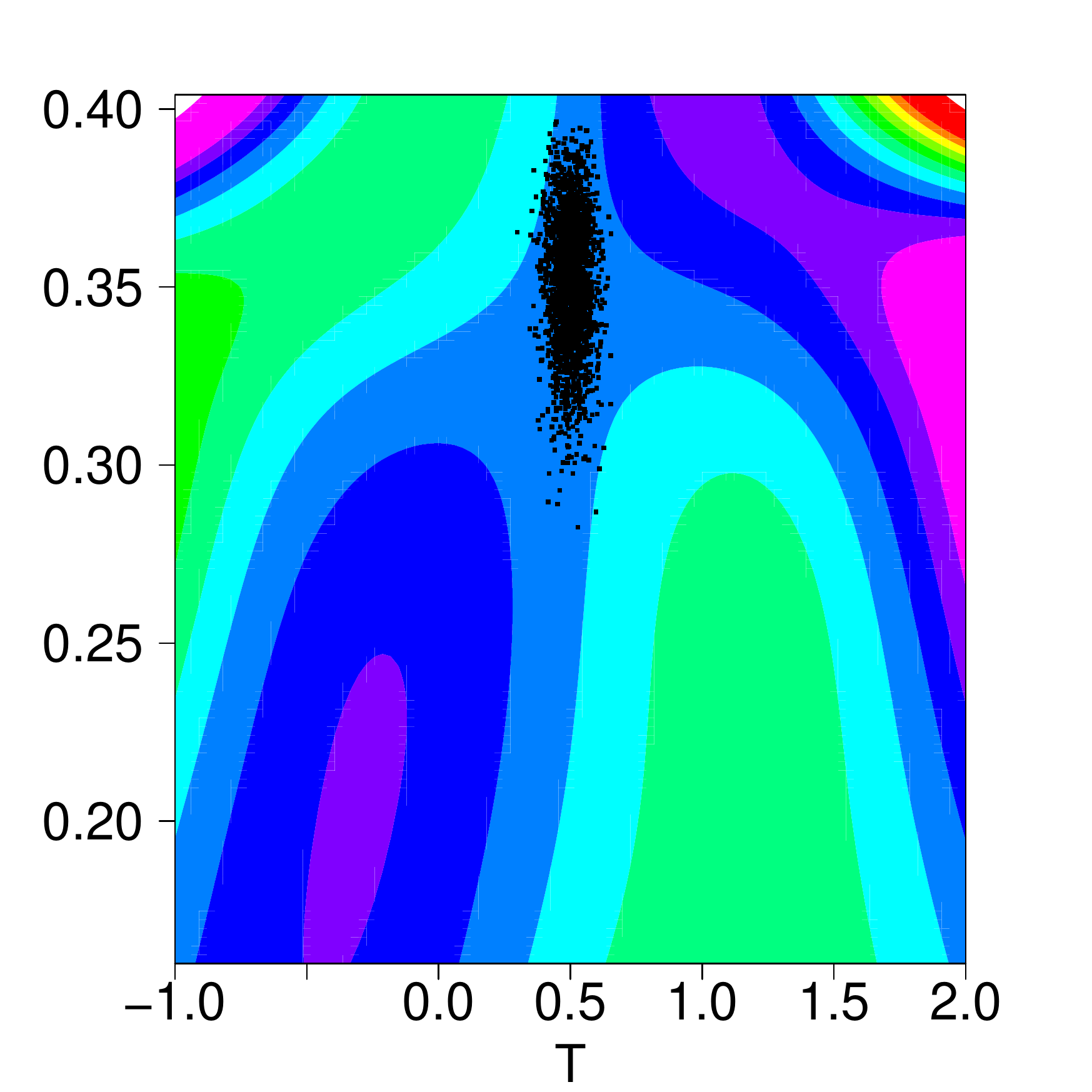}
\includegraphics[width=0.24\textwidth]{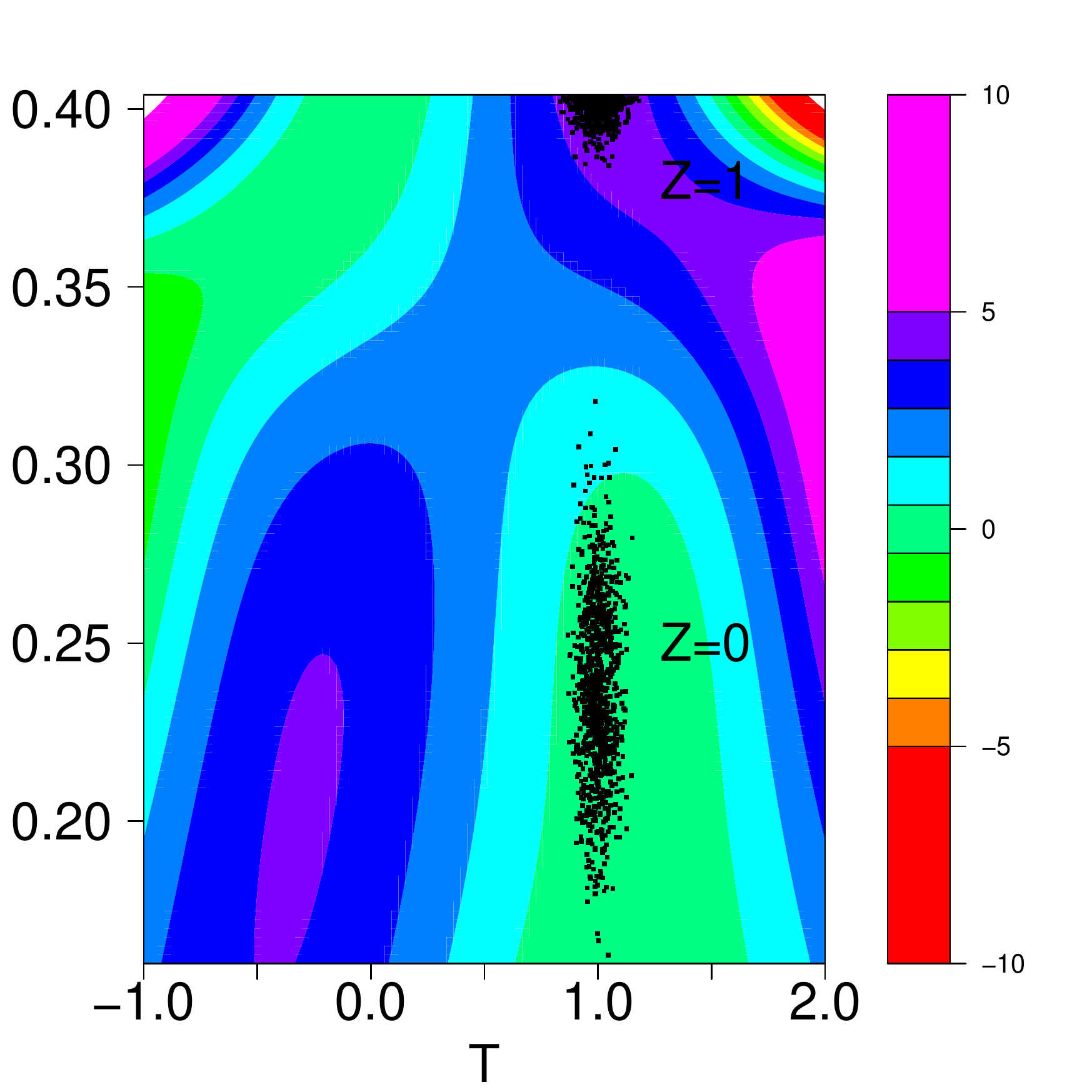}\\
\caption{The scatter plot of $T$ and \gps\ on top of the heat map of
the fitted response model in Simulation IV. Left most column plots the scatter
plot of ($T$,$R$). The rest three columns plots the scatter plot of
($t_i$,$r(t_i,{\bm X}_i$) ), with $t_i$ equal to 0, 0.5, and 1. (We generate
some random noise in the $T$ direction for ease of plotting.) The heat map is
based on the fitted response model $\hat f(T,R)$. }
\label{fig:sim-four-heatmap}
\end{figure}  

To see the source of the cyclic bias, we plot the fitted response
model given in (\ref{eq:scm-gps-model}) as a heat map in the leftmost
panel of Figure 14.  The two bell-shaped curves that appear in the
plotted values of ($T_i, \hat R_i$) stem from the definition of the
\gps\ -- it is the value of the fitted density function of $T$. By the
simulation design, $X$ clusters around the two values of $Z$; these
two clusters correspond to the two bell-shaped curves. Generally
speaking, the overlapping bell-shaped curves induce a cyclic patter in
the fitted response model. To estimate the \drf\ at $t$, the fitted
response model is evaluated and averaged over each $\hat r(t, {\bm
  X}_i)$. As $t$ increases, the cyclic patter in the fitted response
model leads to a corresponding pattern in the fitted DRF (see
Figure~\ref{fig:sim-four-dr}). 

The patterned behavior of the \gps\ means that the response model is
particularly difficult to accurately represent, even with a flexible
non-parametric model. The resulting complexity of the response model
means that extrapolation is especially dangerous. Unfortunately, this
is inevitable: when estimating the DRF we must evaluate the fitted
response model at each value of $\hat r(t, {\bm X}_i)$, including at
unobserved combinations of $t$ and ${\bm X}_i$, see
(\ref{eq:scm-gps-model}) and (\ref{eq:scm-gps-est}).  This is a
difficulty with the underlying response model, regardless of the
choice of fitted response model. Although Simulation IV uses a simple
setting to clearly explain the cyclic bias of \scm(\gps), problems
persist in more complex settings (see Figures~\ref{fig:sim1-1},
\ref{fig:sim_IV}, and~\ref{fig:smoking_sim}).

\section{Estimating the {\large DRF} Using the {\large P-FUNCTION}}
\label{sec:improve}

{In this section, we propose a new method  for robust estimation
of the \drf\ using the {\bf \pfun}. Appendix A discusses another new robust \gps-based method.}

\subsection{Using the {\small P-FUNCTION} in a {\small SCM} to estimate the {\small DRF}}
\label{sec:pfun-drf}

\ivd\ developed the \pfun\ to estimate the average treatment effect,
rather than the full \drf. Nonetheless we use the framework of \ivd\
to compute the \drf\ in Simulation studies~I and II (see
Figures~\ref{fig:sim1-1}~and~\ref{fig:advan_simu}). The method we
employ, however, is constrained by its dependence on the parametric
form of the within subclass model. Practitioners would generally
prefer a robust and flexible \drf, and here we propose a procedure
that allows such estimation. We view this estimate as the best
available for non-binary treatments in an observational study.

We begin by writing the \drf\ as
\begin{equation}  
 E[Y(t)] = \int E[Y(t) \mid {\bm \theta}] \ p({\bm \theta}) d{\bm \theta} = 
\int E[Y(T) \mid {\bm \theta}, T=t] \ p({\bm\theta}) d{\bm \theta}
\label{eq:IvD-SCM2-theory}  
\end{equation}
where the first equality follows from the law of iterated expectation
and the second from the strong ignorability of the treatment
assignment given the \pfun.  We estimate  the \drf\ using the
right-most expression in (\ref{eq:IvD-SCM2-theory}) which we flexibly
model using a \scm,
\begin{equation}  
E[Y(T) \mid {\bm \theta}, T=t] \ = \ f({\bm \theta},T),
\label{eq:IvD-SCM2}  
\end{equation}
where $f(\cdot)$ is a smooth function of  ${\bm \theta}$ and $T$. In practice we
replace $\bm\theta$ by $\hat{\bm\theta}$ from the fitted
treatment model. We approximate the integral in
(\ref{eq:IvD-SCM2-theory}) by averaging over the empirical distribution of
$\hat{\bm\theta}$, to obtain an estimate of the \drf\ using a \scm\ of the
\pfun,
\begin{equation}
 \hat E[Y(t)] \ = \ {1\over n} \sum^n_{i=1} \hat f({\hat {\bm \theta_i}},t),
\label{eq:scm-drf}
\end{equation}
where $\hat f(\cdot)$ is the fitted \scm. {We refer to this method of
estimating the \drf~as the \scm(\pfun) method and typically} evaluate
(\ref{eq:scm-drf}) on a grid of values of $t_1,\ldots, t_D$ evenly spaced in 
range of the observed treatments, as suggested by \hi. Bootstrap standard errors
are computed on the same grid.
 
Comparing (\ref{eq:scm-gps-model})--(\ref{eq:scm-gps-est}) with
(\ref{eq:IvD-SCM2})--(\ref{eq:scm-drf}), \scm(\gps) and \scm(\pfun) are
algorithmically very similar. The primary difference is the choice between the
\pfun\ and \gps\ in the response model. As we shall see, this change has a
siginificant effect on the statistical properties of the estimates.
Simpy put, ${\bm \theta}$ is a much better behaved predictor variable than is
$R$.
When using Gaussian linear regression for the treatment model, for example,
$\bm\theta={\bm X}_i^\top \beta$, whereas $R$ is the Gaussian density evaluated
at $T$. As illustrated in Section~\ref{sec:sim-four}, the dependence of the
\gps\ on $t$ and the non-monotonicity of this dependence both  complicate the
response model and pose challenges to robust estimation.

Computing $\hat E[Y(t_0)]$ with (\ref{eq:scm-drf}) for some particular $t_0$
involves evaluating $\hat f(\cdot, t_0)$ at every observed value of
$\hat\theta_i$. Invariably, the range of $\hat\theta$ observed among units with
$T$ near $t_0$ is smaller than the total range of $\hat\theta$, at least for
some values of $t_0$. Thus, evaluating (\ref{eq:scm-drf}) involves some degree
of extrapolation, at least for some values of $t$. Luckily, this problem is
relatively easy to diagnose with a scatter plot of the observed values of $(T_i,
\hat\theta_i)$. The estimate in (\ref{eq:scm-drf}) may be biased for values of
the treatment where the range of observed $\hat\theta_i$ is relatively small. As
we illustrate in our simulation studies, however, (\ref{eq:scm-drf}) appears
quite robust and this bias is small relative to the biases of other available
methods.

\subsection{Simulation studies I--IV revisited}
\label{sec:sim-one-b}



\begin{figure}
\spacingset{1}
\centering
\includegraphics[width=0.28\textwidth]{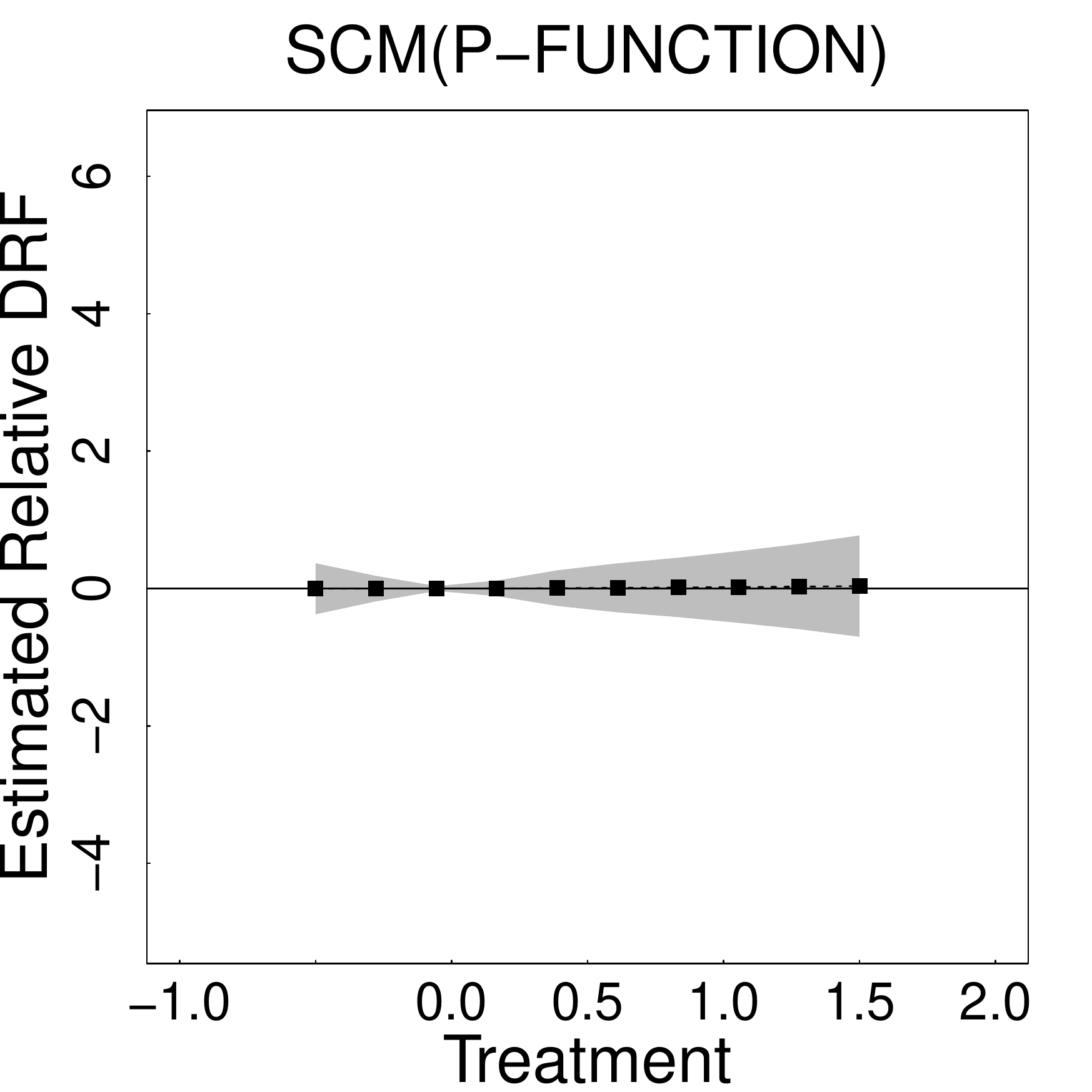}\\
\caption{{Estimated Relative {\small DRF} Using the \scm(\pfun) Method in
Simulation Study I. The solid (dashed) lines represent the true  
(fitted) relative \drf, the 95\% confidence bands are plotted in grey, and the
grid points are identical to those in Figure~\ref{fig:sim1-1}. The fitted relative \drfs\ are much
improved compared with those of \hi, \scm(\gps), and \iw~but without the linear
assumptions of \ivd\ (see Figure~\ref{fig:sim1-1}). }
\label{fig:sim-one-b} }
\end{figure}

We now revisit the simulation studies from Section~\ref{sec:comp},
which illustrate the potentially misleading results {or high variance
  of existing \gps-based methods.} Here, we compare these results with
those of \scm(\pfun). In all cases, \scm(\pfun) was fitted using the
same (correct) treatment assignment model and with the same
equally-spaced grid points.  When fitting the \scm, we continue to use
the penalized cubic regression spline basis for both parameters ($R$
and $T$) and a tensor product to construct a smooth fit of the
continuous function $f(\theta, T)$ (see {\tt mgcv} R-package
documentation). Figure~\ref{fig:sim-one-b} and the right panel of
Figure~\ref{fig:sim-four-dr} show the fitted (relative) \drf\ for
\scm(\pfun) in Simulations I and IV, respectively. The performance of
\scm(\pfun) is a dramatic improvement over that of the \gps-based
methods (see Figures~\ref{fig:sim1-1} and \ref{fig:sim-four-dr}).

Figure~\ref{fig:sim-two-b} presents the results of the \scm(\pfun) method in Simulation study II.
Comparing Figure~\ref{fig:sim-two-b} with Figure~\ref{fig:advan_simu} again
illustrates the advantages of the proposed method. The fits in
Figure~\ref{fig:advan_simu} are quite dependent on the parametric choice of the
response model, whereas the non-parametric fits illustrated in
Figure~\ref{fig:sim-two-b} do not require a parametric form. Among the
non-parametric methods, the advantage of \scm(\pfun) is clear. It essentially
eliminates bias with only a small increase in variance in this simulation.

Finally, the rightmost panel of Figure~\ref{fig:sim_IV} shows that \scm(\pfun) performs very well in Simulation III with no noticable bias and small variance.  Overall, \scm(\pfun) appears to provide more robust estimates of a \drf\ in an observational study than do any other available methods.

  

\section{Example: The effect of smoking on medical expenditures}
\label{sec:num}

\subsection{Background}
\label{sec:smoking-intro}

We now illustrate our proposed methods by estimating the \drf\ of
smoking on annual medical expenditures. The data we use were extracted
from the 1987 National Medical Expenditure Survey (NMES) by
\citet{john:domi:gris:zege:03}.  Its detailed information about
frequency and duration of smoking allows us to continuously
distinguish among smokers and estimate the effects of smoking as a
function of how much they smoke. The response variable, medical costs,
is verified by multiple interviews and additional data from clinicians
and hospitals.  \ivd\ used the propensity function to estimate the
average effect of smoking on medical expenditures. We extend their
analysis and study estimation of the full \drf. Like \ivd, we adjust for the
following subject-level covariates: age at the times of the survey,
age when the individual started smoking, gender, race (white, black,
other), marriage status (married, widowed, divorced, separated, never
married), educational level (college graduate, some college, high
school graduate, other), census region (Northeast, Midwest, South,
West), poverty status (poor, near poor, low income, middle income,
high income), and seat belt usage (rarely, sometimes, always/almost
always).

To measure the cumulative exposure to smoking based on the self-reported
smoking frequency and duration, \citet{john:domi:gris:zege:03} proposed using
the variable of {\tt packyear}, which is defined as
\begin{equation}
{\tt packyear} = \frac{\rm{number~of~cigarettes~per~day}}{20} \times
\rm{number~of~years~smoked}.
\label{eq:def_packyear}
\end{equation} 
We use ${\rm log}({\tt packyear})$ as our treatment variable. We follow
\citet{john:domi:gris:zege:03} and \ivd\ and discard all individuals with
missing values and conduct a complete-case analysis, yielding a sample of 9,073
smokers. Although in general complete-case propensity-score-based analyses
produce biased causal inference unless the data are missing completely at random
\citep{dago:rubi:00}, \citet{john:domi:gris:zege:03} showed that accounting for
the missing data using multiple imputation did not significantly affect their
results.

Because the observed response variable, self-reported medical
expenditure, denoted $Y$, is semicontinuous, we use the two-part model
of \citet{duan:mann:morr:newh:83}. This involves first modeling the
probability of spending some money on medical care, ${\rm Pr}(Y>0\mid
T,\bm{X})$, where $T={\rm log}({\tt packyear})$, and $\bm X$
represents the covariates; and then modeling the conditional
distribution of $Y$ given $T$ and $\bm X$ for those who reported
positive medical expenditure. To illustrate and compare methods for
computing the \drf, we concentrate on the second part of this model.
Because the distribution of $Y$ is skewed, we consider the model
$p({\rm log}(Y)\mid Y>0,T,{\bm X})$.

For our treatment assignment model, we use a Gaussian linear regression adjusted for all available covariates and the second order terms of two age covariates. The model was fitted using sampling weights provided with the original data.  This is the same treatment assignment model used by \ivd\ who demonstrate that it achieves adequate balance.

\subsection{Simulation study based on the smoking data}
\label{sec:smoking-sim}

\begin{figure}[t]
\spacingset{1}
\centering
\includegraphics[width=0.4\textwidth]{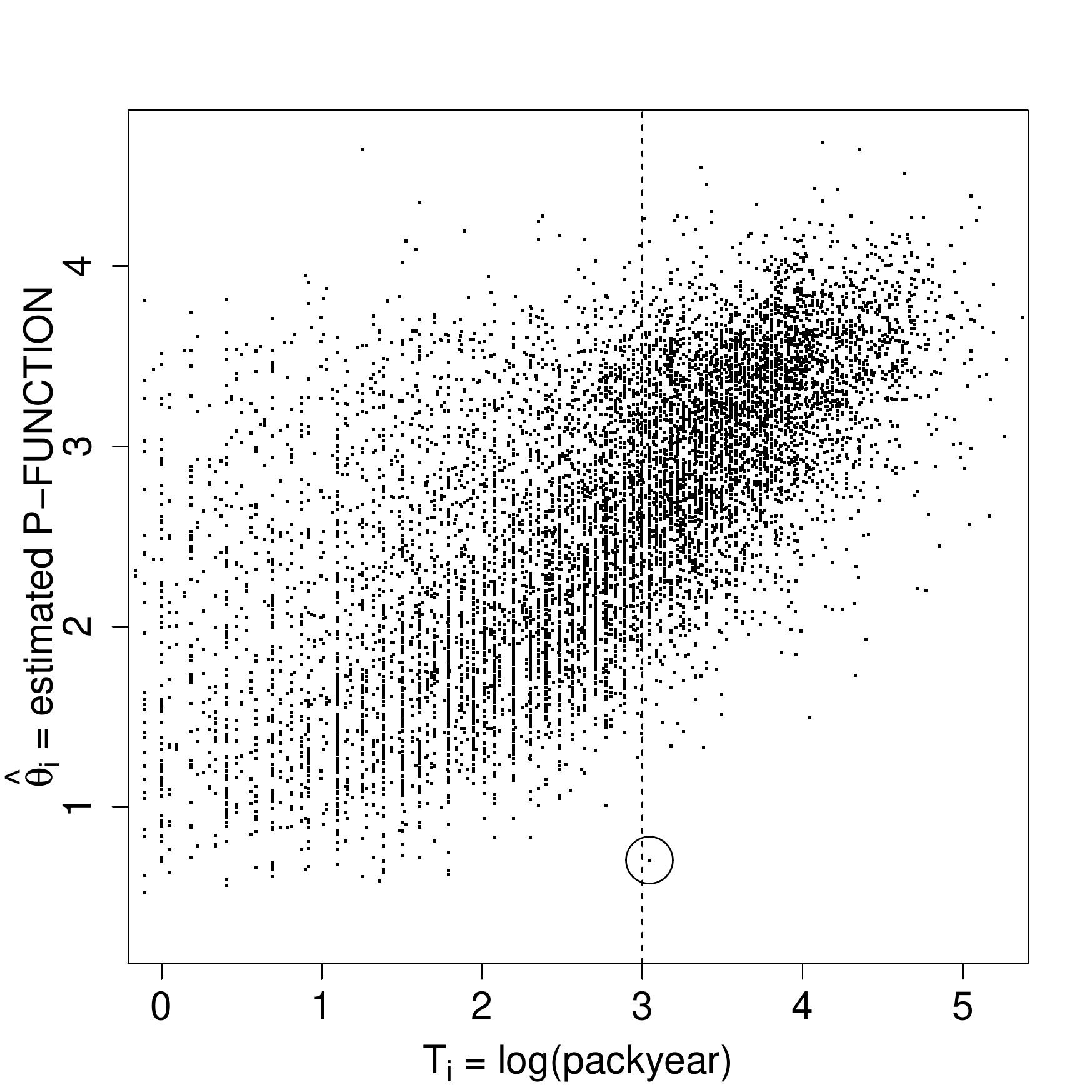}
\caption{Diagnostic for {the \scm(\pfun) method}. Because the range of the
$\hat\theta_i$ when $T_i > 3$ is less  than the overall range of $\hat\theta_i$ estimating the \drf\ for $t>3$
involves extrapolation under the \scm\ and thus possible bias.  The single
individual with $T_i$ slightly larger than three and $\hat\theta_i$ less than
one is circled. To a certain extent this datapoint should mitigate bias for $t$
near three. Nonetheless, the fitted \drf\ for $t>3$ may be seriously biased.}
\label{fig:smoking_sim_pfun}
\end{figure} 

This simulation study aims to mimic the characteristics of the actual
data with the goal of comparing the statistical properties of the
proposed methods in as realistic a setting as possible.  In
particular, we do not alter the observed covariates or treatment and
use the same fitted treatment model used by \ivd. Figure~\ref{fig:smoking_sim_pfun} presents a scatter plot of
the observed treatment variable, $T_i=\log({\tt pack year})$, and the
values of the \pfun\ from the fitted treatment assignment model,
$\hat\theta_i$. As discussed in Section~\ref{sec:pfun-drf}, this plot
can be used as a diagnostic for \scm(\pfun).  Recall that this
estimate requires that we fit a \scm\ to predict the response variable as a function of $T_i$ and
$\hat\theta_i$. To estimate the \drf\ at $t$ we must evaluated the
fitted \scm\ at $(t,\hat\theta_i)$ using each observed $\hat\theta_i$
in the data set. This involves extrapolation and thus possible
bias if the range of $\theta_i$ at a particular value of $t$ is less
than the overall range of $\theta_i$. Judging from
Figure~\ref{fig:smoking_sim_pfun}, this is a concern for $t$ greater
than about three. There is a solitary individual with $T_i$ slightly
above three and $\hat\theta_1 <1$ that is circled in
Figure~\ref{fig:smoking_sim_pfun}.  Even this single point can guard
against significant extrapolation bias for $t$ less than three, but
the concern remains for $t$ in the range of 4 to 5. We emphasize
that this diagnostic is preformed before the response model is 
fit. 

To explore the robustness of the methods to different \drfs, we
simulate the response variable under three known \drfs\ and attempt to
reconstruct them using the \hi, {\bf \scm(\gps), \iw}, covariance
adjustment \gps, and \scm(\pfun) methods. In particular, we assume $\log(Y_i(t))
\sim {\cal N}({\rm E}[\log(Y_i(t))],0.5^2)$ where $t=\log({\tt packyear})$ and
consider three functional forms for ${\rm E}[\log(Y_i(t))]$:
\begin{eqnarray}
{\rm Quadratic \ DRF:} & & 
	E[\log(Y_i(t))] \ = \ \frac{4}{25} \cdot t^2 + [\log({\tt
          age}_i)]^2 \nonumber \\
{\rm Piecewise-Linear \ DRF:} & &  
	E[\log(Y_i(t))] \ = \
	\begin{cases}
	  -4 - 0.5 \cdot t + [\log({\tt age}_i)]^2, &t \leq 2 \cr 
	  -5 - 2.3 \cdot (t-2) + [\log({\tt age}_i)]^2, &t > 2,
	\end{cases} \nonumber \\
{\rm Hockey-Stick \ DRF:} &  &
	E[\log(Y_i(t))] \ = \
	\begin{cases}
	  -8.1 + [\log({\tt age}_i)]^2, &t \leq 3 \cr 
	  -8.1 + 1.5 \cdot (t-3)^2 + [\log({\tt age}_i)]^2, &t > 3,
	\end{cases} \nonumber
\end{eqnarray} 
where  {\tt age} is the age at the time of the survey. We include {\tt age}
because it is the covariate most correlated with $\log({\tt packyear})$ and thus
most able to bias a naive analysis. Each of the response models was fitted using
the sampling weights.\footnote{ It is not obvious how best to incorporate
the sampling weights when using the iw method of FFGN. We construct new
weights by multipling the weights required by the IW method and the sampling weights. Ignoring the
sampling weights leads to similar results. }

\begin{figure}
\spacingset{1}
\centering
\textsf{\textbf{\large Generative Model: Quadratic}}\\ \smallskip
\includegraphics[width=0.24\textwidth]{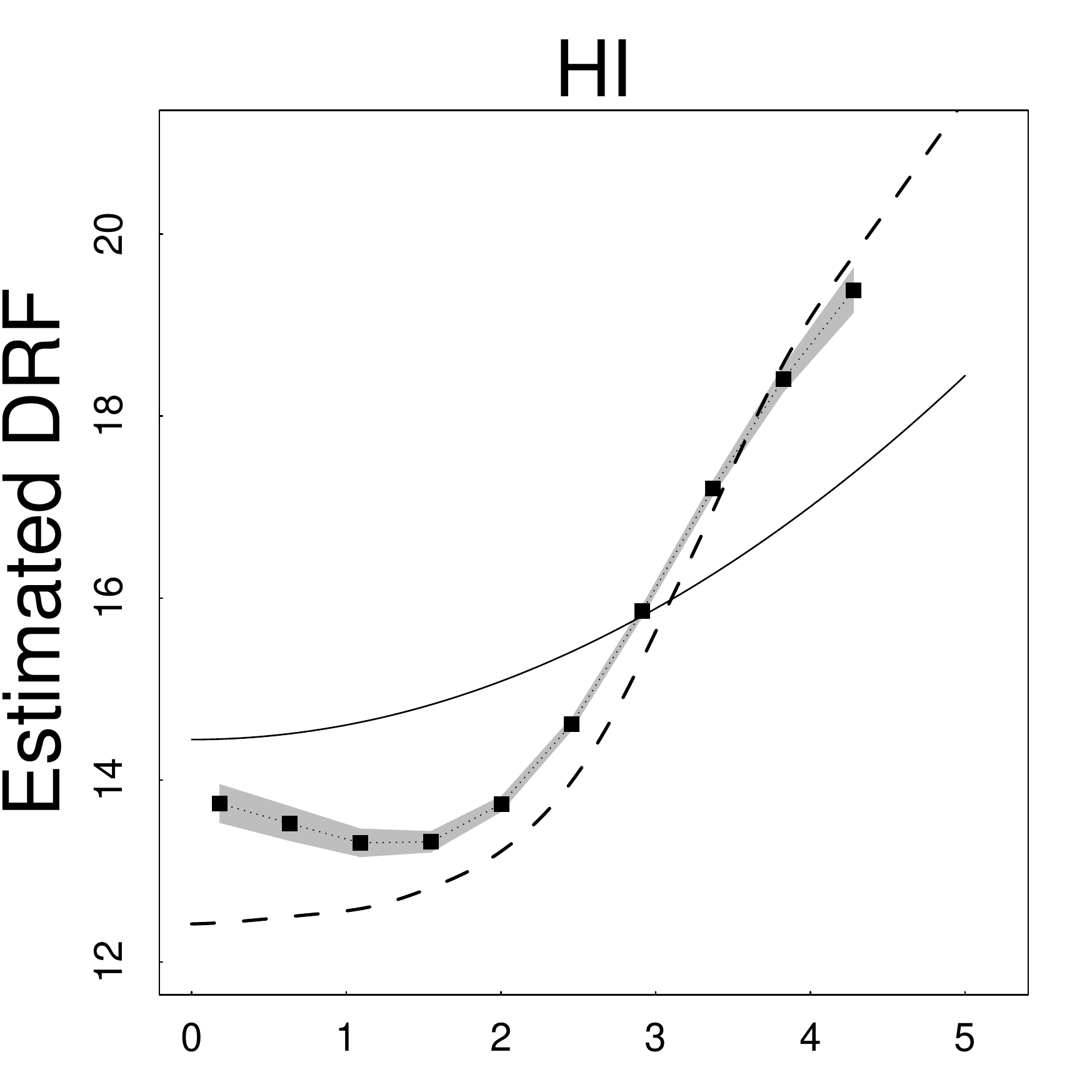}
\includegraphics[width=0.24\textwidth]{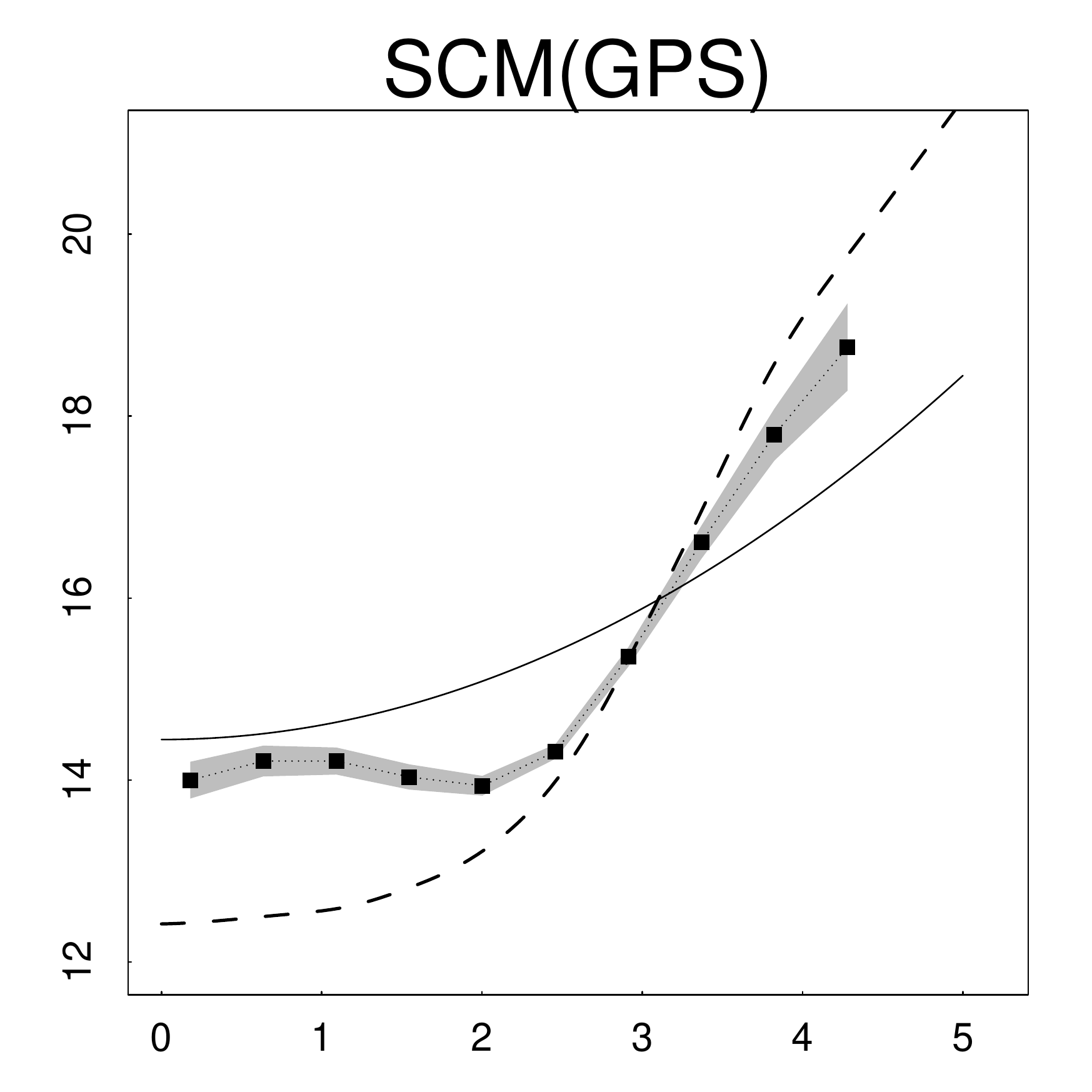}
\includegraphics[width=0.24\textwidth]{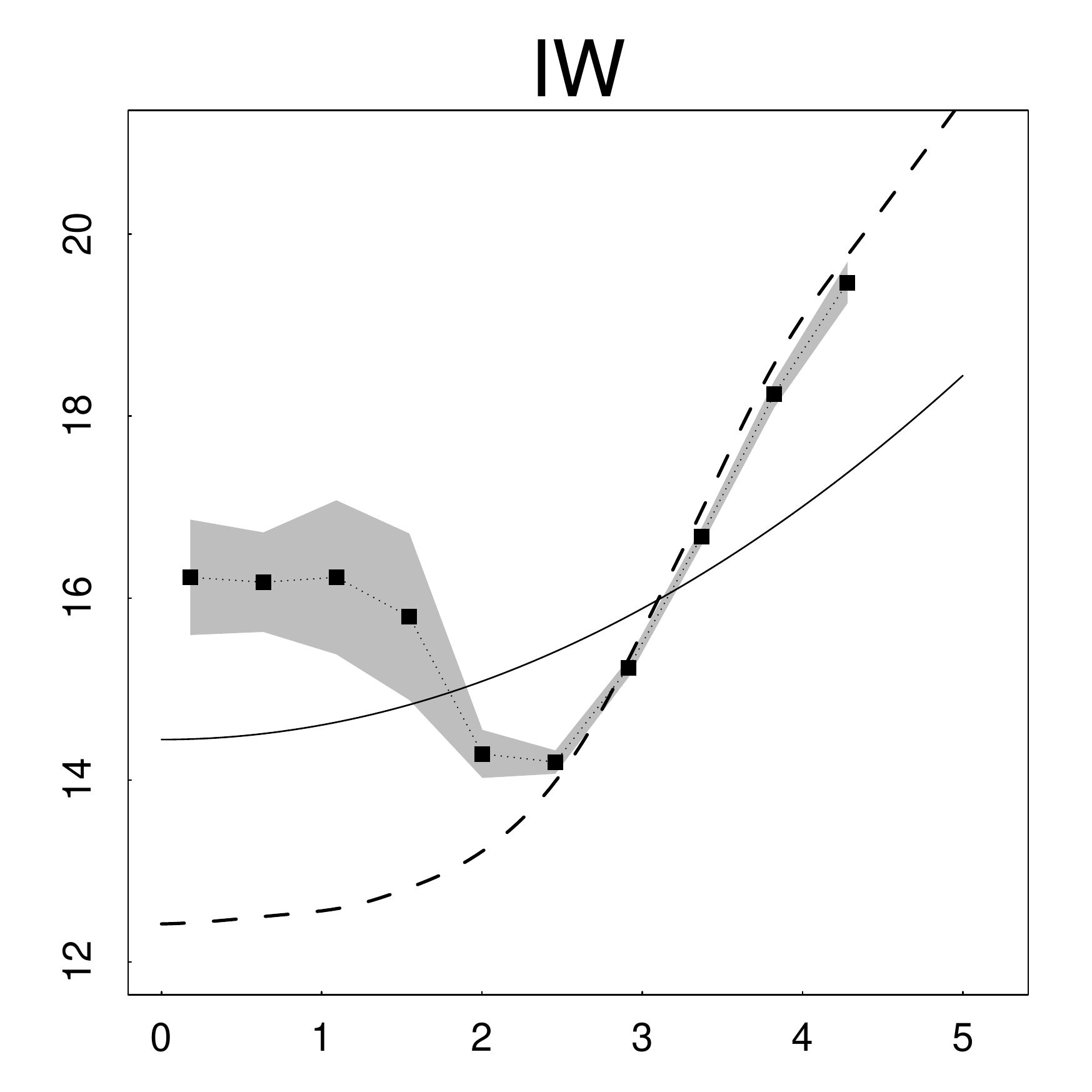}
\includegraphics[width=0.24\textwidth]{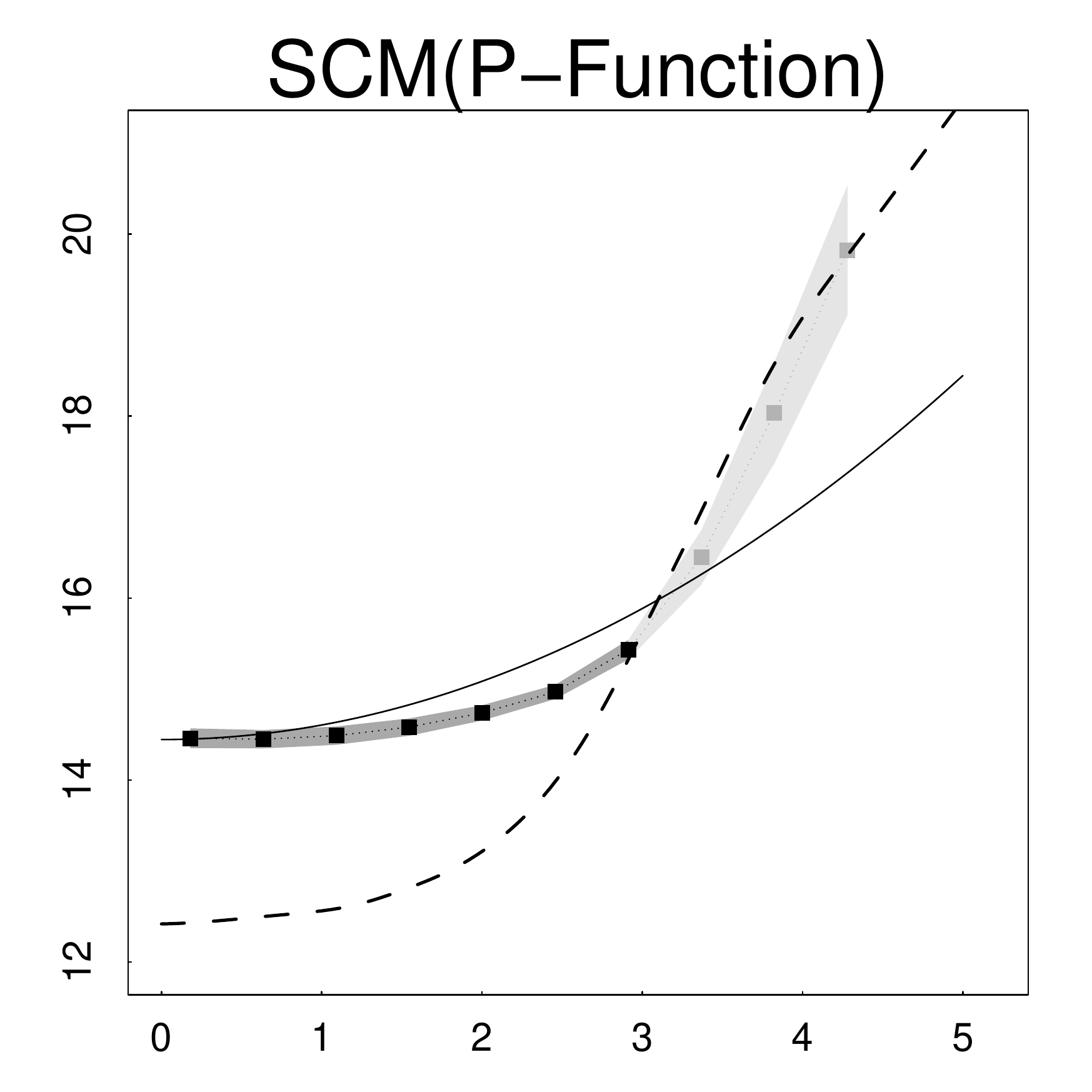}\\   
\textsf{\textbf{\large Generative Model: Piecewise Linear}}\\ \smallskip   
\includegraphics[width=0.24\textwidth]{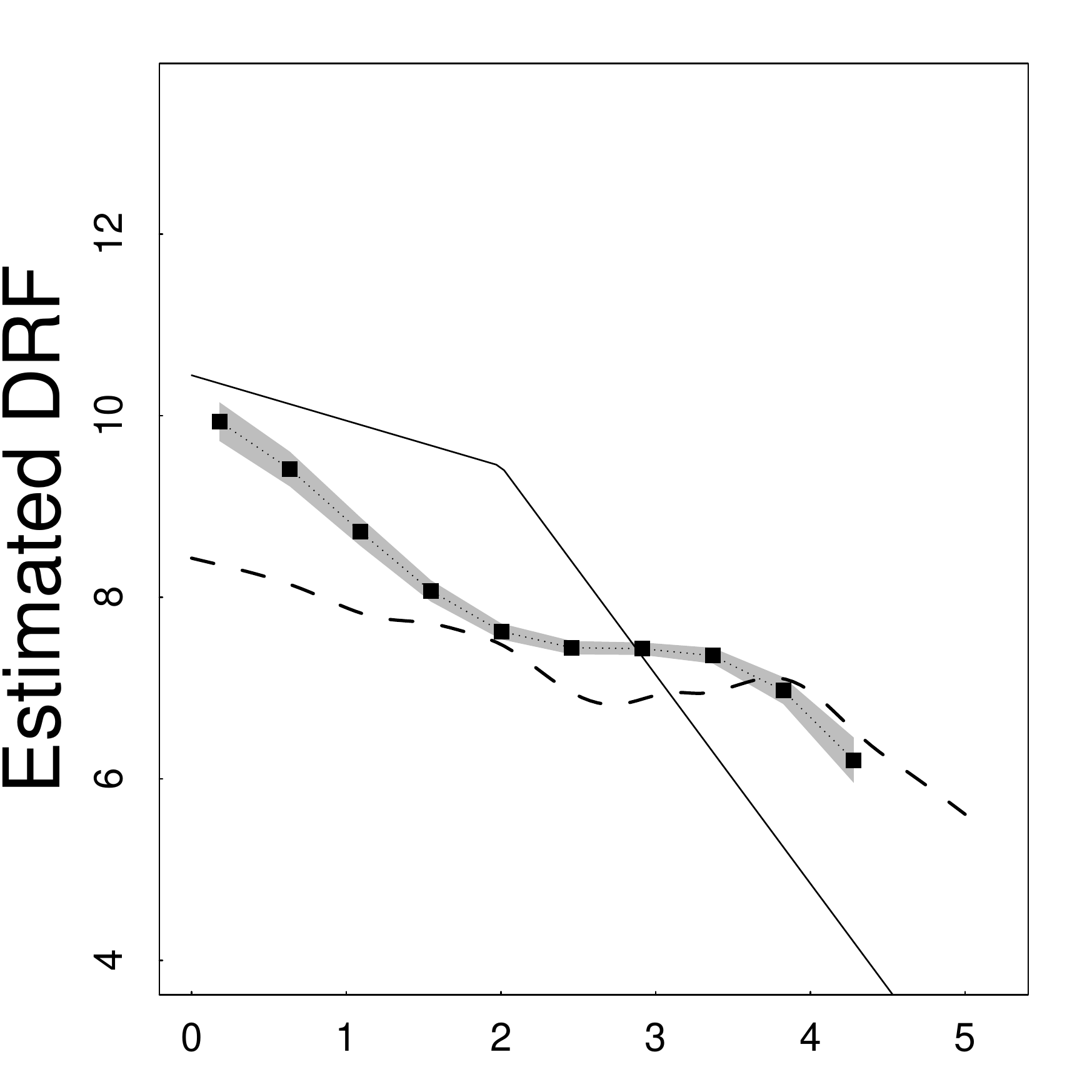}
\includegraphics[width=0.24\textwidth]{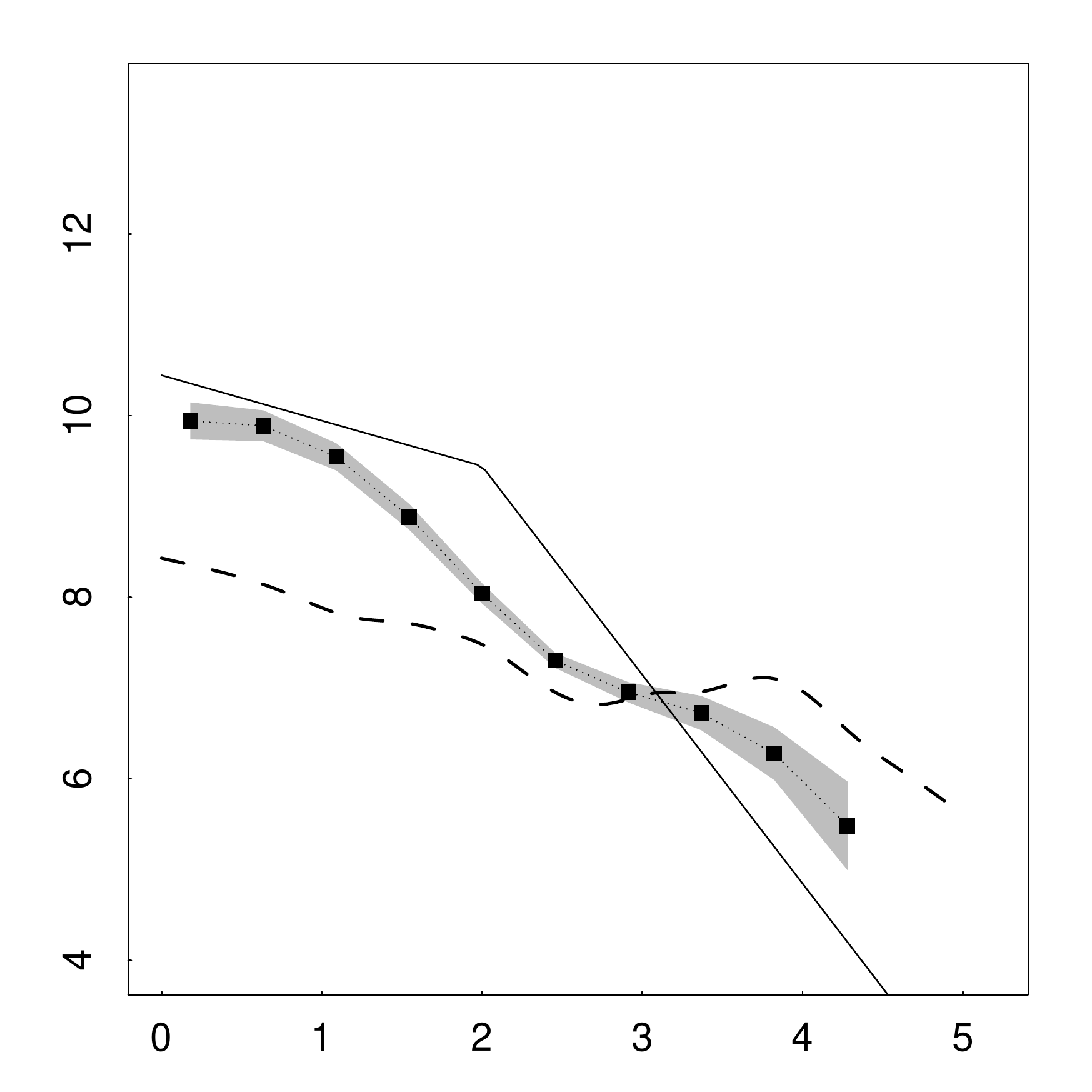}
\includegraphics[width=0.24\textwidth]{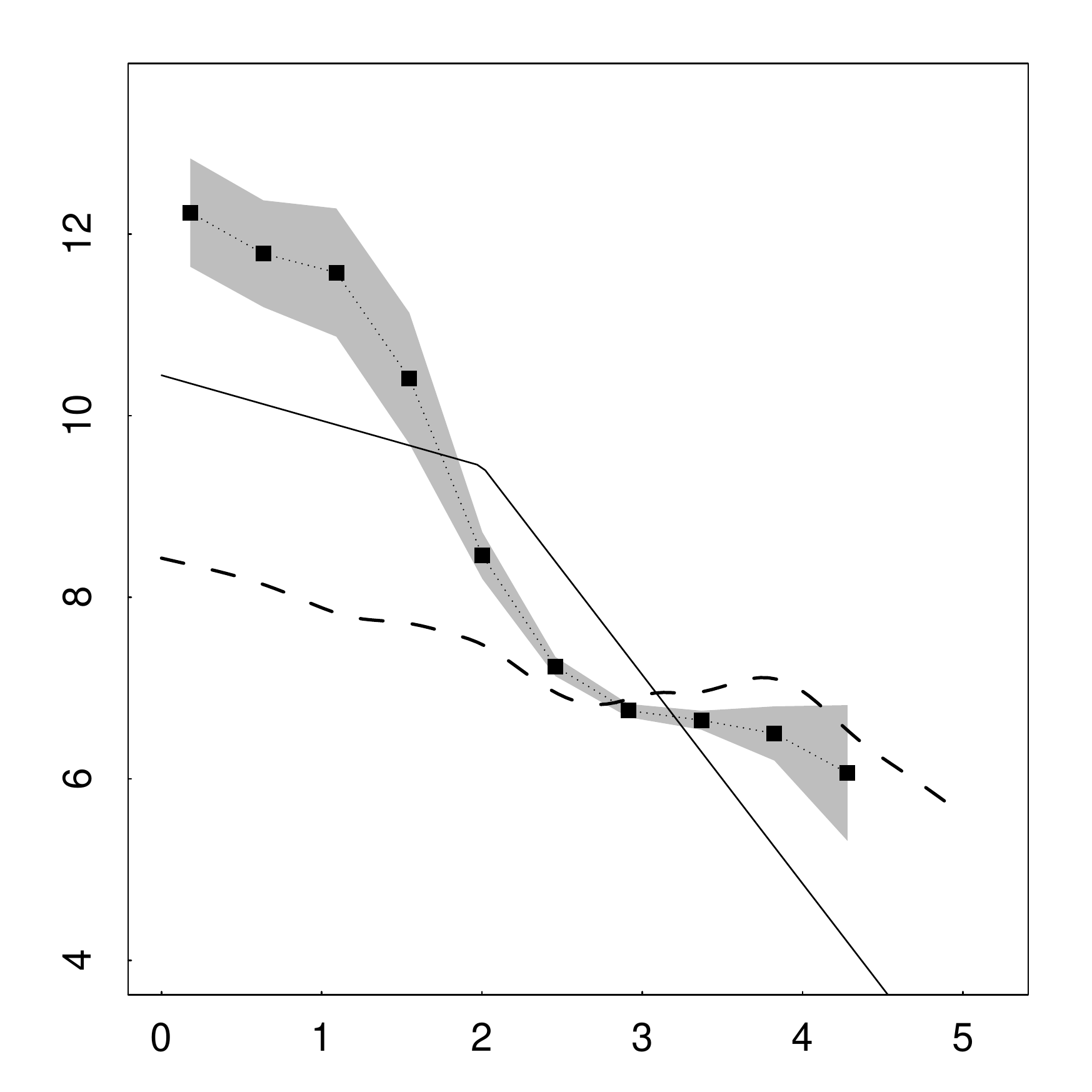}
\includegraphics[width=0.24\textwidth]{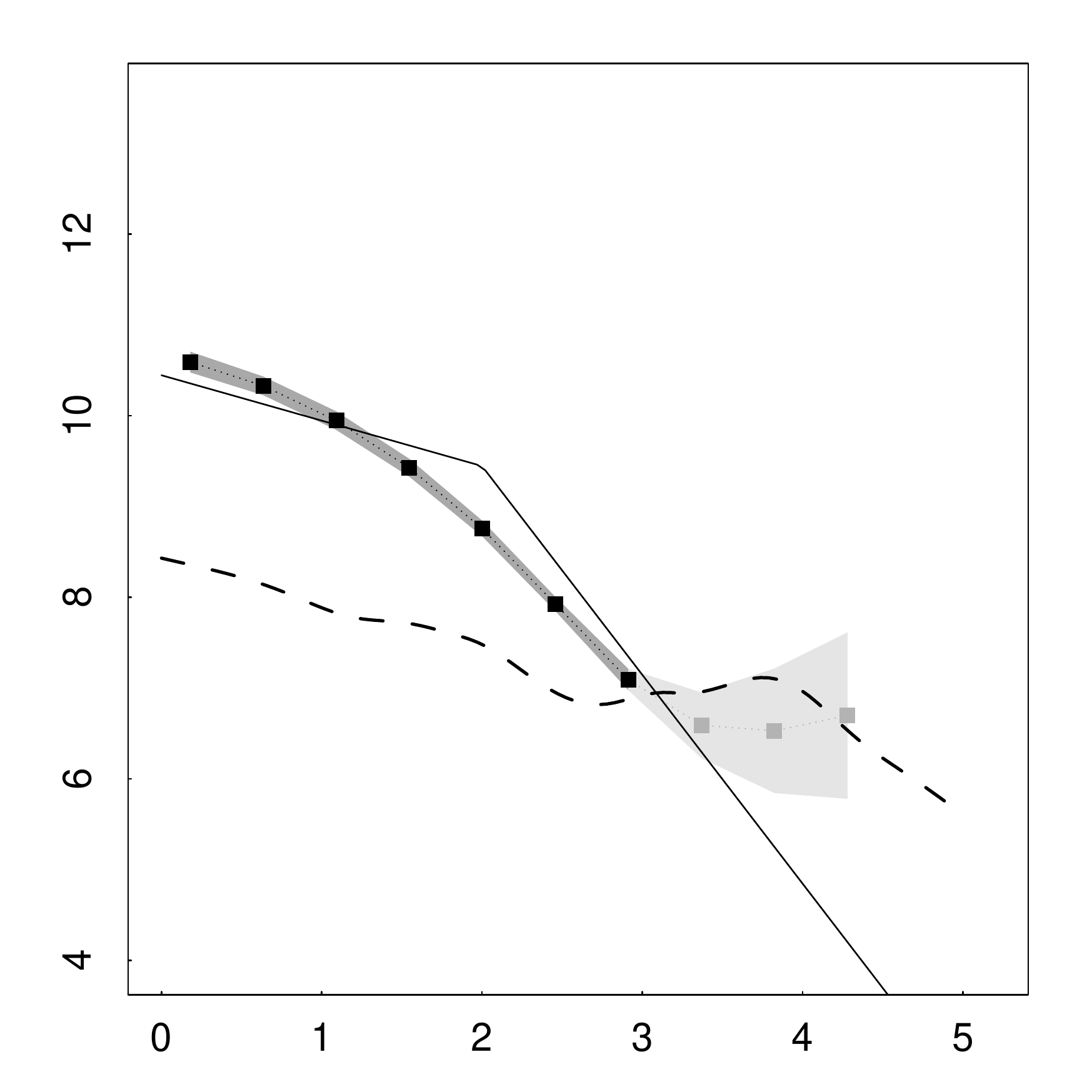}\\
\textsf{\textbf{\large Generative Model: Hockey-stick}}\\ \smallskip
\includegraphics[width=0.24\textwidth]{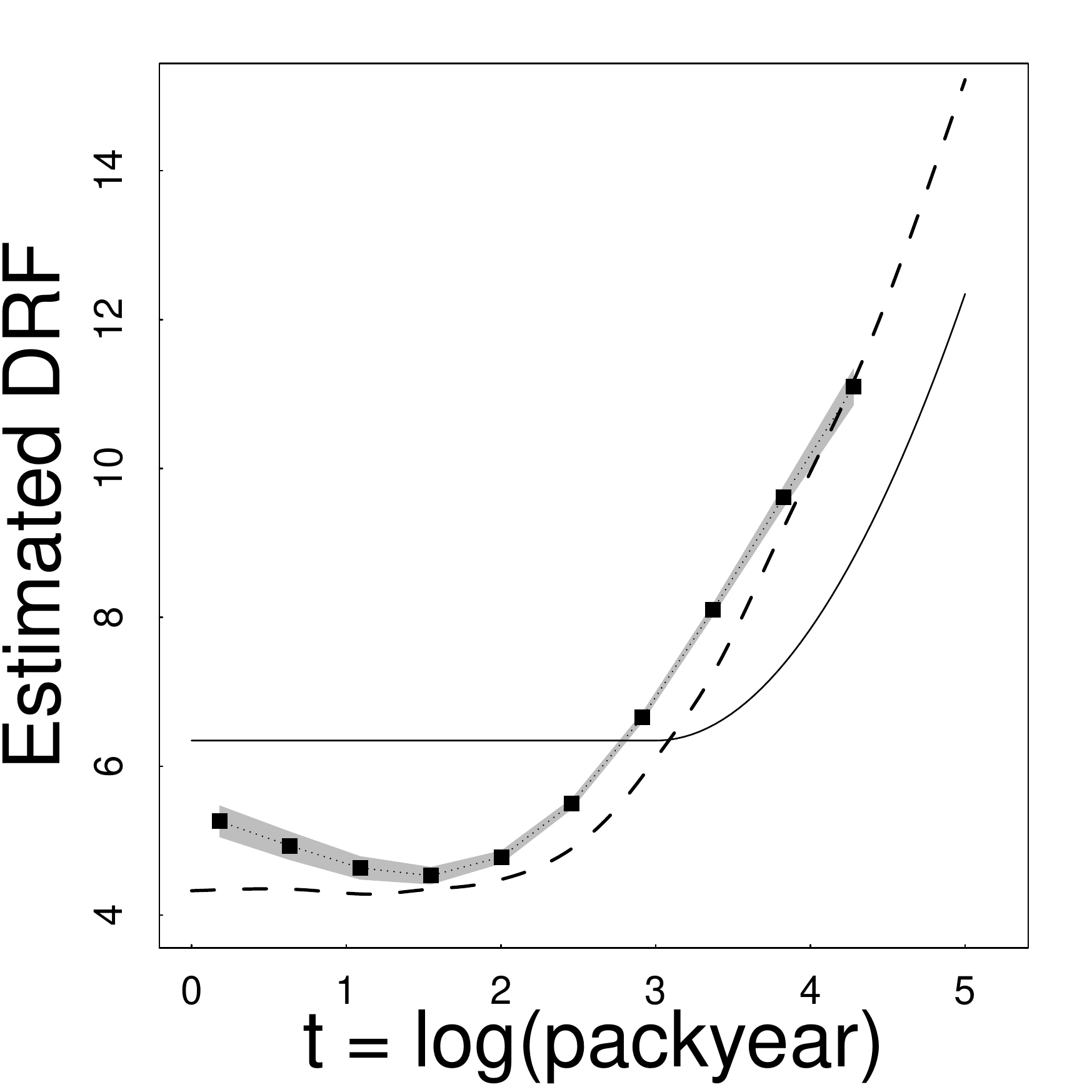}
\includegraphics[width=0.24\textwidth]{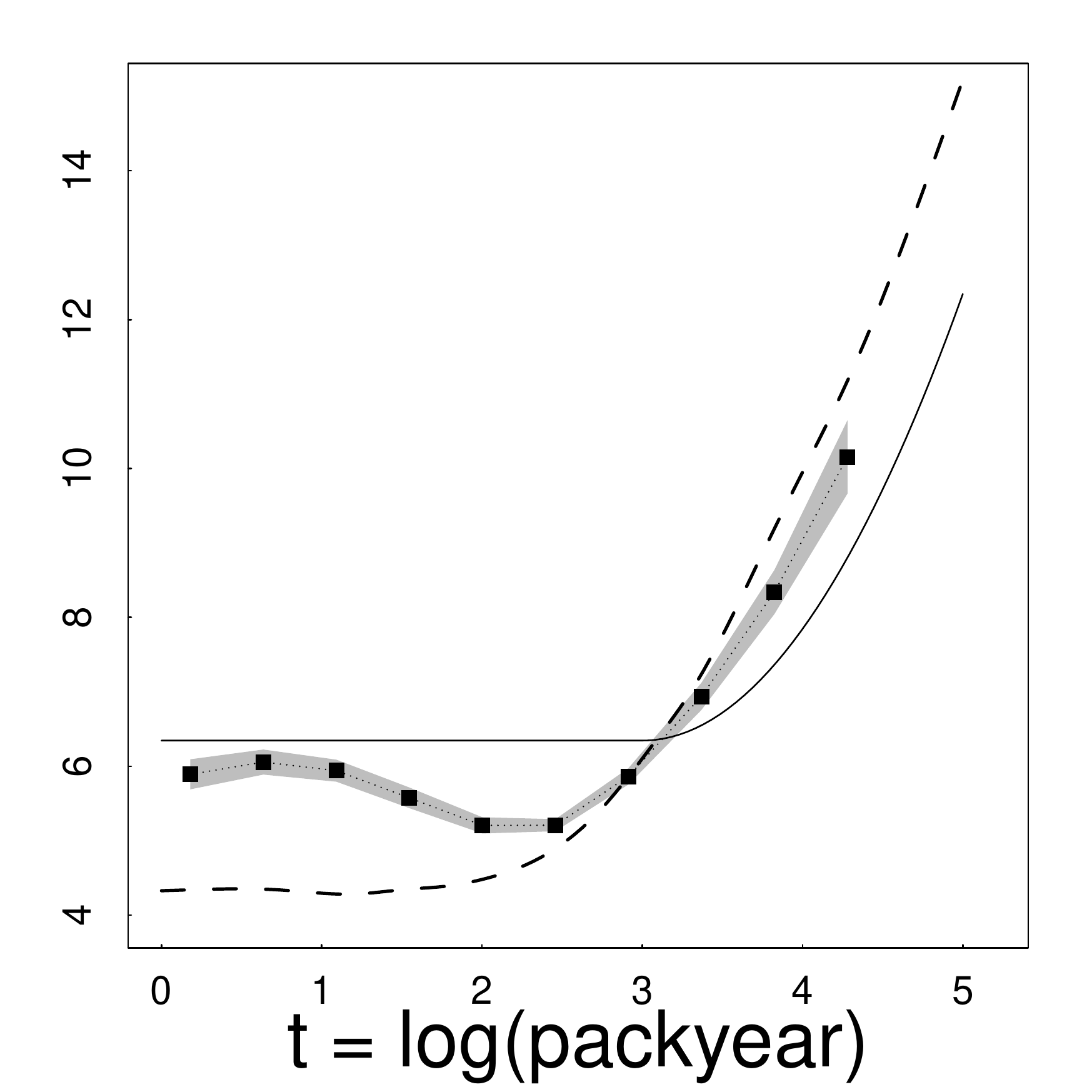}
\includegraphics[width=0.24\textwidth]{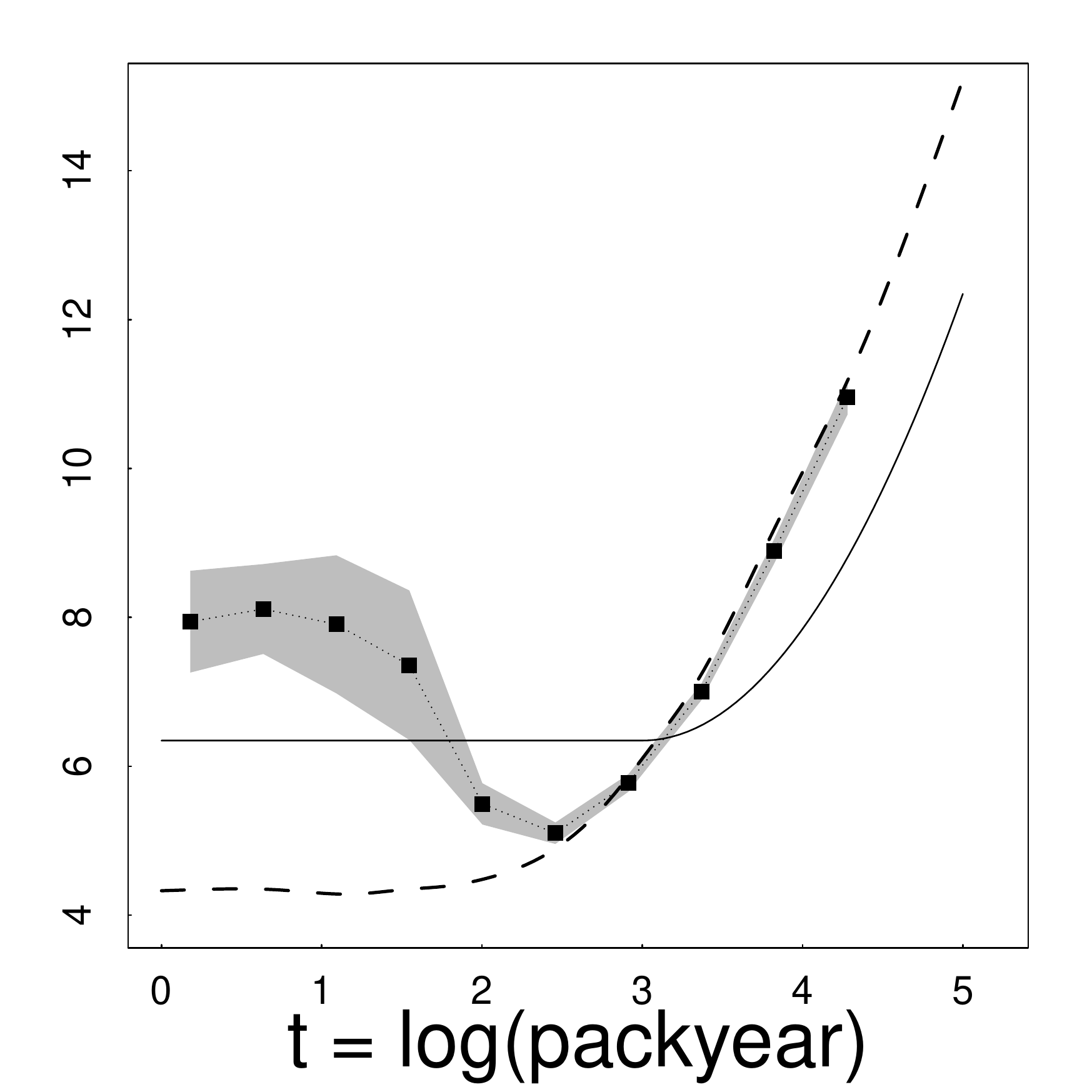}
\includegraphics[width=0.24\textwidth]{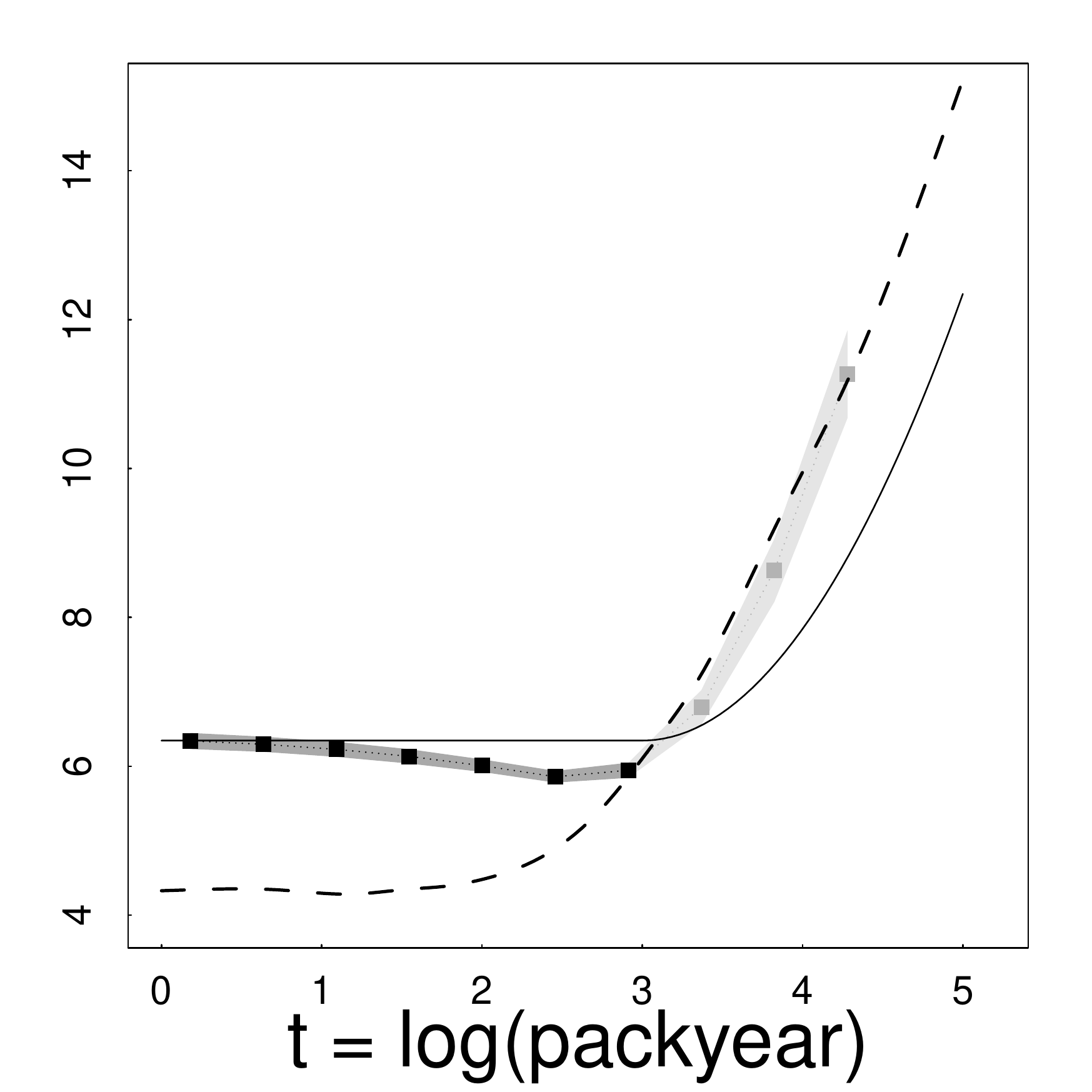}\\
\caption{Estimated \drfs\ for the Simulation Based on Smoking Data. The
four columns correspond to the method of \hi, \scm(\gps), \iw, and \scm(\pfun)
respectively. In all plots the dotted lines with bullets correspond to the
fitted \drf. The true \drf\ is plotted as solid lines while dashed lines
represent the fitted \scm\ of $\log(Y)$ on $T$, unadjusted for the covariates.
The evaluation points are evenly-spaced in $t$. The 95\% asymptotic confidence
bands plotted in grey are based on 1000 bootstrap replications. A lighter shade
of grey is used in the right-most column for $t>3$ because the estimate is less
reliable in this region. The performance of the \scm(\pfun) clearly dominates
the other methods, especially for $t<3$.}
\label{fig:smoking_sim}
\end{figure}

Each of the {four} methods was fitted to one data set generated under
each of the three \drfs. {We evaluate the \drf\ at ten points equally
  spaced between the 5\% and 95\% quantiles of ${\rm log}({\tt
    packyear})$.} The results appear in Figure~\ref{fig:smoking_sim}
where rows correspond to the three generative models and columns
represent the method used to fit the \drf. In all plots, the true
\drf\ is plotted as a solid line and a directly fitted \scm\ of ${\rm
  log}(Y)$ on $T$ as a dashed line. This \scm\ fit is a simple bench
mark; it does not account for covariates in any way, in particular it
does not adjust for any summary of the treatment assignment model.
Dotted lines represent the fitted \drf\ using each of the four
methods; bullets indicate the grid where the estimates are
evaluated. The shaded regions around the fits represent 95\%
point-wise bootstrap confidence intervals. The diagnostic described in
Figure~\ref{fig:smoking_sim_pfun} indicates possible bias in the
\scm(\pfun) method for $t>3$. Thus, we plot the fit in this region in
light grey to emphasize its potential unreliability.

The fit under the \hi\ method misses the true \drf\ under all three generative
models, even the quadratic \drf\ which coincides with the parametric dependence
of ${\rm log}(Y)$ on $T$ under \hi's response model. Instead, \hi's fitted \drf\
tends to follow the unadjusted \scm\ fit of ${\rm log}(Y)$ on $T$. Although
\scm(\gps) exhibits improvement over the \hi\ method,
{it still exhibits a cyclic pattern; notice its cubic-like fits in the first
and third rows. Unfortunately, \iw\ again exhibits instability, although in this
case it takes the form of bias rather than variance.} Finally, {\bf
\scm(\pfun)} closely matches the true \drf\ under all three generative models,
at least for $t<3$. As discussed above, we suspect bias for $t>3$ and see that
the fitted \drf\ reverts to the unadjusted \scm\ in this range. The quality of
the fit can be improved still further by increasing the dimension of the basis
used in the \scm. We do not purse this strategy, however, for fear of over
fitting. Overall, {the \scm(\pfun) estimate} appears to be the most
reliable, especially considering the diagnostic that alerts us the ranges of $t$
where there is the potential for bias.

\section{Concluding Remarks} 
\label{sec:disc}

Propensity score methods have gained wide popularity among applied
researchers in a number of disciplines.  Although propensity score
methods were originally designed exclusively for binary treatment
regimes, the fact that treatment variables of interest are not binary
in many scientific research settings has led to recent proposals for
generalized propensity score methods.  These methods are applicable to
a variety of non-binary treatment regimes, and their applications are
becoming increasingly common.

In this article, we identify {and address limitations of} the two most
frquently used generalized propensity score methods, the \gps\ of
\citet{hira:imbe:04} and \pfun\ of \citet{imai:vand:04}, as well as
the two \gps-based methods of \ffgn.  First, we show {that} the
suggested implementation of the \hi\ method is sensitive to
misspecification of the response model.  Second, we show that while
\scm(\gps) exhibits substantial improvement over \hi's method, it
remains biased and/or can exhibit a cyclic artifact in some
situations. Third, while the \ivd\ method provides a relatively robust
method for average causal effect, its main limitation is its inability
to estimate the \drf.  We show how to obtain an estimate of the \drf\
based on the \pfun\ and empirically compare its performance to that of
the \gps-based estimates. We also give an explanation as to why the
\scm(\pfun) method outperforms the \scm(\gps) method.

There are several important challenges that must still be addressed.
We have assumed throughout the paper that the \gps\ and \pfun\ can be
correctly estimated.  This is an optimistic assumption given that
modeling a multi-valued or continuous treatment in a high-dimensional
covariate space is much more difficult than doing so for a binary
treatment \citep[see e.g.,][for a method that addresses this
issue]{imai:ratk:13a}.  Diagnostic tools developed for the binary
treatment case are also not directly applicable to general treatment
regimes.  Even more challenging is diagnosing misspecification in the
response model. As we have illustrated, this can lead to significant
bias in the estimated \drf. Our proposals rely on implementing more
flexible response models in more natural spaces, but principled
diagnostics for the response model remain elusive. Diagnosing and
correcting for inbalance in either the \pfun~or the \gps\ is another
difficulty. Since the subpopulation that has propensity for treatment
varies with the dose, the estimated dose response function is in
effect the treatment effect on a varying subpopulation.  Future
research must develop methods for estimating the \gps\ and \pfun\ in
the presence of possible misspecification of the treatment assignment
model and the \drf\ in the presence of possible misspecification of
the response model, as well as diagnostics for both models.

\bibliography{pscore}

\begin{thebibliography}{}

\bibitem[Bia \emph{et~al.}(2011)Bia, Flores, and Mattei]{bia:flor:matt:01}
Bia, M., Flores, A.~C., and Mattei, A. (2011).
\newblock Nonparametric estimators of dose-response functions.
\newblock \emph{CEPS/INSTEAD} .

\bibitem[Cochran(1968)]{coch:68}
Cochran, W.~G. (1968).
\newblock The effectiveness of adjustment by subclassification in removing bias
  in observational studies.
\newblock \emph{Biometrics} \textbf{24}, 295--313.

\bibitem[D'Agostino and Rubin(2000)]{dago:rubi:00}
D'Agostino, Ralph~B., J. and Rubin, D.~B. (2000).
\newblock Estimating and using propensity scores with partially missing data.
\newblock \emph{Journal of the American Statistical Association} \textbf{95},
  451, 749--759.

\bibitem[Duan \emph{et~al.}(1983)Duan, Manning, Morris, and
  Newhouse]{duan:mann:morr:newh:83}
Duan, N., Manning, W.~G., Morris, C.~N., and Newhouse, J.~P. (1983).
\newblock A comparison of alternative models for the demand for medical care.
\newblock \emph{Journal of Business \& Economics Statistics} \textbf{1}, 2,
  115--126.

\bibitem[Ertefaie and Stephens(2010)]{erte:step:10}
Ertefaie, A. and Stephens, D.~A. (2010).
\newblock Comparing approaches to causal inference for longitudinal data:
  Inverse probability weighting versus propensity scores.
\newblock \emph{The International Journal of Biostatistics} \textbf{6}.

\bibitem[Fan and Gijbels(1996)]{fan:gijb:1996}
Fan, J. and Gijbels, I. (1996).
\newblock \emph{Local Polynomial Modelling and its Applications}.
\newblock Chapman and Hall, London.

\bibitem[Flores \emph{et~al.}(2012)Flores, Flores-Lagunes, Gonzalez, and
  Neumann]{flor:flor:gonz:Neum:2012}
Flores, C.~A., Flores-Lagunes, A., Gonzalez, A., and Neumann, T.~C. (2012).
\newblock Estimating the effects of length of exposure to instruction in a
  training program: The case of job corps.
\newblock \emph{The Review of Economics and Statistics} \textbf{94}, 1,
  153--171.

\bibitem[Hirano and Imbens(2004)]{hira:imbe:04}
Hirano, K. and Imbens, G.~W. (2004).
\newblock The propensity score with continuous treatments.
\newblock In \emph{Applied Bayesian Modeling and Causal Inference from
  Incomplete-Data Perspectives: An Essential Journey with Donald Rubin's
  Statistical Family}, chap.~7. Wiley.

\bibitem[Imai \emph{et~al.}(2008)Imai, King, and Stuart]{imai:king:stua:08}
Imai, K., King, G., and Stuart, E.~A. (2008).
\newblock Misunderstandings among experimentalists and observationalists about
  causal inference.
\newblock \emph{Journal of the Royal Statistical Society, {Series A}
  (Statistics in Society)} \textbf{171}, 2, 481--502.

\bibitem[Imai and Ratkovic(2013)]{imai:ratk:13a}
Imai, K. and Ratkovic, M. (2013).
\newblock Covariate balancing propensity score.
\newblock \emph{Journal of the Royal Statistical Society, {Series B}
  (Statistical Methodology)}  Forthcoming.

\bibitem[Imai and van Dyk(2004)]{imai:vand:04}
Imai, K. and van Dyk, D.~A. (2004).
\newblock Casual inference with general treatment regimes: Generalizing the
  propensity score.
\newblock \emph{Journal of the American Statistical Association} \textbf{99},
  854--866.

\bibitem[Imbens(2000)]{imbe:00}
Imbens, G.~W. (2000).
\newblock The role of the propensity score in estimating dose-response
  functions.
\newblock \emph{Biometrika} \textbf{87}, 3, 706--710.

\bibitem[Johnson \emph{et~al.}(2003)Johnson, Dominici, Griswold, and
  Zeger]{john:domi:gris:zege:03}
Johnson, E., Dominici, F., Griswold, M., and Zeger, S.~L. (2003).
\newblock Disease cases and their medical costs attributable to smoking: An
  analysis of the national medical expenditure survey.
\newblock \emph{Journal of Econometrics} \textbf{112}, 135--151.

\bibitem[Kang and Schafer(2007)]{kang:scha:07}
Kang, J. D.~Y. and Schafer, J.~L. (2007).
\newblock Demystifying double robustness: A comparison of alternative
  strategies for estimating a population mean from incomplete data.
\newblock \emph{Statistical Science} \textbf{22}, 4, 523--539.

\bibitem[Moodie and Stephens(2012)]{mood:step:12}
Moodie, E.~E. and Stephens, D.~A. (2012).
\newblock Estimation of dose-response functions for longitudinal data using the
  generalised propensity score.
\newblock \emph{Statistical Methods in Medical Research} \textbf{21}, 2,
  149--166.

\bibitem[Robins(1998)]{robi:98}
Robins, J.~M. (1998).
\newblock Marginal structural models.
\newblock \emph{1997 Proceedings of the American Statistical Association,
  Section on Bayesian Statistical Science}  1--10.

\bibitem[Robins \emph{et~al.}(2000)Robins, Hern$\acute{a}$n, and
  Brumback]{robi:hern:brum:00}
Robins, J.~M., Hern$\acute{a}$n, M.~A., and Brumback, B. (2000).
\newblock Marginal structural models and causal inference in epidemiology.
\newblock \emph{Epidemiology} \textbf{11}, 550--560.

\bibitem[Rosenbaum(1987)]{rose:87}
Rosenbaum, P.~R. (1987).
\newblock Model-based direct adjustment.
\newblock \emph{Journal of the American Statistical Association} \textbf{82},
  398, 387--394.

\bibitem[Rosenbaum and Rubin(1983)]{rose:rubi:83}
Rosenbaum, P.~R. and Rubin, D.~B. (1983).
\newblock The central role of the propensity score in observational studies for
  causal effects.
\newblock \emph{Biometrika} \textbf{70}, 41--55.

\bibitem[Rubin(1990)]{rubi:90}
Rubin, D.~B. (1990).
\newblock Comments on ``on the application of probability theory to
  agricultural experiments. essay on principles. section 9," by j.
  splawa-neyman, translated from the polish and edited by d. m. dabrowska and
  t. p. speed.
\newblock \emph{Statistical Science} \textbf{5}, 472--480.

\bibitem[Rubin(2004)]{rubi:04}
Rubin, D.~B. (2004).
\newblock On principles for modeling propensity scores in medical research.
\newblock \emph{Pharmacoepidemiology And Drug Safety} \textbf{13}, 855--857.

\end{thebibliography}

\newpage 
\appendix
\centerline{\bf\LARGE ONLINE SUPPLEMENTAL MATERIALS}

\section{Appendix: Covariance Adjustment GPS}
\label{sec:appA}

\subsection{Covariance adjustment for catagorical treatments}
\label{sec:cat-cov}

One of the response models suggested by \rr\ for a binary treatment in
an observational study involves {\it covariance adjustment}.  With
this method, the response variable is regressed on the fitted \pscore\
separately for the treatment and control groups.  Suppose we use the
\gps\ in place of the \pscore\ in the context of a binary treatment. Specifically, for units in the
treatment group, we use the ordinary \pscore, $R_i = r(1, \bX_i) =
p_\psi(T=1 \mid \bX_i)$, but for units assigned to the control group,
we use the probability of control rather than the probability of
treatment, $R_i = r(0, \bX_i) = p_\psi(T=0 \mid \bX_i)$.  Because the
\gps~is equal to the \pscore~for treatment units and is equal to one
minus the \pscore~for control units \citep{imbe:00}, it is easy to see
that the usual covariance adjustment is equivalent to fitting the
following regression model,
\begin{equation}
  Y_i \ \sim \  \alpha_t + \beta_t \hat R_i, 
  \label{eq:bi-cov-adj}
\end{equation}
separately for the treatment and control units, i.e., $t=0$ and
$1$. The linear transformation of the predictor variable does not
effect the predicted value of the response for the control group.

After fitting the model given in equation~\eqref{eq:bi-cov-adj}, the
average of the two potential outcomes can be estimated by averaging
the fitted values over all units in the sample. That is, we compute
\begin{equation}
\hat E\{ Y(t)\} \ = \ {1\over n}\sum_{i=1}^n\l\{ \hat\alpha_t +
\hat\beta_t\ \hat r(t, \bX_i)\r\}, 
\label{eq:bi-mean-resp}
\end{equation}
for $t=0,1$. The estimated average causal effect is simply the
difference $\hat E\{ Y(1)\} - \hat E\{Y(0)\}$, which is equivalent to
the estimate reported in equation~\eqref{eq:cov-adj}. Thus, with a
binary treatment, the method of \hi\ is equivalent to \rr's covariate
adjustment, except that \hi\ propose a quadratic rather than a linear
response model. 

Suppose now that the treatment variable is categorical with more than
two levels.  In principle, exactly the same procedure can be
applied. Namely, the regression model given in
equation~\eqref{eq:bi-cov-adj} can be fitted separately for units in
each treatment group and the average potential outcome can be computed
using the formula of equation~\eqref{eq:bi-mean-resp} for each level
of the treatment.  We refer to this procedure as {\sl covariate
  adjustment \gps\ for categorical treatments}.  The relative \drf\ can be estimated as $\hat E\{ Y(t)\} - \hat E\{Y(0)\}$ for each level of $t$. The validity of this procedure follows
directly from the theory of \rr\ because we only consider two
treatments at a time.

If the categorical treatment variable is ordinal with a meaningful
numerical scale, we can use the quadratic regression model of
equation~\eqref{eq:gps-r-model} suggested by \hi.  However, such a
model is restrictive because the slope for \gps\ in the model changes
in a particular way across the treatment levels. Figure~{\ref{fig:sim-one-b-hi-prob}} shows
that this assumption may be too strong to justify in practice.

The usefulness of the covariance adjustment \gps\ for
categorical variables is limited by our ability to fit multiple
regression models with limited data.  When the treatment takes a large
number of values, the method may be infeasible. This
problem is even more acute for continuous treatments where it is
simply impossible to fit a separate regression model for each observed
treatment level. We now discuss the covariate adjustment for
continuous treatments. 

\subsection{Covariance adjustment {\large GPS} for continuous treatments}
\label{sec:cont-cov}

To use covariance adjustment with a continuous treatment variable, we
propose to subclassify the data on the treatment variable rather than
on the \gps\ or the \pfun. To facilitate the computation of standard
errors via bootstrap (see below), we form the subclasses using the
theoretical quantiles of the fitted treatment assignment model. This
is typically easy to accomplish via Monte Carlo. We draw a large
sample from the fitted treatment assignment model with parameters
fixed at their fitted values and covariates sampled from their
observed values and estimate the theoretical quantiles based on this
sample.  We also compute the theoretical median, or its Monte Carlo
approximation, within each subclass and denote it as $t_s$ for
$s=1,\dots,S$ with $S$ the number of subclasses.

With the subclassifed data in hand, we fit the model defined in
equation~\eqref{eq:bi-cov-adj} separately for each subclass. Alternatively, we
can use a more flexible model.  Here, we consider both quadratic regression,
i.e., $Y_i \sim \alpha_t + \beta_t \hat R_i + \gamma_t \hat R_i^2$, and the
\scm\ given in equation~\eqref{eq:IvD-smooth-coef} with $T$ replaced by $\hat
R$.  We then compute the \gps\ for each unit at the median treatment value within each
subclass, i.e., $\hat r(t_s, \bX_i)$ for $i=1,\ldots, n$ and $s=1,\ldots, S$. Finally, we estimate
the \drf\ by computing $\hat E\{ Y(t_s)\}$ for each $t_s$ using
equation~\eqref{eq:bi-mean-resp} or an appropriate generalization of it if a
different response model is used.  The derivative of the \drf\ at $t_s$ can be
estimated as in (\ref{eq:der}).  Notice that the grid values at which we compute
the \drf\ are different than those advocated by \hi. Ours are based on
percentiles of the fitted treatment assignment model, whereas theirs are equally
spaced in the range of observed treatments.

 
The standard bias-variance tradeoff arises when selecting the number
of subclasses, $S$.  We generally defer to Cochran's advice and use about
five \citep{coch:68}. Sensitivity to the choice of $S$ can be
quantified by repeating the entire procedure with $S$ equal to
approximately three and ten.  One source of bias in this procedure
results from using units with a range of treatment values to fit the
model given in equation~\eqref{eq:bi-cov-adj} (or a more flexible
version of it). This bias will be especially acute in subclasses with
a relatively wide range of the treatment value. If the distribution of
the treatment has tails in either direction this correspond to extreme
evaluation points of the \drf, $t_1$ and $t_S$. Thus, in some cases,
we might want to increase the number of subclasses, especially when
the extremities of the \drf\ are of interest. This point is
illustrated in Sections~\ref{sec:appA_perf}.

We approximate the standard errors of the estimated \drf\ and its
estimated derivative via bootstrap resampling. We resample the data,
fit the treatment model, subclassify, and compute the \drf\ and its
derivative for each resampled data as described above.  We use the
same evaluation points, $t_1, \ldots, t_S$ for each resampled data
set.  Because both the treatment assignment model and the response
model are fitted to each bootstrap sample, this procedure accounts for
both sources of uncertainty.

\subsection{The numerical performance of Covariance Adjustment GPS}
\label{sec:appA_perf}
\begin{figure}
\spacingset{1}
\centering
\includegraphics[width=0.29\textwidth]{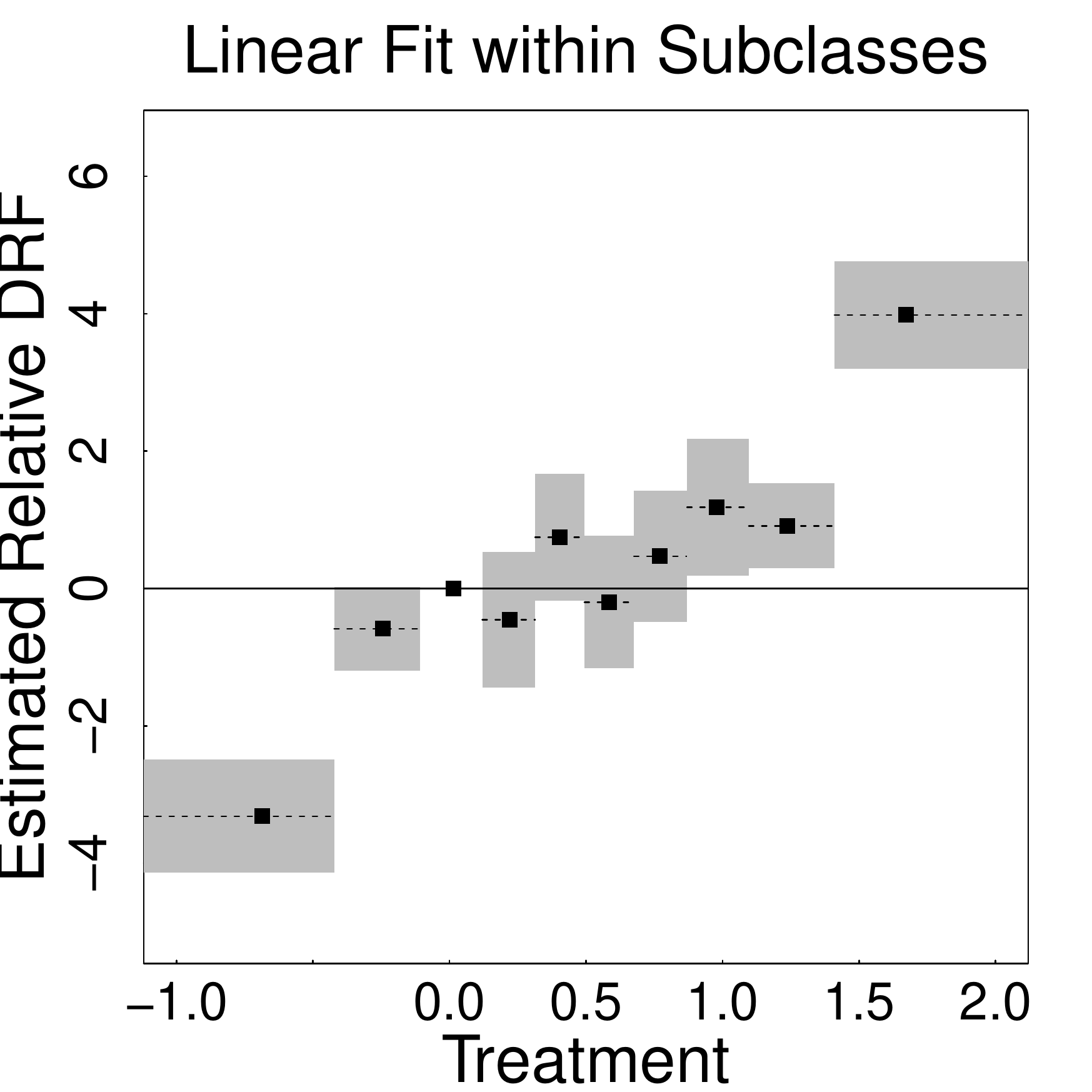}
\includegraphics[width=0.29\textwidth]{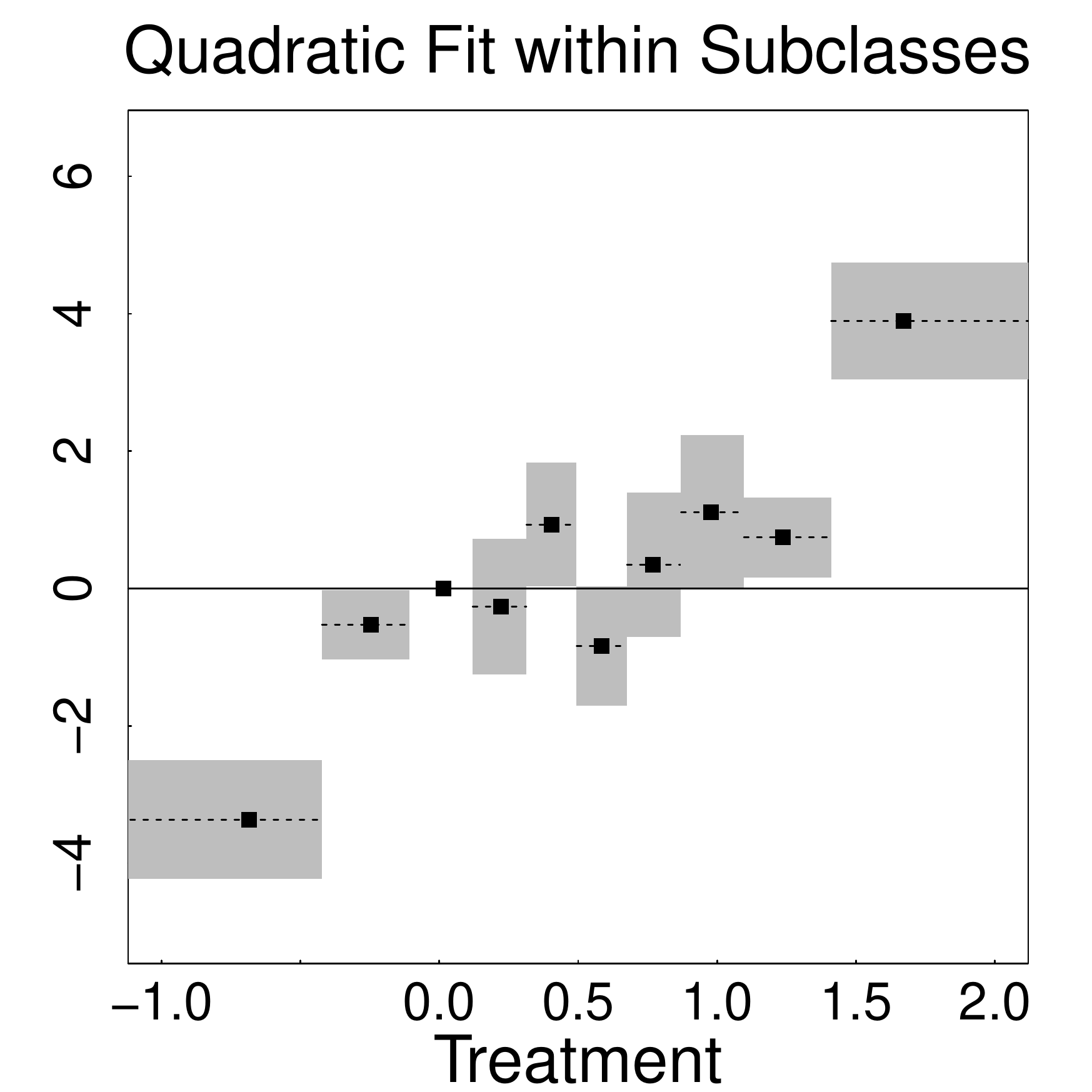}
\includegraphics[width=0.29\textwidth]{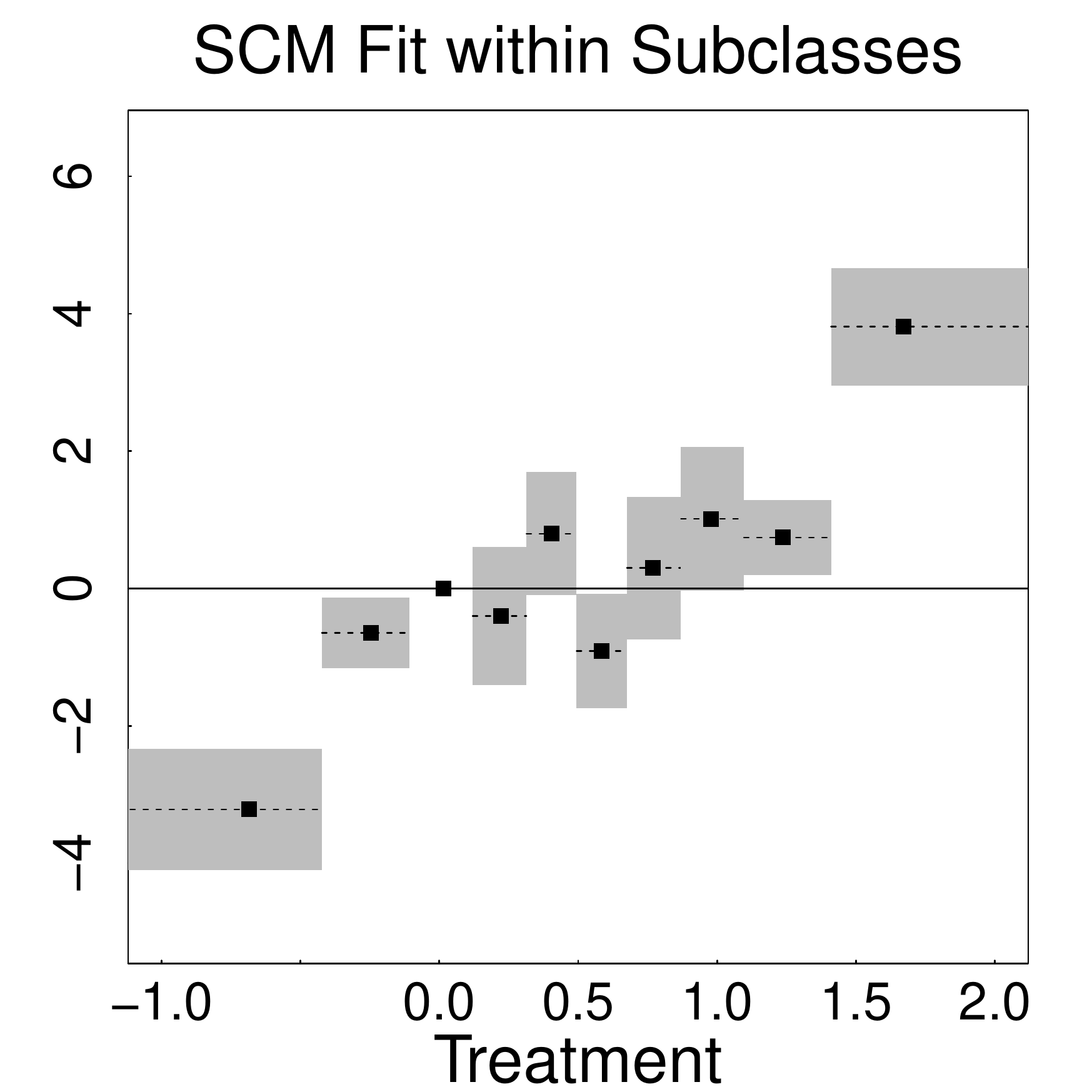}\\
\caption{{Estimated Relative \drfs~in Simulation Study I for the
Covariance Adjustment \gps\ Method. The three plots correspond to the three
within subclass models. In all plots the solid (dashed) lines represent the true
(fitted) relative \drf\ and 95\% confidence bands based on 1000 bootstrap
replications are plotted in grey.}
\label{fig:sim-one-app} }
\end{figure}

{We now examine the performance of covariance adjustment \gps\ in Simulations
I and II, as well as in the estimation of the \drf\ of smoking on annual medical
expenditures.} In Simulation I, we again use the the correct treatment
assignment model, $S=10$ subclasses with grid points at the $5\%, 15\%, \ldots,$ and $95\%$ quantiles of $T$, (Using $S=5$ or 15 gives similar
results.), and three within subclass response models: (i) $Y \sim R$, (ii) $Y \sim R+R^2$, and
(iii) $Y \sim f(R)$, where $f(\cdot)$ is a \scm. The results are shown in
Figure~\ref{fig:sim-one-app}. The three response models are labelled linear,
quadratic, and SCM fit within subclasses, respectively. The response models are
conditional on $R$, rather than on $T$ as in Section~\ref{sec:sim-one} because
covariance adjustment \gps\ subclassifies on $T$. As mentioned in
Section~\ref{sec:cont-cov}, the fitted  {relative} \drf\ exhibit bias in
extreme subclasses owing to the relatively large range of treatment levels in these classes. Because the three
within subclass models used with covariance adjustment \gps\ lead to very
similar fits, we only present results for the quadratic model in the rest of
this article.

\begin{figure}
\spacingset{1} 
\centering
\includegraphics[width=0.29\textwidth]{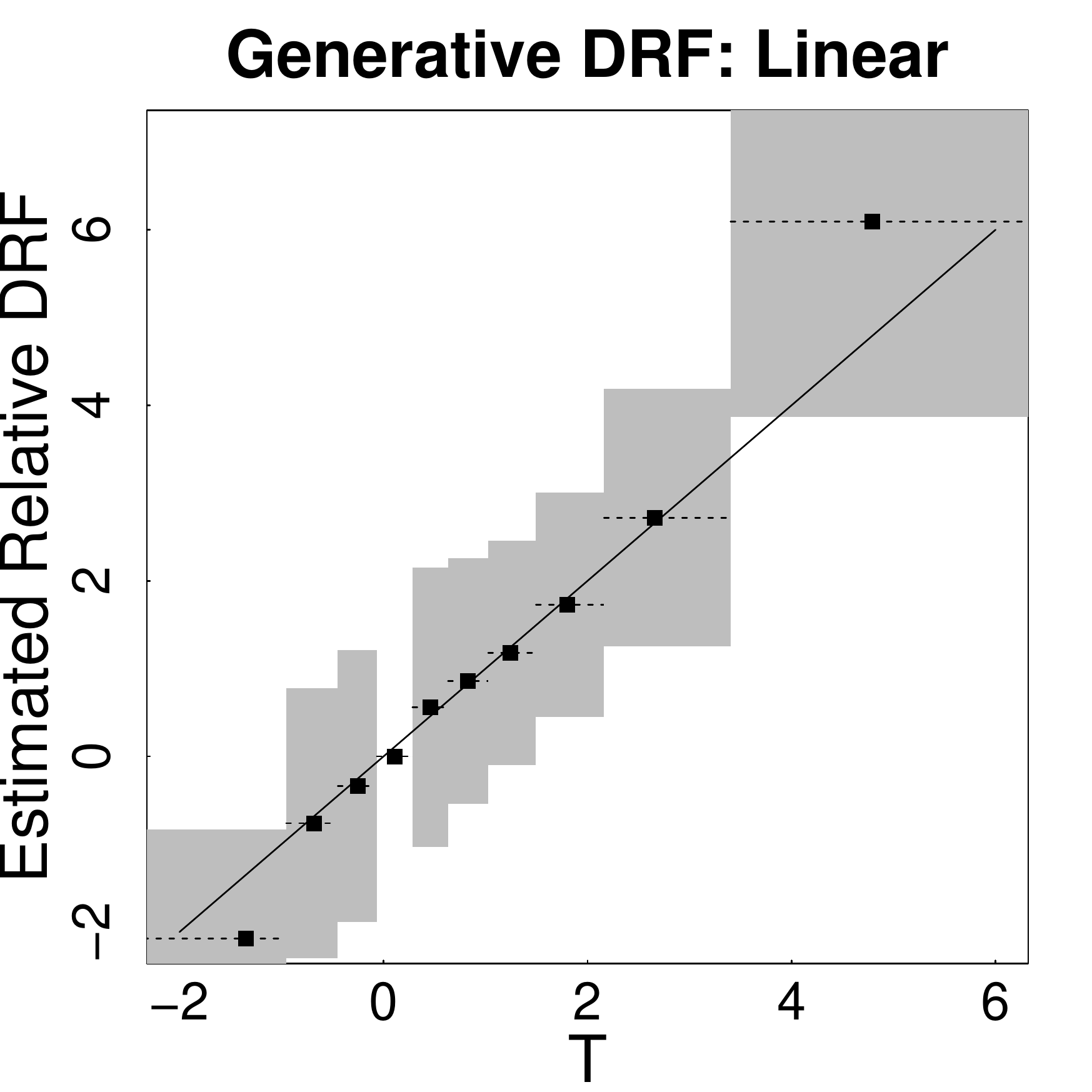}
\includegraphics[width=0.29\textwidth]{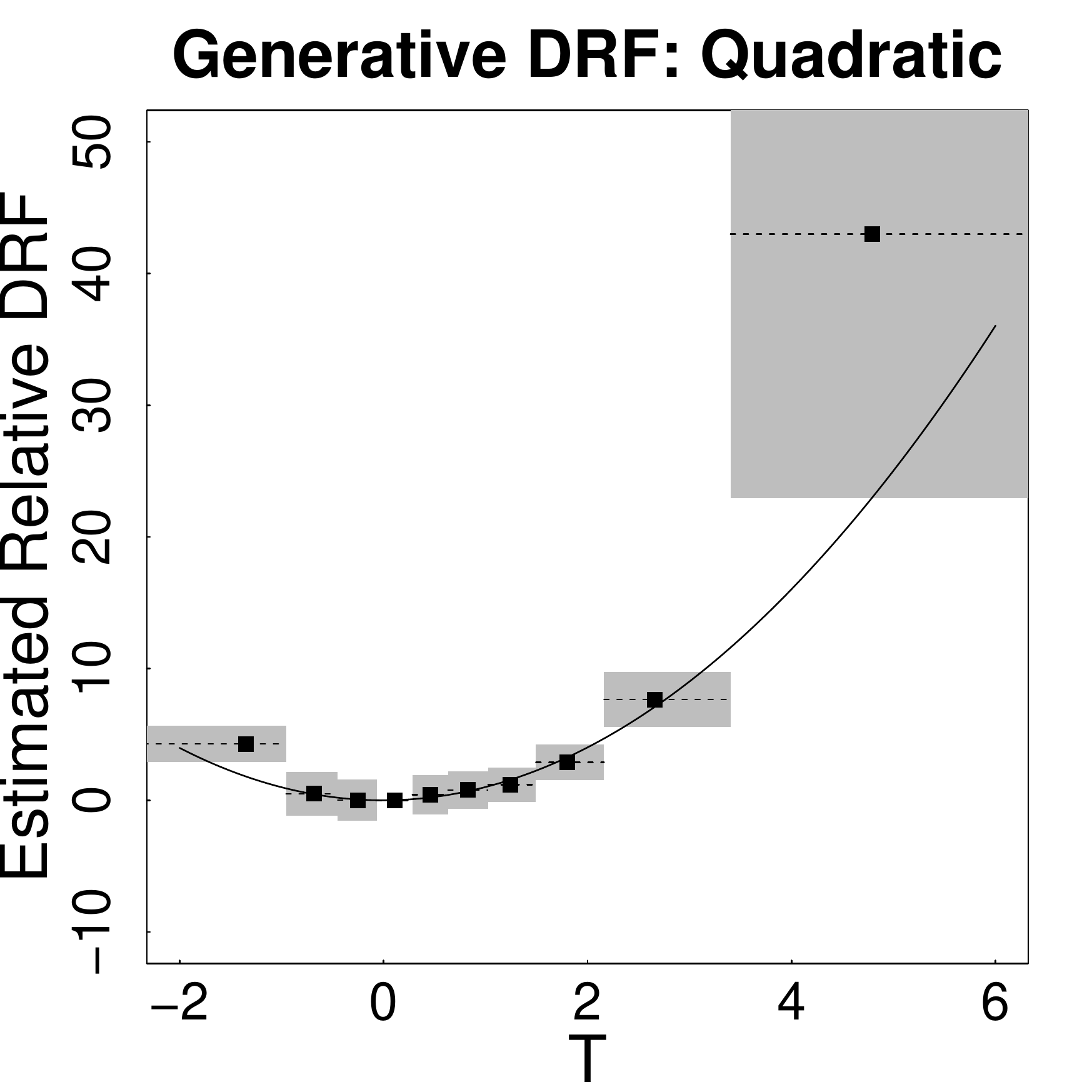}\\
\caption{ {Estimated Relative \drfs~in Simulation
  Study~II for the Covariance Adjustment \gps\ Method. Solid lines represent the
  true relative \drf\ and dashed lines the average of the fitted relative \drfs\
  across 1000 simulations. Points represent the theoretical quantiles of $T$ used to construct
  the subclasses. The grey shaded regions represent pointwise intervals
  containing 95\% of the 1000 fitted relative \drfs. Except in the extreme
  subclasses, the estimated \drf\ appears to be essentially unbiased. }
\label{fig:sim-two-app} }
\end{figure}          

In simulation II, we use the correct treatment assignment model and 
a quadratic model within each of $S=10$ subclasses as the response model. (Using
$S=7$ or 13 and/or the other two within subclass models yields similar results.)
Results are shown in Figure~\ref{fig:sim-two-app}. 
{Except in the two most extreme subclasses, the estimated \drf\ appears
to be essentially unbiased.}
As in Simulation study~I, the
fitted relative \drf\ deteriorates in the extreme, more heterogeneous treatment
subclasses.  Because the distribution of treatment is right
skewed, this is less of a problem for the left-most than for the right-most subclass.
This along with the blocky nature of the fitted \drf\ may lead many users to
prefer the smooth fitted {relative \drf\ obtained with \scm(\pfun)}.

\begin{figure}
\spacingset{1}
\centering
\includegraphics[width=0.32\textwidth]{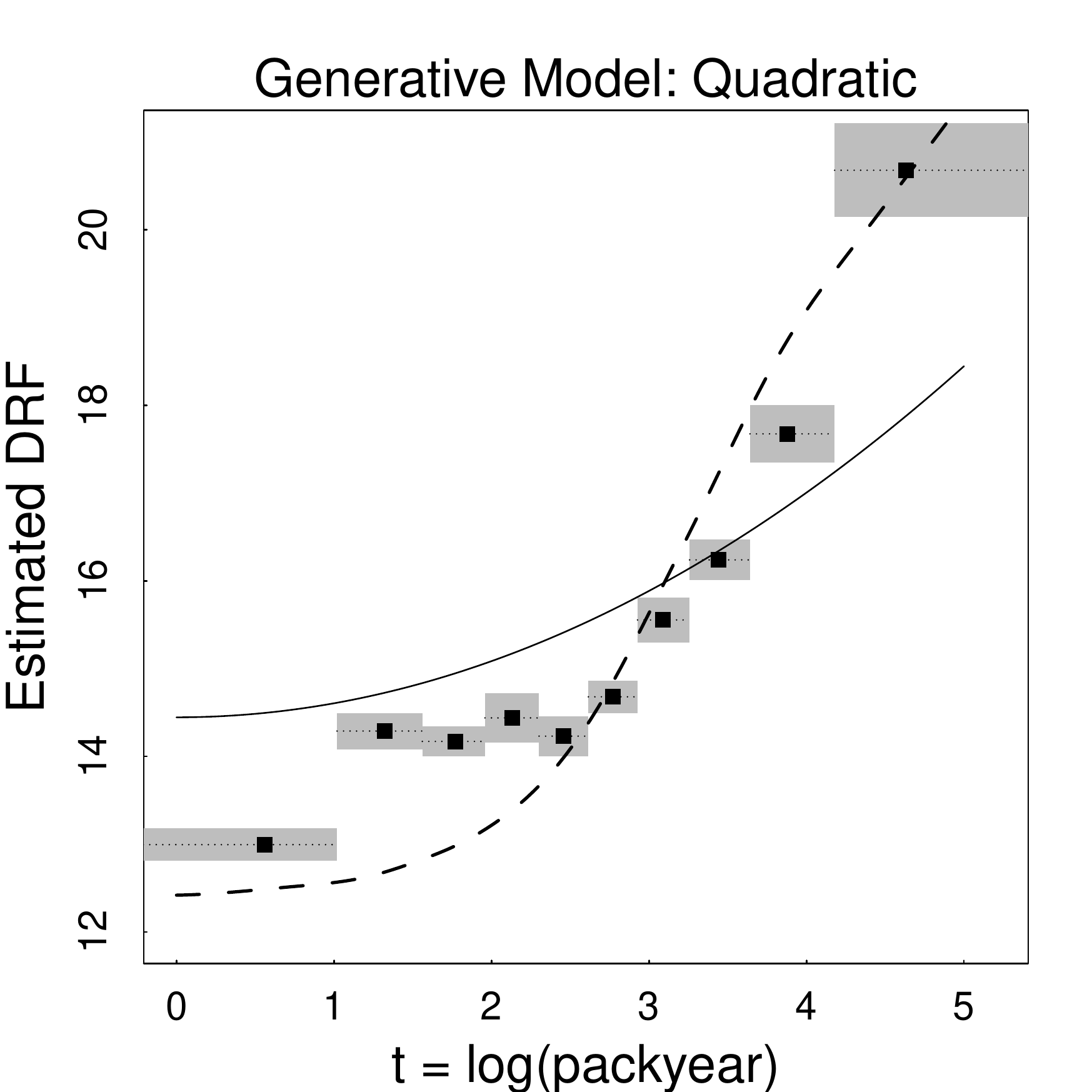}
\includegraphics[width=0.32\textwidth]{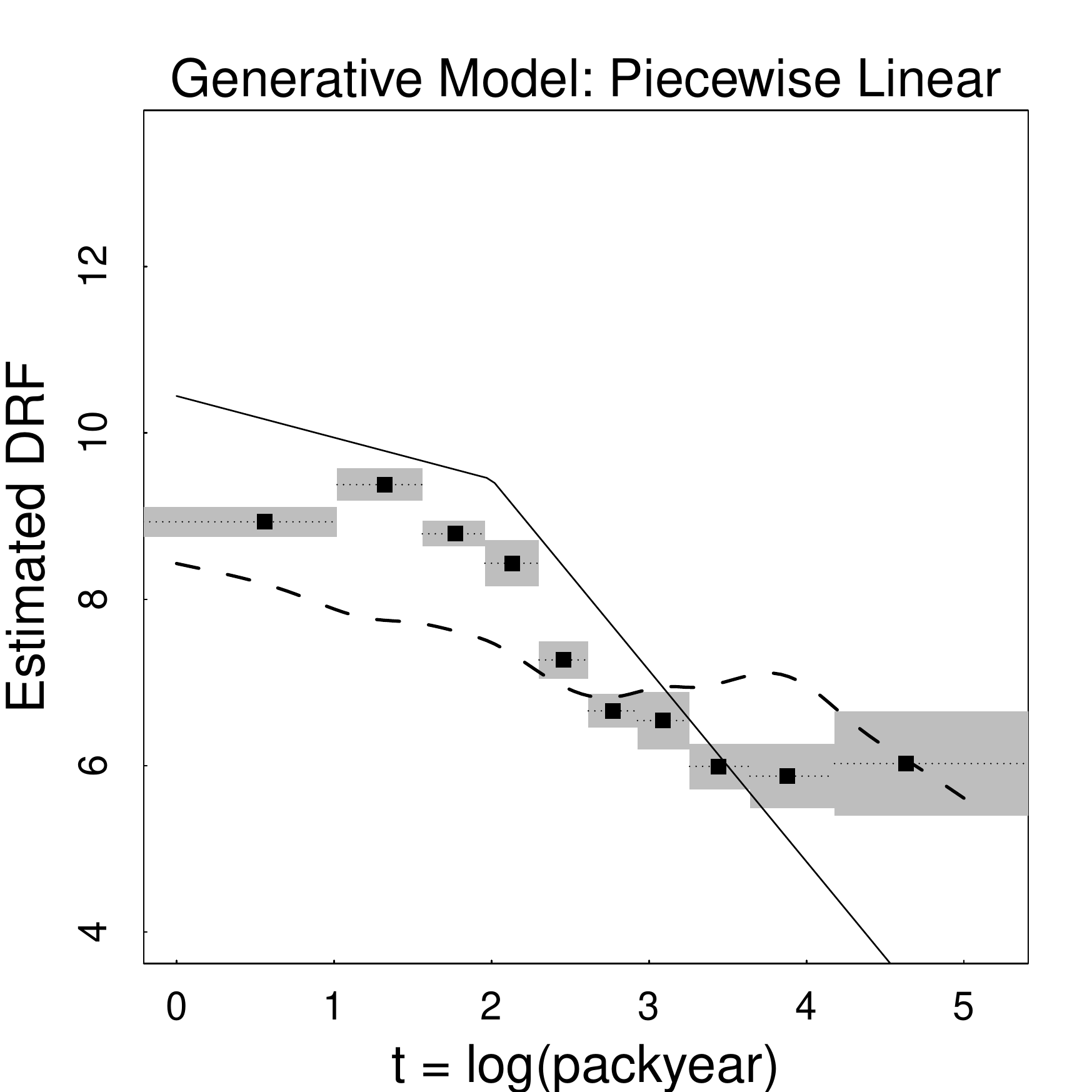}
\includegraphics[width=0.32\textwidth]{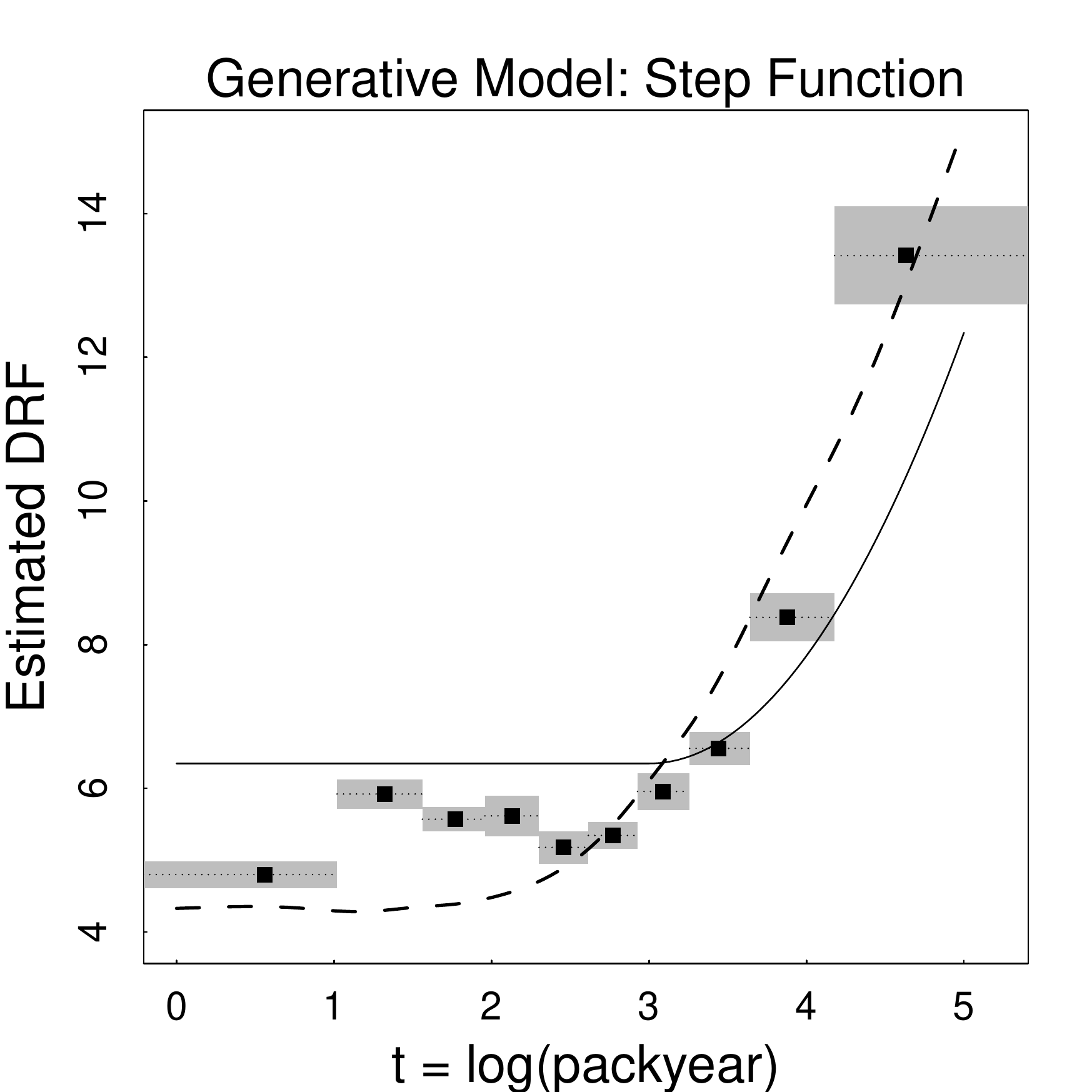}\\
\caption{ Estimated \drf\ for the Simulation Based on Smoking Data Using the
Covariance Adjustment \gps\ Method. The three columns correspond to the
quadratic, piecewise linear, and step function generative model. In all plots
the dotted lines with bullets correspond to the fitted \drf. The true \drf\ is
plotted as solid lines while dashed lines represent the fitted \scm\ of
$\log(Y)$ on $T$, unadjusted for the covariates. Evaluation points are based on
the theoretical quantiles of $\log({\tt packyear})$. The 95\% asymptotic
confidence bands plotted in grey are based on 1000 bootstrap replications. }
\label{fig:smoking_sim_app}
\end{figure}

Figure~\ref{fig:smoking_sim_app} shows the estimated \drf\ for the simulation
based on the applied-example in Section~{\ref{sec:num}}. We used
$S=7,10,$ and $13$ subclasses using linear, quadratic, and SCM fits of $\log(Y)$ on $\hat R$
within each subclass. The results are all similar and we only present those with
$S=10$ using a quadratic model within subclass fit. For this fit, the \drf\ is
evaluated at the midpoint of each subclass, corresponding to the theoretical
5\%, 15\%, \ldots, 95\% quantiles of $\log({\tt packyear})$.
The covariance adjustment \gps\ method exhibits a marked
improvement over the \hi\ method, especially for subclasses that do not
contain the extreme values of $t$. {The general shape of the estimated
\drfs\ is very similar to those estimated with \scm(\gps), see
Figure~{\ref{fig:smoking_sim}}. }

\end{document}